\documentclass[conference]{IEEEtran}
\usepackage{algpseudocode}
\usepackage{amsfonts}
\usepackage{amsmath}
\usepackage{amssymb} 
\usepackage{amsthm}
\usepackage{booktabs}
\usepackage[hypcap = false]{caption}
\usepackage{comment}
\usepackage{color}
\usepackage{colortbl}
\usepackage{enumerate}
\let\labelindent\relax
\usepackage{enumitem}
\usepackage{float}
\usepackage{graphicx}
\usepackage{hyperref}
\hypersetup{
	colorlinks   = true,
	citecolor    = blue
}
\usepackage{makecell}
\usepackage{multirow}
\usepackage{nicematrix}
\usepackage{pgfplots}
\usepackage{ragged2e}
\usepackage{stmaryrd}
\usepackage{subcaption}
\usepackage{tabu}
\usepackage{tabularx}
\usepackage{threeparttable}
\usepackage{tikz}
\usetikzlibrary{patterns}
\usepackage{wrapfig}
\usepackage{xcolor}
\usepackage[framemethod=tikz]{mdframed} 

\newtheorem{theorem}{Theorem}[section]
\newtheorem{lemma}[theorem]{Lemma}

\newcommand{\xor}{\oplus}

\newcommand{\band}{\odot}

\newcommand{\Order}{\mathcal{O}}

\newcommand{\MatMul}{\bigodot}

\newcommand{\Prob}{\ensuremath{\mathsf{Pr}}}

\newcommand{\Z}[1]{\ensuremath{\mathbb{Z}}_{2^{#1}}}

\newcommand{\bitb}{\ensuremath{\mathsf{b}}} 
\newcommand{\arval}[1]{\ensuremath{#1^{\sf R}}} 

\newcommand{\nf}{\ensuremath{\mathsf{n}}} 

\setlist[description]{style=unboxed,leftmargin=0cm}

\newenvironment{myitemize}{
	\begin{list}{{$\bullet$}}{
			\setlength\partopsep{0pt}
			\setlength\parskip{0pt}
			\setlength\parsep{3pt}
			\setlength\topsep{2pt}
			\setlength\itemsep{1pt}
			\setlength{\itemindent}{5pt}
			\setlength{\leftmargin}{1pt}
		}
	}{
	\end{list}
}

\newenvironment{usenix_item_text}{
	\begin{list}{{$\bullet$}}{
			\setlength\partopsep{2pt}
			\setlength\parskip{0pt}
			\setlength\parsep{4pt}
			\setlength\topsep{2pt}
			\setlength\itemsep{0pt}
			\setlength{\itemindent}{12pt}
			\setlength{\leftmargin}{1pt}
		}
	}{
	\end{list}
}

\newenvironment{ccsitemize}{
	\begin{list}{{$\bullet$}}{
			\setlength\partopsep{0pt}
			\setlength\parskip{0pt}
			\setlength\parsep{2pt}
			\setlength\topsep{2pt}
			\setlength\itemsep{0.5pt}
			\setlength{\itemindent}{8pt}
			\setlength{\leftmargin}{1pt}
		}
	}{
	\end{list}
}

\newenvironment{spitemize}{
	\begin{list}{{$\bullet$}}{
			\setlength\partopsep{0pt}
			\setlength\parskip{0pt}
			\setlength\parsep{2pt}
			\setlength\topsep{2pt}
			\setlength\itemsep{2pt}
			\setlength{\itemindent}{22pt}
			\setlength{\leftmargin}{1pt}
		}
	}{
	\end{list}
}

\newenvironment{ccsitem}{
	\begin{list}{{$\bullet$}}{
			\setlength\partopsep{0pt}
			\setlength\parskip{0pt}
			\setlength\parsep{2pt}
			\setlength\topsep{2pt}
			\setlength\itemsep{1pt}
			\setlength{\itemindent}{14pt}
			\setlength{\leftmargin}{1pt}
		}
	}{
	\end{list}
}

\newenvironment{inneritemize}{
	\begin{list}{{$\bullet$}}{
			\setlength\partopsep{0pt}
			\setlength\parskip{0pt}
			\setlength\parsep{1pt}
			\setlength\topsep{2pt}
			\setlength\itemsep{0.5pt}
			\setlength{\itemindent}{2pt}
			\setlength{\leftmargin}{9pt}
		}
	}{
	\end{list}
}

\newlist{myenumlist}{enumerate}{2}
\setlist[myenumlist]{leftmargin=*,label={\arabic*)}}

\newcounter{itemcount}

\newcommand{\tabref}[1]{Table~\protect\ref{tab:#1}}

\newcommand{\figlab}[1]{\label{fig:#1}}
\newcommand{\refeqn}[1]{Eq.~\eqref{#1}}

\newenvironment{boxfig*}[2]{
	\begin{figure*}[h!]		
		\fontsize{5}{5}\selectfont
		\newcommand{\FigCaption}{#1}
		\newcommand{\FigLabel}{#2}
		\vspace{-.05cm}
		\begin{center}
			\begin{small}			 
				\begin{adjustbox}{max width=\textwidth}
					\begin{tabular}{@{}|@{~~}l@{~~}|@{}}
						\hline
						\rule[-1ex]{0pt}{1ex}\begin{minipage}[b]{.95\linewidth}
							\vspace{1ex}	
						}{%
						\end{minipage}\\
						\hline
					\end{tabular}	
				\end{adjustbox}		
			\end{small}
			\vspace{-0.1cm}
			\caption{\FigCaption}
			\figlab{\FigLabel}
		\end{center}
		\vspace{-.38cm}
	\end{figure*}
}

\newenvironment{myboxfig*}[2]{
	\begin{figure*}[!htb]		
		\fontsize{5}{5}\selectfont
		\newcommand{\FigCaption}{#1}
		\newcommand{\FigLabel}{#2}
		\vspace{-.10cm}
		\begin{center}
			\caption{\FigCaption}
			\begin{small}			 
				\begin{adjustbox}{max width=\textwidth}
					\begin{tabular}{@{}|@{~~}l@{~~}|@{}}
						\hline
						\rule[-1ex]{0pt}{1ex}\begin{minipage}[b]{.95\linewidth}
							\vspace{1ex}	
						}{%
						\end{minipage}\\
						\hline
					\end{tabular}	
				\end{adjustbox}		
			\end{small}
			\vspace{-0.25cm}
			\figlab{\FigLabel}
		\end{center}
		\vspace{-.38cm}
	\end{figure*}
}

\newcommand{\boxref}[1]{Fig.~\ref{#1}}

\RequirePackage{mdframed}

\newenvironment{titlebox}[5]
{\mdfsetup{
		style=#2,
		innertopmargin=1.1\baselineskip,
		skipabove={\dimexpr0.2\baselineskip+\topskip\relax},
		skipbelow={1em},needspace=3\baselineskip,
		singleextra={\node[#3,right=10pt,overlay] at (P-|O){~{\sffamily\bfseries #1 }};},%
		firstextra={\node[#3,right=10pt,overlay] at (P-|O) {~{\sffamily\bfseries #1 }};},
		frametitleaboveskip=9em,
		innerrightmargin=5pt
	}
	\newcommand{\TitleCaption}{#4}
	\newcommand{\TitleLabel}{#5}
	\begin{mdframed}[font=\small]
		\setlist[itemize]{leftmargin=13pt}\setlist[enumerate]{leftmargin=13pt}\raggedright%
	}
	{\end{mdframed}
	\vspace{-1.5em} 
	{\captionof{figure}{\small \TitleCaption}\label{\TitleLabel}}
	\smallskip
}

\tikzstyle{normal} = [thick, fill=white, text=black, draw, rounded corners, rectangle, minimum height=.5cm, inner sep=2pt]
\tikzstyle{gray} = [thick, fill=gray!90, text=white, rounded corners, rectangle, minimum height=.7cm, inner sep=3pt]

\mdfdefinestyle{commonbox}{%
	align=center, middlelinewidth=1.1pt,userdefinedwidth=\linewidth
	innerrightmargin=0pt,innerleftmargin=2pt,innertopmargin=5pt,
	splittopskip=7pt,splitbottomskip=3pt
}

\mdfdefinestyle{roundbox}{style=commonbox,roundcorner=5pt,userdefinedwidth=\linewidth}

\tikzstyle{mygray} = [fill=gray!20, text=black, draw, rounded corners, rectangle, minimum height=.7cm, inner sep=3pt]

\mdfdefinestyle{commonboxgray}{%
    tikzsetting={fill=gray!10,fill opacity=0.5},
	align=center, middlelinewidth=1.5pt,userdefinedwidth=\linewidth
	innerrightmargin=0pt,innerleftmargin=2pt,innertopmargin=5pt,
	splittopskip=7pt,splitbottomskip=3pt
}

\mdfdefinestyle{roundboxgray}{
style=commonboxgray,roundcorner=5pt,userdefinedwidth=\linewidth}

\newenvironment{protboxgray}[3]
{\begin{titlebox}{Protocol \normalfont #1}{commonboxgray}{mygray}{#2}{#3}}
	{\end{titlebox}}
	
\newenvironment{systemboxgray}[3]
{\begin{titlebox}{Functionality \normalfont #1}{roundboxgray}{mygray}{#2}{#3}}
	{\end{titlebox}}
	
\newenvironment{simulatorboxgray}[3]
{\begin{titlebox}{Simulator \normalfont #1}{commonboxgray}{mygray}{#2}{#3}}
	{\end{titlebox}}
	
\mdfdefinestyle{commonsplitboxgray}{%
    tikzsetting={fill=gray!10,fill opacity=0.5},
	align=center, middlelinewidth=1.5pt,userdefinedwidth=\linewidth
	innerrightmargin=0pt,innerleftmargin=2pt,innertopmargin=5pt,
	splittopskip=7pt,splitbottomskip=3pt
}

\mdfdefinestyle{roundboxgray}{
style=commonboxgray,roundcorner=5pt,userdefinedwidth=\linewidth}

\mdfdefinestyle{roundsplitboxgray}{
style=commonsplitboxgray,roundcorner=5pt,userdefinedwidth=\linewidth}

\newenvironment{splittitlebox}[5]
{\mdfsetup{
		style=#2,
		innertopmargin=1.1\baselineskip,
		skipabove={\dimexpr0.2\baselineskip+\topskip\relax},
		skipbelow={1em},needspace=3\baselineskip,
		singleextra={\node[#3,right=10pt,overlay] at (P-|O){~{\sffamily\bfseries #1 }};},%
		firstextra={\node[#3,right=10pt,overlay] at (P-|O) {~{\sffamily\bfseries #1 }};},
		frametitleaboveskip=9em,
		innerrightmargin=5pt
	}
	\newcommand{\TitleCaption}{#4}
	\newcommand{\TitleLabel}{#5}
	\begin{mdframed}[font=\small]
		\setlist[itemize]{leftmargin=13pt}\setlist[enumerate]{leftmargin=13pt}\raggedright%
	}
	{\end{mdframed}
	\vspace{-0.5em} 
	{\captionof{figure}{\small \TitleCaption}\label{\TitleLabel}}
	\smallskip
}

\mdfdefinestyle{commonsplitbox}{%
	align=center, middlelinewidth=1.1pt,userdefinedwidth=\linewidth
	innerrightmargin=0pt,innerleftmargin=2pt,innertopmargin=5pt,
	splittopskip=7pt,splitbottomskip=3pt
}

\newenvironment{systembox*}[3]
{\begin{strip}
\vspace{\baselineskip}\begin{titlebox}{Functionality \normalfont #1}{roundbox}{normal}{#2}{#3}}
	{\end{titlebox}
\end{strip}}

\newenvironment{gsystembox*}[3]
{\begin{strip}
\vspace{\baselineskip}\begin{titlebox}{Global Functionality \normalfont #1}{roundbox}{normal}{#2}{#3}}
	{\end{titlebox}
\end{strip}}

\newenvironment{protocolbox*}[3]
{\begin{strip}
\begin{titlebox}{Protocol \normalfont #1}{commonbox}{normal}{#2}{#3}}
	{\end{titlebox}
\end{strip}}

\newenvironment{algobox*}[3]
{\begin{strip}
\begin{titlebox}{Algorithm \normalfont #1}{commonbox}{normal}{#2}{#3}}
	{\end{titlebox}
\end{strip}}

\newenvironment{reductionbox*}[3]
{\begin{strip}
\begin{titlebox}{Reduction \normalfont #1}{commonbox}{normal}{#2}{#3}}
	{\end{titlebox}
\end{strip}}

\newenvironment{gamebox*}[3]
{\begin{strip}
\begin{titlebox}{Game \normalfont #1}{commonbox}{gray}{#2}{#3}}
	{\end{titlebox}
\end{strip}}

\newenvironment{simulatorbox*}[3]
{\begin{strip}
\begin{titlebox}{Simulator \normalfont #1}{commonbox}{normal}{#2}{#3}}
	{\end{titlebox}
\end{strip}}

\mdfdefinestyle{mycommonbox}{%
	align=center, middlelinewidth=1.1pt,userdefinedwidth=\linewidth
	innerrightmargin=0pt,innerleftmargin=2pt,innertopmargin=3pt,
	splittopskip=7pt,splitbottomskip=3pt
}
\mdfdefinestyle{myroundbox}{style=mycommonbox,roundcorner=5pt,userdefinedwidth=\linewidth}

\newenvironment{titlebox*}[5]
{\mdfsetup{
		style=#2,
		innertopmargin=0.3\baselineskip,
		skipabove={0.4em},
		skipbelow={1em},needspace=3\baselineskip,
		frametitleaboveskip=5em,
		innerrightmargin=5pt
	}
	\newcommand{\TitleCaption}{#4}
	\newcommand{\TitleLabel}{#5}
	\begin{mdframed}[font=\small]
		\setlist[itemize]{leftmargin=13pt}\setlist[enumerate]{leftmargin=13pt}\raggedright%
	}
	{\end{mdframed}
	\vspace{-1.2em}
	{\captionof{figure}{\normalfont \TitleCaption}\label{\TitleLabel}}
}

\newenvironment{mysystembox*}[3]
{\begin{strip}
		\vspace{\baselineskip}\begin{titlebox*}{Functionality \normalfont #1}{myroundbox}{normal}{#2}{#3}}
		{\end{titlebox*}
\end{strip}}

\newenvironment{mygsystembox*}[3]
{\begin{strip}
		\vspace{\baselineskip}\begin{titlebox*}{Global Functionality \normalfont #1}{myroundbox}{normal}{#2}{#3}}
		{\end{titlebox*}
\end{strip}}

\newenvironment{myprotocolbox*}[3]
{\begin{strip}
		\begin{titlebox*}{Protocol \normalfont #1}{mycommonbox}{normal}{#2}{#3}}
		{\end{titlebox*}
\end{strip}}

\newenvironment{myalgobox*}[3]
{\begin{strip}
		\begin{titlebox*}{Algorithm \normalfont #1}{mycommonbox}{normal}{#2}{#3}}
		{\end{titlebox*}
\end{strip}}

\newenvironment{myreductionbox*}[3]
{\begin{strip}
		\begin{titlebox*}{Reduction \normalfont #1}{mycommonbox}{normal}{#2}{#3}}
		{\end{titlebox*}
\end{strip}}

\newenvironment{mygamebox*}[3]
{\begin{strip}
		\begin{titlebox*}{Game \normalfont #1}{mycommonbox}{gray}{#2}{#3}}
		{\end{titlebox*}
\end{strip}}

\newenvironment{mysimulatorbox*}[3]
{\begin{strip}
		\begin{titlebox*}{Simulator \normalfont #1}{mycommonbox}{normal}{#2}{#3}}
		{\end{titlebox*}
\end{strip}}

\newenvironment{mytbox}[5]
{\mdfsetup{
		style=#2,
		innertopmargin=1.8\baselineskip,
		skipabove={\dimexpr0.2\baselineskip+\topskip\relax},
		skipbelow={1em},needspace=3\baselineskip,
		singleextra={\node[#3,right=10pt,overlay] at (P-|O){~{\sffamily\bfseries #1 }};},%
		firstextra={\node[#3,right=10pt,overlay] at (P-|O) {~{\sffamily\bfseries #1 }};},
		frametitleaboveskip=9em,
		innerrightmargin=5pt
	}
	\newcommand{\TitleCaption}{#4}
	\newcommand{\TitleLabel}{#5}
	\begin{mdframed}[font=\small]
		\setlist[itemize]{leftmargin=13pt}\setlist[enumerate]{leftmargin=13pt}\raggedright%
	}
	{\end{mdframed}
	\vspace{-1.5em}
	{\captionof{figure}{\small \TitleCaption}\label{\TitleLabel}}
}

\newenvironment{mytsplitbox}[5]
{\mdfsetup{
		style=#2,
		innertopmargin=1.8\baselineskip,
		skipabove={\dimexpr0.2\baselineskip+\topskip\relax},
		skipbelow={1em},needspace=3\baselineskip,
		singleextra={\node[#3,right=10pt,overlay] at (P-|O){~{\sffamily\bfseries #1 }};},%
		firstextra={\node[#3,right=10pt,overlay] at (P-|O) {~{\sffamily\bfseries #1 }};},
		frametitleaboveskip=9em,
		innerrightmargin=5pt
	}
	\newcommand{\TitleCaption}{#4}
	\newcommand{\TitleLabel}{#5}
	\begin{mdframed}[font=\small]
		\setlist[itemize]{leftmargin=13pt}\setlist[enumerate]{leftmargin=13pt}\raggedright%
	}
	{\end{mdframed}
	\vspace{-0.5em}
	{\captionof{figure}{\small \TitleCaption}\label{\TitleLabel}}
}

\newcommand{\algoHead}[1]{\vspace{0.2em} \underline{\textbf{#1}} \vspace{0.3em}}

\makeatletter
\algnewcommand{\ExtendedState}[1]{\State
	\parbox[t]{\dimexpr\linewidth-\ALG@thistlm}{\hangindent=\algorithmicindent\strut\hangafter=3#1\strut}}
\makeatother

\algnewcommand\algorithmicinput{\textbf{Input:}}
\algnewcommand\Input{\item[\algorithmicinput]}

\algrenewcommand{\algorithmiccomment}[1]{{\color{gray}// #1}}

\newcommand{\ckt}{\ensuremath{\mathsf{ckt}}}

\newcommand{\MS}{\ensuremath{\mathsf{M}}}

\newcommand{\wx}{\mathsf{x}}
\newcommand{\wy}{\mathsf{y}}
\newcommand{\wz}{\mathsf{z}}

\newcommand{\negl}{\ensuremath{\mathsf{negl}}}
\newcommand{\csec}{\kappa}

\newcommand{\abort}{\ensuremath{\mathtt{abort}}}

\newcommand{\continue}{\ensuremath{\mathtt{continue}}}
\newcommand{\flag}{\ensuremath{\mathsf{flag}}}

\newcommand{\Adv}{\ensuremath{\mathcal{A}}}
\newcommand{\Advsh}{\ensuremath{\mathcal{A}^{\mathsf{sh}}}}
\newcommand{\Advmal}{\ensuremath{\mathcal{A}^{\mathsf{mal}}}}
\newcommand{\Sim}{\ensuremath{\mathcal{S}}}

\newcommand{\Hash}{\ensuremath{\mathsf{H}}}
\newcommand{\commit}{\ensuremath{\mathsf{Com}}}

\newcommand{\msg}{\mathsf{msg}}

\newcommand{\maxv}{\ensuremath{\mathsf{max}}}

\newcommand{\Key}[1]{\ensuremath{k_{#1}}}

\newcommand{\rtt}{\ensuremath{\mathsf{rtt}}}
\newcommand{\TP}{\ensuremath{\mathsf{TP}}}

\newcommand{\INPUT}{\ensuremath{\mathsf{Input}}}
\newcommand{\OUTPUT}{\ensuremath{\mathsf{Output}}}
\newcommand{\SIGNAL}{\ensuremath{\mathsf{Signal}}}

\newcommand{\FSETUP}{\ensuremath{\mathcal{F}_{\mathsf{setup}}}} 

\newcommand{\vd}{\ensuremath{\mathsf{d}}}
\newcommand{\vc}{\ensuremath{\mathsf{c}}}
\newcommand{\vb}{\ensuremath{\mathsf{b}}}
\newcommand{\va}{\ensuremath{\mathsf{a}}}
\newcommand{\ve}{\ensuremath{\mathsf{e}}}

\newcommand{\vr}{\ensuremath{\mathsf{r}}}
\newcommand{\vx}{\ensuremath{\mathsf{x}}}
\newcommand{\vy}{\ensuremath{\mathsf{y}}}
\newcommand{\vz}{\ensuremath{\mathsf{z}}}

\newcommand{\sfu}{\ensuremath{\mathsf{u}}}
\newcommand{\sfv}{\ensuremath{\mathsf{v}}}

\newcommand{\Mat}[1]{\ensuremath{\mathbf{#1}}}

\newcommand{\trunc}[1]{\ensuremath{{#1}^{d}}}

\newcommand{\sgr}[1]{\ensuremath{\left[#1\right]}}
\newcommand{\sgre}[2]{\ensuremath{\left[#1\right]^{#2}}}

\newcommand{\sqr}[1]{\ensuremath{\langle #1 \rangle}}
\newcommand{\sqrB}[1]{\ensuremath{\langle #1 \rangle}^{\bf B}}

\newcommand{\sqre}[2]{\ensuremath{\langle #1 \rangle^{#2}}}

\newcommand{\tsgr}[1]{\ensuremath{{}^{\Evlset}{\left[ #1 \right]}}}
\newcommand{\tsgra}[2]{\ensuremath{{}^{#2}{\left[ #1 \right]}}}

\newcommand{\tsgrB}[1]{\ensuremath{{}^{\Evlset}{\left[ #1 \right]}^{\bf B}}}

\newcommand{\shrd}{\ensuremath{\llangle \cdot \rrangle}}

\newcommand{\shr}[1]{\ensuremath{\llangle #1 \rrangle}} 
\newcommand{\shrB}[1]{\ensuremath{\llangle #1 \rrangle }^{\bf B}} 

\newcommand{\lv}[1]{\ensuremath{\lambda_{#1}}}  

\newcommand{\val}{\ensuremath{\mathsf{v}}}

\newcommand{\Mult}{\ensuremath{\mathsf{mult}}}

\newcommand{\BitExt}{\ensuremath{\mathsf{bitext}}}
\newcommand{\BitA}{\mathsf{bit2A}}
\newcommand{\BitInj}{\mathsf{BitInj}}

\newcommand{\ReLU}{\ensuremath{\mathsf{ReLU}}}

\newcommand{\MSB}{\ensuremath{\mathsf{msb}}}

\newcommand{\piMult}{\ensuremath{\Pi_{\Mult}}}

\newcommand{\piBitExt}{\ensuremath{\Pi_{\BitExt}}}
\newcommand{\PiBitInj}{\ensuremath{\Pi_{\BitInj}}}

\newcommand{\FMulPre}{\ensuremath{\mathcal{F}_{\mathsf{MulPre}}}}
\newcommand{\FDotPPre}{\ensuremath{\mathcal{F}_{\mathsf{DotPPre}}}}

\makeatletter
\newsavebox{\@brx}
\newcommand{\llangle}[1][]{\savebox{\@brx}{\(\m@th{#1\langle}\)}%
	\mathopen{\copy\@brx\kern-0.5\wd\@brx\usebox{\@brx}}}
\newcommand{\rrangle}[1][]{\savebox{\@brx}{\(\m@th{#1\rangle}\)}%
	\mathclose{\copy\@brx\kern-0.5\wd\@brx\usebox{\@brx}}}
\makeatother

\newcommand{\Evlset}{\ensuremath{\mathcal{E}}}
\newcommand{\Hlpset}{\ensuremath{\mathcal{D}}}
\newcommand{\Pking}{\ensuremath{P_{\mathsf{king}}}}		
\newcommand{\PiRecfair}{\ensuremath{\Pi_{\mathsf{Rec}}^{\mathsf{fair}}}}

\newcommand{\Fmpc}{\ensuremath{\mathcal{F}_{\mathsf{n-PC}}}}
\newcommand{\Fmpcmal}{\ensuremath{\mathcal{F}_{\mathsf{n-PC}}^{\mathsf{mal}}}}
\newcommand{\Fzk}{\ensuremath{\mathcal{F}^{\abort}_{\mathsf{proveDeg2Rel}}}}

\newcommand{\Pisqtosh}{\ensuremath{\mathrm{\Pi}_{\tiny {\sqr{\cdot} \rightarrow \shr{\cdot}}}}}
\newcommand{\Pishtosq}{\ensuremath{\mathrm{\Pi}_{\tiny{\shr{\cdot} \rightarrow \sqr{\cdot}}}}}
\newcommand{\Piptosh}{\ensuremath{\mathrm{\Pi}_{\tiny{\cdot \rightarrow \shr{\cdot}}}}}
\newcommand{\Piptoshf}{\ensuremath{\Pi_{\tiny{\cdot \rightarrow \shr{\cdot}}}}}

\newcommand{\T}{\ensuremath{{\mathcal{T}}}}

\newcommand{\A}{\ensuremath{\mathsf{A}}}
\newcommand{\h}{\ensuremath{\mathsf{q}}}
\newcommand{\g}{\ensuremath{\mathsf{g}}}
\newcommand{\e}{\ensuremath{\mathsf{e}}}

\newcommand{\Picta}{\ensuremath{\Pi_{\tiny{\sqr{\cdot} \rightarrow \tsgr{\cdot}}}}}
\newcommand{\Pictaa}{\ensuremath{\mathrm{\Pi}_{\tiny{\sqr{\cdot} \rightarrow \tsgra{\cdot}{\T}}}}}
\newcommand{\Pictaf}{\ensuremath{\mathrm{\Pi}_{\tiny{\sqr{\cdot} \rightarrow \sgr{\cdot}}}}}
\newcommand{\Pictashr}{\ensuremath{\Pi_{\tiny{\shr{\cdot} \rightarrow \tsgr{\cdot}}}}}
\newcommand{\Pictashra}{\ensuremath{\mathrm{\Pi}_{\tiny{\shr{\cdot} \rightarrow \tsgra{\cdot}{\T}}}}}
\newcommand{\Pictashrf}{\ensuremath{\mathrm{\Pi}_{\tiny{\shr{\cdot} \rightarrow \sgr{\cdot}}}}}
\newcommand{\Picprod}{\ensuremath{\mathrm{\Pi}_{\tiny{\sqr{\cdot} \cdot \sqr{\cdot} \rightarrow \sgr{\cdot}}}}}
\newcommand{\Picon}{\ensuremath{\mathrm{\Pi}_{\mathsf{agree}}}}

\newcommand{\Cor}{\ensuremath{\mathcal{C}}}

\newcommand{\PiSh}{\ensuremath{\Pi_{\mathsf{Sh}}}}
\newcommand{\PiRSh}{\ensuremath{\mathrm{\Pi}_{\sqr{\cdot}}}}
\newcommand{\PiMult}{\ensuremath{\Pi_{\mathsf{mult}}}}
\newcommand{\PitMult}{\ensuremath{\Pi_{\mathsf{3\mbox{-}mult}}}}
\newcommand{\PifMult}{\ensuremath{\Pi_{\mathsf{4\mbox{-}mult}}}}

\newcommand{\PiDotP}{\ensuremath{\Pi_{\mathsf{dp}}}}

\newcommand{\PiRandR}{\ensuremath{\mathrm{\Pi}_{\mathsf{rand}}}}
\newcommand{\PiRandRP}{\ensuremath{\mathrm{\Pi}_{\mathsf{pRand}}}}

\newcommand{\concat}{||}

\newcommand{\PiDSBits}{\ensuremath{\Pi_{\mathsf{dsBits}}}}

\newcommand{\PiED}{\ensuremath{\Pi_{\mathsf{ED}}}}
\newcommand{\PiSSQ}{\ensuremath{\Pi_{\mathsf{SSQ}}}}

\newcommand{\seq}{\ensuremath{\mathsf{s}}}
\newcommand{\query}{\ensuremath{\mathsf{q}}}

\newcommand{\seql}{\ensuremath{\omega}}
\newcommand{\seqn}{\ensuremath{m}}

\newcommand{\lut}{\ensuremath{\mathsf{LUT}}}
\newcommand{\dist}{\ensuremath{\mathsf{d}}}

\newcommand{\PiEq}{\ensuremath{\Pi_{\mathsf{Eq}}}}

\newcommand{\graph}{\ensuremath{\mathsf{G}}}
\newcommand{\vertices}{\ensuremath{\mathsf{V}}}
\newcommand{\edges}{\ensuremath{\mathsf{E}}}

\newcommand{\alive}{\ensuremath{\mathsf{alive}}}

\newcommand{\vu}{\ensuremath{\mathsf{u}}}

\newcommand{\PiMultPre}{\ensuremath{\Pi_{\mathsf{multPre}}}}
\newcommand{\PiMultMal}{\ensuremath{\Pi_{\mathsf{mult}}^{\mathsf{M}}}}
\newcommand{\PiShMal}{\ensuremath{\Pi_{\mathsf{Sh}}^}{\mathsf{M}}}
\newcommand{\PiDotPre}{\ensuremath{\Pi_{\mathsf{dotPre}}}}

\newcommand{\PiDSBitsMal}{\ensuremath{\Pi_{\mathsf{dsBits}}^{\mathsf{M}}}}

\newcommand{\Pivrfy}{\ensuremath{\Pi_{\mathsf{Vrfy}}}}

\newcommand{\istr}{\ensuremath{\mathsf{isTr}}}

\newcommand{\this}{\ensuremath{\boldsymbol{\mathsf{This}}}}

\newcommand{\simsh}{\ensuremath{\Sim^{\mathsf{sh}}}}
\newcommand{\simmal}{\ensuremath{\Sim^{\mathsf{mal}}}}
\newcommand{\kbits}{\ensuremath{\mathsf{k}}}

\newcommand{\PictashrB}{\ensuremath{\Pi_{\tiny{\shr{\cdot} \rightarrow \tsgr{\cdot}}}^{\bf B}}}
\newcommand{\PiptoshB}{\ensuremath{\Pi_{\tiny{\cdot \rightarrow \shr{\cdot}}}^{\bf B}}}

\newcommand{\D}{\ensuremath{\mathsf{D}}}
\newcommand{\E}{\ensuremath{\mathsf{E}}}
\newcommand{\Ab}{\ensuremath{\mathsf{P}}}

\newcommand{\Ftrgen}{\ensuremath{\mathcal{F}_{\mathsf{TrGen}}}}
\newcommand{\Ftrgenmal}{\ensuremath{\mathcal{F}_{\mathsf{TrGen}}^{\mathsf{M}}}}

\newcommand{\clv}[1]{\ensuremath{\Lambda_{#1}}}  
\newcommand{\cmv}[1]{\ensuremath{{\text{M}}_{#1}}}  

\newcommand{\mynum}[1]{\ensuremath{\mathsf{(#1)}}}

\newcommand{\ED}[2]{\ensuremath{\mathbf{EuD}_{#1#2}}}

\newcommand{\DV}{\ensuremath{\mathbf{DV}}}

\newcommand{\Partyset}{\ensuremath{\mathcal{P}}}

\newcommand{\abs}[1]{| #1 |}

\newcommand{\bool}[1][\relax]{\ensuremath{{\{0,1\}}^{#1}}}

\newcommand{\poly}{\ensuremath{\mathsf{poly}}}

\newcommand{\piab}{\ensuremath{\mathrm{\Pi}_{\mathsf{A2B}}}}

\newcommand{\PiBitA}{\ensuremath{\mathrm{\Pi}_{\BitA}}}

\newcommand{\pizero}{\ensuremath{\mathrm{\Pi}_{[0]}}}

\newcommand{\mv}[1]{\ensuremath{\mathsf{m}_{#1}}}    

\definecolor{UniBlau}{cmyk}{1,0.7,0,0}
\definecolor{UniGruen}{cmyk}{0.6,0,1,0}
\definecolor{UniOrange}{cmyk}{0,0.3,1,0}
\definecolor{UniRot}{cmyk}{0.4,1,0,0}
\definecolor{darkred}{rgb}{.6,0,0}
\definecolor{darkgreen}{rgb}{0,.4,0}
\definecolor{darkblue}{rgb}{0,0,.6}

\definecolor{LightGray}{gray}{0.97}

\newcommand{\circled}[1]{\tikz[baseline=(char.base)]{
            \node[shape=circle,draw,inner sep=0.5pt] (char) {#1};}}

\newcommand{\bfcircled}[1]{\circled{\textbf{#1}}}

\newcommand{\negspace}[1]{\indent\vspace{#1}}
\newcommand{\negspaceqtr}{\negspace{-0.25cm}}

\newif\ifsubmission
\submissionfalse
\submissiontrue

\ifsubmission
	\newcommand{\commentA}[1] {}
	\newcommand{\commentAJ}[1]{}
	\newcommand{\commentS}[1] {} 
	\newcommand{\commentN}[1] {}
	\newcommand{\EXTRALINES}[1] {}

\else
	\newcommand{\commentA}[1] {\textcolor{blue}  {{\sf (}{\sl{#1}} {\sf - Arpita)}}}
	\newcommand{\commentAJ}[1]{\textcolor{violet}{{\sf (}{\sl{#1}} {\sf - Ajith)}}}
	\newcommand{\commentS}[1] {\textcolor{olive} {{\sf (}{\sl{#1}} {\sf - Shravani)}}}
	\newcommand{\commentN}[1] {\textcolor{cyan} {{\sf (}{\sl{#1}} {\sf - Nishat)}}}
	\newcommand{\EXTRALINES}[1] {\textcolor{darkblue} {#1}}

\fi

\begin{document}
\date{}
\title{MPClan: Protocol Suite for Privacy-Conscious Computations}
%
\author{
	\IEEEauthorblockN{Nishat Koti\IEEEauthorrefmark{1}, 
	Shravani Patil\IEEEauthorrefmark{1},
	Arpita Patra\IEEEauthorrefmark{1},
		Ajith Suresh\IEEEauthorrefmark{2}, 
		}
	\IEEEauthorblockA{\IEEEauthorrefmark{1}Indian Institute of Science, Bangalore, Email: \{kotis, shravanip, arpita\}@iisc.ac.in}
	\IEEEauthorblockA{\IEEEauthorrefmark{2}Technical University of Darmstadt, Germany, Email: suresh@encrypto.cs.tu-darmstadt.de}
}

\maketitle

\begin{abstract}
The growing volumes of data being collected and its analysis to provide better services are creating worries about digital privacy. To address privacy concerns and give practical solutions, the literature has relied on secure multiparty computation. However, recent research has mostly focused on the small-party honest-majority setting of up to four parties, noting efficiency concerns. In this work, we extend the strategies to support a larger number of participants in an honest-majority setting with efficiency at the center stage.

Cast in the preprocessing paradigm, our semi-honest protocol improves the online complexity of the decade-old state-of-the-art protocol of Damg\aa rd and Nielson (CRYPTO'07). In addition to having an improved online communication cost, we can shut down almost half of the parties in the online phase, thereby saving up to 50$\%$ in the system's operational costs. Our maliciously secure protocol also enjoys similar benefits and requires only half of the parties, except for one-time verification, towards the end.

To showcase the practicality of the designed protocols, we benchmark popular applications such as deep neural networks, graph neural networks, genome sequence matching, and biometric matching using prototype implementations. Our improved protocols aid in bringing up to 60-80$\%$ savings in monetary cost over prior work. 
\end{abstract}

\section{Introduction} 
\label{sec:intro}
Today's world is seeing a visible transition from offline services to a heavy dependency on online platforms for banking, socializing, healthcare, etc. This is leading to an increased user presence online, which leaves a trail of online activity and personal data over the Internet. 
The availability of such user-specific data opens up possibilities for its misuse. For instance, there has been a lot of concern raised regarding advertisement service providers such as Google, Facebook breaching user privacy for targeted advertisement services~\cite{signal}. In the process of providing enhanced targeted advertisement services, service providers are allegedly learning more information about their users than they are entitled to (e.g., user's shopping activity, browsing history) from various data collection entities. These entities collect user data via website cookies, loyalty cards, etc.~\cite{datacollection}. While such targeted advertisements offer a personalized online experience, they may come at the cost of revealing unauthorized user data to these service providers. Such a challenge is also encountered in the healthcare sector. Collaborative analysis among healthcare institutes over patient data is known to facilitate better diagnosis and improved treatment. However, laws such as GDPR, which prevent sharing of patient records, hinder such collaborations, thereby re-emphasizing the need for mechanisms that enable privacy-preserving computations. 

Such mechanisms that ensure privacy-preserving computations can be facilitated 
via several privacy-enhancing technologies such as homomorphic encryption~\cite{HE1,HE2}, differential privacy~\cite{dp}, secure multiparty computation~\cite{Yao82,Ben-OrGW88,GoldreichMW87}, to name a few. We focus on secure multiparty computation (MPC) as it has been the cornerstone of research lately, showcasing its effectiveness in various applications such as privacy-preserving machine learning~\cite{SWIFT, MR18, Falcon}, secure collaborative analytics~\cite{senate}, secure genome matching~\cite{ST19, AHLR18}, etc.  Essentially, it offers a solution to the potential privacy issues which may arise in collaborative computation scenarios such as targeted advertisements described earlier.
MPC allows mutually distrusting parties to perform computations on their private inputs such that they learn nothing beyond the output of the computation. The distrust among the parties is captured by the notion of a centralized adversary, which is said to corrupt up to $t$ out of the $n$ participating parties. Depending on its behaviour, the adversary can be categorized as either {\em semi-honest} or {\em malicious}~\cite{goldreich09}. Semi-honest adversary models the corruption scenario where the corrupt parties are restricted to follow the protocol and cannot deviate arbitrarily, as in the stronger notion of malicious corruption. 

\begin{figure}[htb!]
    \centering
	\includegraphics[width=0.8\columnwidth]{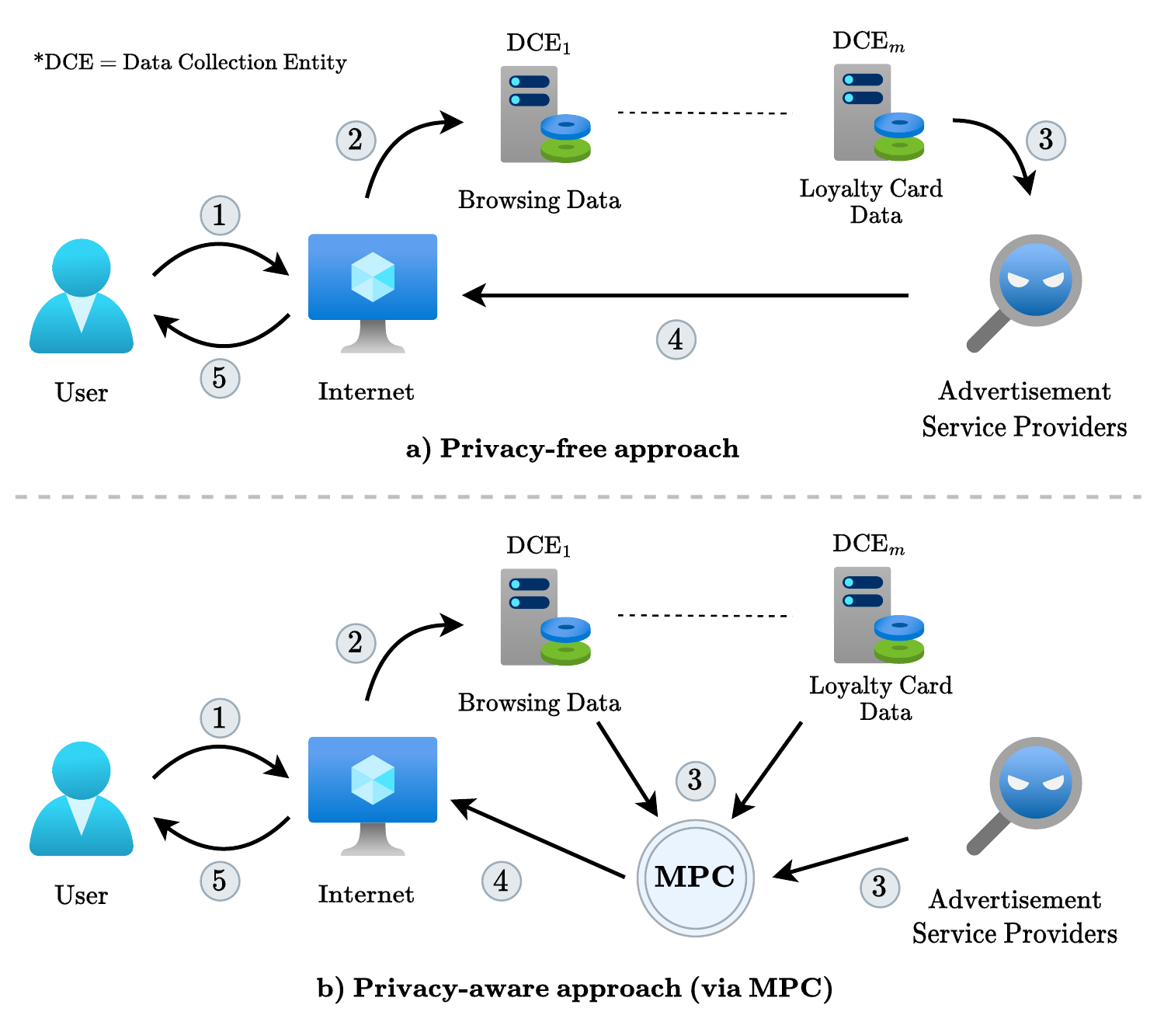}
	\vspace{-3mm}
	\caption{\small Use case for privacy-conscious solutions} \label{fig:usecase}
	\vspace{-7mm}
\end{figure}

MPC with honest majority, where only a minority of the parties are corrupt, enables construction of efficient protocols for multiple parties~\cite{BonehBCGI19,DamgaardN07,AbspoelDEN19,GoyalS20,blanton2020improved,senate}. The recent concretely efficient protocols have only considered small number of parties~\cite{SWIFT, BLAZE, Trident, Falcon, MR18,DEK20, MLRG20,MPCLeague}, which restricts the number of corruptions to at most one ($t = 1$). Although the small-party setting has found application in the outsourced computation paradigm too, the generic {\em multiparty} setting is a better fit for real-world deployments due to its resiliency to a higher number of corruptions ($t < n/2$). Thus, for larger $n$, the number of corruptions that can be tolerated is also higher, thereby increasing the trust in the system. Moreover, multiparty setting allows for privacy-conscious computations even in a non-outsourced deployment scenario, such as in providing targeted advertisement services (described in \boxref{fig:usecase} and elaborated below), when outsourcing the computation is not feasible/preferable. Hence, to design efficient protocols, we focus on honest majority multiparty computation. 

\paragraph{Use Case} Consider the scenario of targeted advertisement services depicted in \boxref{fig:usecase}(a). Typically, data collection entities track a user's online activities via website cookies while browsing the Internet (\bfcircled{1}). Also known as cookie profiling, such data collection allows the entities to create a ``profile'' for each user, which may contain information such as browsing habits, gender, marital status, and age, to name a few, as shown in \bfcircled{2}. These profiles can facilitate targeted advertisements via specialized algorithms (\bfcircled{4}), which is leveraged by the advertisement service providers such as Google and Facebook. While such services offer a personalized experience, it comes at the expense of users' private data being revealed to the service providers, as indicated in \bfcircled{3}. A feasible solution (\boxref{fig:usecase}(b)) instead is to place a solution box at the interface between these service providers and the data collection entities such that it provides mechanisms to ensure the privacy of user data while also facilitating the required computations over the same (to provide targeted advertisement services). MPC being a technology that supports privacy-preserving computations, lends itself well to such tasks. Instead of the data collection entities directly revealing the user data to the advertisement service providers, they can engage in an instance of MPC protocol (\bfcircled{3}) which securely runs the required algorithm on the user data while maintaining its privacy. Moreover, such a computation does not require the data collection entities to reveal their data to each other, thus offering a viable solution. Furthermore, as studied in \cite{guido2011targeting}, the effectiveness of targeted advertisements can greatly benefit from the use of machine learning algorithms. In particular, neural networks, and more recently graph neural networks~\cite{nnad1,nnad2,gnnad1,gnnad2,gnnad3} have shown the potential to better analyse the data available via user profiles, in turn allowing for a refined personalized experience. We thus focus on protocols for securely evaluating the standard neural networks such as VGG16 \cite{vgg16} (deep neural network) and graph neural network, and provide benchmarks for the same in Section~\ref{sec:Implementation}. 

\subsection{Related work}
\label{subsec:relatedwork}
We restrict related work to MPC protocols in honest-majority setting. 
%
Despite the interest in MPC for small population~\cite{AFLNO16, ABFLLNOWW17, FLNW17, CGHIKLN18, AbspoelDEN19, ASTRA, BLAZE, Trident, FLASH, SWIFT, Falcon, DEK20}, MPC protocols for arbitrary number of parties have been studied largely~\cite{ED20,DamgaardN07,GoyalS20,BBY20,BLO16,BLO17,BGIN20,BonehBCGI19,motion,RT19,zaphod,manticore,blanton2020improved,GSY21}. In the honest majority ($t<n/2$) semi-honest setting,~\cite{DamgaardN07, GIPST14} forms the state of the art MPC protocols over fields in the information theoretic setting.  This was further optimized in the computational setting in~\cite{BonehBCGI19} using a one-time setup for correlated randomness. 
We will often refer to this optimized honest-majority semi-honest protocol of~\cite{DamgaardN07} as DN07. In the information-theoretic setting, the work of~\cite{GLOPS21}, improves upon the communication and round complexity of~\cite{DamgaardN07}.  
The work of ~\cite{ED20} recently demonstrates MPC protocols in the honest majority setting in the preprocessing model with malicious security, which requires communicating $3t$ field elements in the online as well as the preprocessing phase. We observe that the semi-honest protocol derived from this requires communicating $2t$ elements in the online and $3t$ elements in the preprocessing phase. 
The recent work of \cite{blanton2020improved, BBY20} provides semi-honest MPC protocols which require {\em each} party to communicate roughly $t$ elements per multiplication gate, resulting in quadratic communication in the number of parties. 
DN07 has served as the basis for obtaining malicious security for free (i.e. amortized communication cost of $3t$ elements per multiplication gate) in the computational setting~\cite{BonehBCGI19, BGIN20} as well as in the information-theoretic setting~\cite{GoyalS20, GLOPS21}. Both \cite{GoyalS20} and \cite{BGIN20} follow the approach of executing a semi-honest protocol, followed by a verification phase to check the correctness of multiplication which involves heavy polynomial interpolation operations. 
As mentioned earlier, the recent work of~\cite{ED20} focuses on maliciously secure protocols for honest-majority setting in the preprocessing model. Their protocol relies on an instantiation of \cite{GoyalS20} in the preprocessing phase that requires communicating $3t$ elements while requiring another $3t$ element communication in the online phase. However, their protocol is inefficient due to a consistency check required after each level of multiplication and introduces depth-dependent overhead in communication complexity. The absence of this check results in a privacy breach as described in~\cite{GoyalLS19} and is elaborated in \S\ref{appsec:multattack}.

\subsection{Towards practically efficient protocols}
\label{subsec:designchoices}

Before stating our contributions, we elaborate on the choices made in designing a practically efficient protocol. 
\begin{usenix_item_text}
	\item[1.] \textit{Preprocessing paradigm.} With the goal of attaining as fast a response time as possible, the protocols are cast in the preprocessing paradigm~\cite{DPSZ12,SPDZ2,SPDZ3,KOS16,BaumDTZ16,DamgardOS18,CramerDESX18,RiaziWTS0K18,KellerPR18, BLAZE, Trident}. Here, expensive {\em data-independent} computations are carried out in a preprocessing phase, thereby making way for a fast and efficient {\em data-dependent} online phase. We thus focus on improving the online phase without hampering the overall protocol complexity. 
	\item[2.] \textit{Algebraic structure.} To further enhance efficiency by utilizing the underlying CPU architecture, several protocols work over rings~\cite{MR18, SWIFT, Trident, Falcon, Tetrad, MLRG20}. We follow this approach and design MPC protocols operating over the ring $\Z{\ell}$ and rely on replicated secret sharing (RSS). Note that usage of RSS inherently results in exponential blow-up in the number of shares for an arbitrary number of parties. Hence, it is well-suited for the practically-oriented scenarios comprising of a constant number of parties~\cite{BonehBCGI19, BGIN20}, which we restrict to for benchmarking our protocols.
	\item[3.] \textit{Masked evaluation.}  To make our protocols efficient in the preprocessing paradigm, we use the masked evaluation paradigm, a variant of the replicated secret sharing scheme. The secret data is masked using a masking value in this case, and the mask is RSS shared. The computation is done on the publicly available masked values and the shared masks. This technique was first introduced in the context of circuit garbling schemes (see~\cite{LPNY15, WRK17}), and was then adapted to secret sharing-based protocols in dishonest majority (see~\cite{KKW18, turbospdz19}). It was later applied to small-population honest-majority settings such as~\cite{GRW18, ASTRA, BLAZE, DEK20, SWIFT} and~\cite{Tetrad} to aid in the development of practically efficient protocols.
	%
	\item[4.] \textit{Adversarial strategy.} Based on the deployment scenario, different levels of security may be desired. While semi-honest security suffices for several applications as shown in~\cite{AFLNO16, LVBJV16, ASTRA, MohasselZ17, AHLR18, ST19, CDGOSS21, SGA21}, malicious security is always desirable. Thus, to cater to different scenarios, our protocols are designed to provide semi-honest and malicious security, where each security goal has its merit.
	\item[5.] \textit{Monetary cost.} To reduce the operational costs in the online phase, several recent works~\cite{BLAZE, SWIFT, Trident, Tetrad} reduce the number of (online) computing parties. This is useful in long computations such as those involved in privacy-preserving machine learning (PPML) applications, which span several days or even weeks. Reducing the number of online parties is especially advantageous for protocols deployed in the secure outsourced computation (SOC) setting since one has to pay for the up-time of every hired server. Shutting down even a single server significantly helps in reducing the monetary cost~\cite{MP0SY20, Tetrad} of the system. We thus focus on ensuring the participation of a minimal number of parties during the online computation in our protocols. This is achieved for the first time in generic $n$-party protocols
	\footnote{A recent work~\cite{ED20} also claims to achieve this reduction in online parties. However, their protocol suffers from a privacy breach as explained in \S\ref{appsec:multattack}.}. Specifically, all the protocols for the semi-honest setting in our framework benefit from using only $t+1$ parties in the online phase. The protocols in the malicious setting also enjoy this benefit except that the remainder $t$ parties are required to come online for a short verification phase at the end. The reduction in online parties aids in improving the operational cost of the framework by almost 50$\%$. This is unlike prior works~\cite{DamgaardN07, BonehBCGI19, BGIN20, GoyalS20, GLOPS21} which require active participation from all parties throughout the computation. 
\end{usenix_item_text}

\subsection{Our Contributions}
We begin with a quick overview of the contributions of this work, followed by the details.
\begin{myitemize}
    \item We construct an $n$-party semi-honest protocol in the preprocessing paradigm which offers an improved online phase than the decade-old state-of-the-art protocol of \cite{DamgaardN07}, without inflating its total cost. Moreover, our protocol reduces the number of active parties in the online phase, thereby improving the system's operational cost when deployed in SOC setting.
    \item We extend our semi-honest protocol to the malicious setting, while retaining the benefits of requiring reduced number of parties in online phase for majority of the computation. Our offer over state-of-the-art protocol of \cite{ED20} is a stronger security guarantee of fairness, and $\Order(d)$ improvement in round complexity. Here, $d$ denotes depth of the circuit to be evaluated.
    \item We provide support for $3$ and $4$ input multiplication, at the same online complexity as that of the $2$ input multiplication. In addition to improving the communication cost over the approach of sequential multiplications, multi-input multiplication offers a $2\times$ improvement in the round complexity which is beneficial for high latency networks.
    \item We design building blocks for a range of applications such as deep neural networks, graph neural networks, genome sequence matching and biometric matching. When the applications are benchmarked, our semi-honest protocol witnesses a saving of up to $69\%$ in monetary cost, and has $3.5\times$ to $4.6\times$ improvements in online run time and throughput over \cite{DamgaardN07}. Interestingly, our maliciously secure protocols outperforms the semi-honest protocol of \cite{DamgaardN07} in terms of online run time and throughput for the applications under consideration, achieving the goal of fast online phase.
\end{myitemize}

We now elaborate on the contributions and highlight the technical details and novelty of our work.

\begin{figure}[htb!]
	\includegraphics[width=\columnwidth]{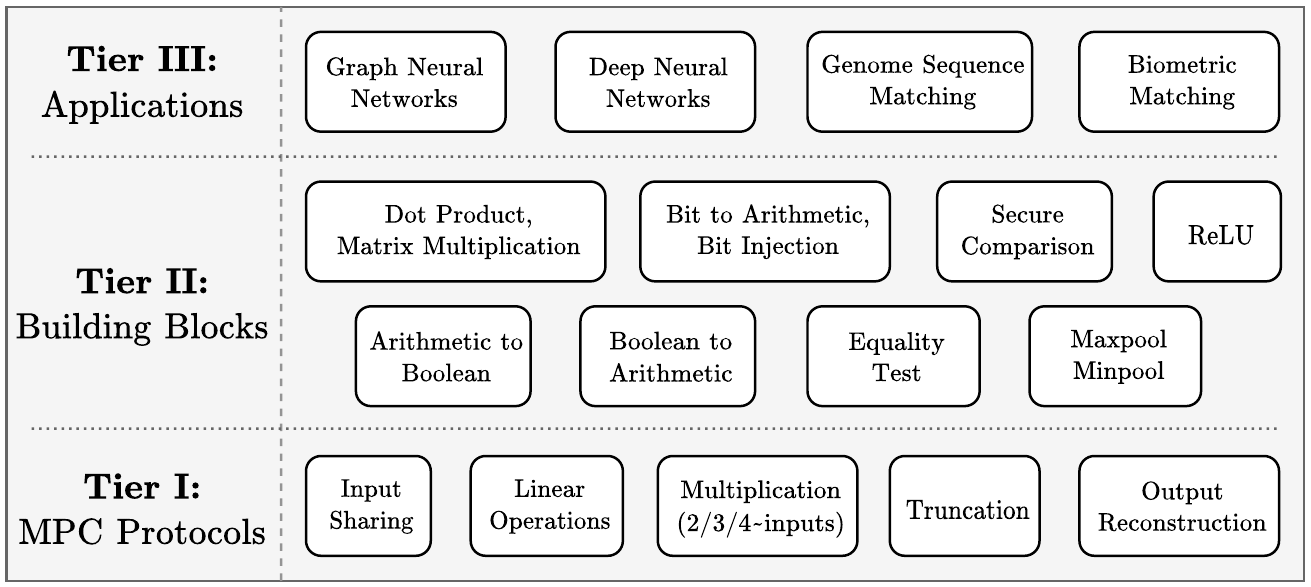}
	\vspace{-5mm}
	\caption{\small Hierarchy of primitives in our 3-tier framework} \label{fig:architecture}
	\vspace{-5mm}
\end{figure}

Our protocol suite follows a 3-tier architecture (\boxref{fig:architecture}) to attain the final goal of privacy-conscious computations. The first tier comprises fundamental primitives such as input sharing, reconstruction, multiplication (with truncation), and multi-input multiplication. The second tier includes building blocks such as dot product, matrix multiplication, conversion between Boolean and arithmetic worlds, comparison, equality, non-linear activation functions, to name a few, as required in the applications considered. Finally, the third tier is applications. Our main contribution lies in Tier I and is detailed below.

\subsubsection{Tier I - MPC protocols}
Our goal is to design protocols with a fast online phase. Thus, working over $\Z{\ell}$ and relying on RSS, we design a semi-honest MPC protocol in the computational setting assuming a one-time shared-key setup for correlated randomness. 

Note that the straightforward extension of semi-honest multiplication protocol of \cite{DamgaardN07} to the preprocessing model, which can also be derived from the recent work of \cite{ED20}, incurs a communication of $3t$ elements in the preprocessing phase while communicating $2t$ elements in the online. This amounts to a $1.6\times$ overhead in the total cost over \cite{DamgaardN07}.
Our contribution lies in ensuring a fast online phase, without inflating the total communication cost of the protocol. Specifically, our protocol requires communicating only $2t$ ring elements in the online phase and $t$ in the preprocessing, for a multiplication gate. We are the first to achieve a communication cost of $2t$ in the online phase (unlike $3t$ in the prior works~\cite{DamgaardN07, GIPST14}), without incurring any overhead in the total cost, i.e., our total cost still matches that of the best known (optimized) semi-honest honest-majority protocol~\cite{DamgaardN07, GIPST14}. 

We extend our protocol to provide malicious security with {\em fairness}\footnote{Guarantees either all parties receive the output or none do.} at the cost of additionally communicating $t$ elements in the online phase and $2t$ in the preprocessing phase. 
Although ({\em abort}\footnote{Honest parties may not receive the output while corrupt parties do.}) protocol of \cite{ED20} has the same communication as our maliciously secure protocol, we achieve a stronger security notion of fairness. Moreover, \cite{ED20} requires an additional round of communication for consistency checks after each level, the absence of which results in a privacy breach (described in \cite{GoyalLS19} and elaborated in \S\ref{appsec:multattack}), and necessitates participation from all parties. However, by relying on a variant of RSS, our protocol avoids the consistency check after each level of circuit evaluation and ensures privacy. Notably, we only require participation from all parties for a one-time verification at the end of evaluation, thus reducing the number of rounds by $d$ ($d$ denotes circuit depth). 

\paragraph*{3 and 4 input multiplications}
Following \cite{aby2, ohata20, Tetrad}, to reduce the online communication cost and round complexity, we design protocols to enable the multiplication of 3 and 4 inputs in a single shot. Compared to the naive approach of performing sequential multiplications to multiply 3/4 inputs, the \textit{multi-input multiplication} protocol enjoys the benefit of having the same online phase complexity as that of the 2-input multiplication protocol. This brings in a $2\times$ improvement in the online round complexity and improves the online communication cost. Support for multi-input multiplication enables usage of optimized adder circuits~\cite{aby2} for secure comparison and Boolean addition, thereby resulting in a faster online phase. 
The recent work of~\cite{GLOPS21} also proposes a method to improve the round complexity of circuit evaluation by evaluating all gates in two consecutive layers in a circuit in parallel. We observe that their method can be viewed as a variant of multi-input multiplication with 3 and 4 inputs. Thus, our protocols need not be limited to facilitate faster comparison and Boolean additions alone (as described above), but can be used to reduce the round and communication complexity of any general circuit evaluation. Note that~\cite{GLOPS21} only improves the round complexity (2$\times$) without inflating the communication cost when compared to~\cite{DamgaardN07}. However, we focus on improving round complexity (2$ \times$) {\em as well as} communication of the online phase by trading off an increase in the preprocessing. 

\subsubsection{Tier II - Building Blocks}
We design efficient protocols for several building blocks in semi-honest and malicious settings, which are stepping stones for Tier III applications. These are extensions from the small party setting~\cite{MR18,BLAZE,SWIFT,aby2}, and hence we defer the details to \S\ref{BB_semi} (semi-honest) and \S\ref{BB_mal} (malicious).

\subsubsection{Tier III - Applications}
To showcase the practicality of our framework and improvements of our protocols, we benchmark a range of applications such as neural networks (NN), which also includes the popular deep NN called VGG16~\cite{vgg16}, graph neural network, genome sequence matching, and biometric matching, and are considered for the first time in the $n$-party honest-majority setting. We benchmark the applications in the WAN setting using Google Cloud instances. As mentioned, owing to the inherent restrictions of RSS and keeping the focus on practical scenarios, we showcase the performance of our protocols for $n = 5, 7$, and $9$ and compare with the state-of-the-art semi-honest protocol of \cite{DamgaardN07}. 
  
\begin{usenix_item_text}
	\item[1.] {\em Deep neural networks.} We benchmark inference phases of deep neural networks such as LeNet~\cite{lenet} and VGG16~\cite{vgg16}. We observe savings of up to $69\%$ in monetary cost, and improvements of up to $4.3\times$ in online run-time 
	and throughput, in comparison to \cite{DamgaardN07}.
	\item[2.] {\em Graph neural network.} We benchmark the inference phase of graph neural network~\cite{defferrard,SCSDF20} on MNIST~\cite{MNIST10} data set. In comparison to \cite{DamgaardN07}, our protocol improves up to $3.5\times$ in online run-time, and sees up to $15\%$ savings in monetary cost.
	\item[3.] {\em Genome sequence matching.} We demonstrate an efficient protocol for similar sequence queries (SSQ), which can be used to perform secure genome matching. Our protocol is based on the protocol of ~\cite{ST19} which works for $2$ parties and uses an edit distance approximation~\cite{AHLR18}. We extend and optimize the protocol for the multiparty setting. In comparison to \cite{DamgaardN07}, we witness improvements of up to $4\times$ in online run-time and throughput, and savings of $66\%$ in monetary cost.
	\item[4.] {\em Biometric matching.} We propose efficient protocols for computing Euclidean distance (ED), which forms the basis for performing biometric matching. Continuing the trend, we witness a $4.6\times$ improvement in online run-time and throughput compared to \cite{DamgaardN07}, and savings of up to $85\%$ in monetary cost.
\end{usenix_item_text}

\section{Preliminaries}
\label{sec:prelim}
We cast our protocols in the (function-dependent) preprocessing paradigm to enable a fast online phase. Parties rely on a one-time shared key setup (see \S\ref{app:prelims})~\cite{BLAZE, Trident,SWIFT,MR18,AFLNO16,FLASH} to enable generation of correlated randomness, non-interactively. 
Our protocols are designed for rings ($\Z{\ell}$).
We use fixed-point arithmetic (FPA)~\cite{MohasselZ17,MR18,ASTRA,Trident,BLAZE, SWIFT} representation to operate over decimal values. Here, a decimal value is represented as an $\ell$-bit integer in signed 2's complement representation. The most significant bit ($\MSB$) represents the sign bit, and $d$ least significant bits are reserved for the fractional part. The $\ell$-bit integer is then treated as an element of $\Z{\ell}$, and operations are performed modulo $2^{\ell}$. We let $\ell = 64$, $d = 13$, with $\ell - d - 1$ bits for the integer part.

This work considers both semi-honest and malicious adversarial models with static and at most $t<n/2$ corruptions. The security of constructions is proved using the real-world/ ideal-world simulation paradigm~\cite{Lindell17}, and the details are provided in \S\ref{app:SecurityMPC}.
Let $\Partyset = \{P_1, P_2, \ldots, P_n\}$ denote the set of $n$ parties which are connected by pair-wise private and authentic channels in a synchronous network. 
Set $\Evlset = \{P_1, P_2, \ldots, P_{t+1}\}$, termed as the evaluator set, comprises parties that are active during the online phase. Set $\Hlpset = \{P_{t+2}, P_{t+3}, \ldots, P_n\}$, termed as the helper set, comprises parties which help in the preprocessing phase, and in the online verification in the malicious setting.
Parties agree on a $\Pking \in \Evlset$. Without loss of generality, let $\Pking = P_{t+1}$.

	\begin{table*}[htb!]
		\centering
		\resizebox{0.99\textwidth}{!}{
			\begin{tabular}{l l l}
				\toprule 
				\multicolumn{1}{l}{Helper primitive} 
				&  
				\multicolumn{1}{l}{Input}
				&  
				\multicolumn{1}{l}{Output} \\
				\midrule
				$\pizero$	
				&	
				$\mbox{-}$
				&
				$\sgr{\cdot}$-sharing of $0$ \\
		    	$\PiRandR$ 
		    	&
		    	$\mbox{-}$
		    	& 
		    	$\sqr{\cdot}$-sharing of a random value $\vr \in \Z{\ell}$ \\
		        $\PiRandRP$
		    	&
		    	Identity of a party $P_s$
		    	& 
		    	$\sqr{\cdot}$-sharing of a random value $\vr \in \Z{\ell}$ such that $P_s$ learns all shares\\
		        $\Piptosh$		    	
		        &
		    	$\va \in \Z{\ell}$ held by at least $t+1$ parties 
		    	& 
		    	$\shr{\va}$-sharing\\
		        $\Pictaa$		    	
		        &
		    	$\sqr{\va}$-sharing, $\T \subset \Partyset$ such that $|\T| = t+1$
		    	& 
		    	$\tsgra{\va}{\T}$-sharing\\
		       $\Pictaf$
		        &
		    	$\sqr{\va}$-sharing 
		    	& 
		    	$\sgr{\va}$-sharing\\
		        $\Pictashra$
		        &
		    	$\shr{\va}$-sharing, $\T \subset \Partyset$ such that $|\T| = t+1$    
		    	& 
		    	$\tsgra{\va}{\T}$-sharing \\
		        $\Pictashrf$
		        &
		    	$\shr{\va}$-sharing    
		    	& 
		    	$\sgr{\va}$-sharing \\
		        $\Pishtosq$
		        &
		    	$\shr{\va}$-sharing   
		    	& 
		    	$\sqr{\va}$-sharing\\
		        $\Picprod$
		        &
		    	$\sqr{\va}$-sharing, $\sqr{\vb}$-sharing   
		    	& 
		    	$\sgr{\va \vb}$-sharing\\
		        $\Picon$
		        &
		    	$\Partyset, \vec{\val_1}, \ldots, \vec{\val_n}$
		    	& 
		    	`$\continue$' if $\vec{\val_i} = \vec{\val_j}$  for all $P_i, P_j \in \Partyset$, `$\abort$' otherwise \\
		        $\PiRSh$
		        &
		    	$\va$, identity of a party $P_s$
		    	& 
		    	$\sqr{\va}$-sharing\\
				\bottomrule
			\end{tabular}
		}
		\vspace{-1mm}
		\caption{Description of helper primitives~--~all primitives are non-interactive, except $\Picon$ (see \S\ref{appsec:helper} for details) 
		}\label{tab:helper}
		\vspace{-3mm}
	\end{table*}


\paragraph{Sharing semantics}
We use the following sharing semantics, based on RSS \& additive sharing schemes, which facilitate a fast online phase.

\smallskip
\begin{usenix_item_text}
\item {\em $\sqr{\cdot}$-sharing:} This denotes the replicated secret sharing (RSS) of a value with threshold $t$. 
A value $\va \in \Z{\ell}$ is said to be RSS-shared with threshold $t$ if for every subset $\T \subset \Partyset$ of $n-t$ parties there exists $\sqr{\va}_{\T} \in \Z{\ell}$ possessed by all $P_i \in \T$ such that $\va = \sum_{\T} \sqr{\va}_{\T}$. 

Alternatively, for every set of $t$ parties, the residual $h = n-t$ parties forming the set $\T$, hold the share $\sqr{\va}_{\T}$.
Let $\T_1, \T_2, \ldots, \T_{\h} \subset \Partyset$ be the distinct subsets of size $h$, where $\h = \binom n {h}$ represents the total number of shares. 
Since $P_i$ belongs to $\binom{n-1}{h-1}$ such sets, the tuple of shares $\{\sqr{\va}_{\T}\}$ that it possesses are denoted as $\sqr{\va}_i$.

\item {\em $\sgr{\cdot}$-sharing:} A value $\va \in \Z{\ell}$ is said to be $\sgr{\cdot}$-shared (additively shared) among parties in $\Partyset$ if $P_i \in \Partyset$ possesses $\sgr{\va}_i \in \Z{\ell}$ such that $\va  = \sgr{\va}_1 + \sgr{\va}_2 + \ldots + \sgr{\va}_n$. 
%

\item {\em $\tsgra{\cdot}{\T}$-sharing:} A value $\va \in \Z{\ell}$ is said to be $\tsgra{\cdot}{\T}$-shared among $t+1$ parties in $\T$,  
if each $P_i \in \T$ holds $\tsgra{\va}{\T}_i$ such that $\va  = \sum_{P_i \in \T} \tsgra{\va}{\T}_{i}$. We refer to this sharing scheme as $(t+1)$-additive sharing, and use $\tsgr{\va}$ to denote such a sharing among parties in $\Evlset$.

\item {\em $\shr{\cdot}$-sharing:} A value $\va \in \Z{\ell}$ is said to be $\shr{\cdot}$-shared in the semi-honest setting if there exist values $\lv{\va}, \mv{\va} \in \Z{\ell}$ such that $\mv{\va} = \va + \lv{\va}$ where $\lv{\va}$ is $\sqr{\cdot}$-shared among $\Partyset$ and every $P_i \in \Evlset$ holds $\mv{\va}$. We denote the shares of $P_i \in \Hlpset$ by $\shr{\va}_{i} = \sqr{\lv{\va}}_{i}$ and that of $P_i \in \Evlset$ as  $\shr{\va}_{i} = (\mv{\va}, \sqr{\lv{\va}}_{i})$.
In the malicious setting, $\mv{\va}$ is held by all parties, and $\shr{\va}_{i} = (\mv{\va}, \sqr{\lv{\va}}_{i})$ for all $P_i \in \Partyset$. 
\end{usenix_item_text}

It is trivial to see that all the sharing schemes mentioned above are linear. This allows parties to compute linear operations such as addition and multiplication with constants locally.
The Boolean world operates over $\Z{}$, and we denote the corresponding Boolean sharing with a superscript {\bf B}. Notations are summarized in \tabref{notations}.

\begin{footnotesize}
	\begin{table}[htb!]
		\centering
		\resizebox{0.49\textwidth}{!}{
			\begin{tabular}{p{1.3cm} l }
				\toprule
				Notation & Description\\
				\midrule 
				$n = 2t+1$ & Total number of parties with $t$ corrupt and $h = t+1$ honest \\[3pt]
				$\T_1, \ldots, \T_{\h}$ & $\h = \binom{n}{h}$ distinct subsets of $\Partyset$ with $t+1$ parties each\\[3pt]
				$\h$ & Number of replicated secret shares (RSS) of a value \\[3pt]
				$\g = \binom{n-1}{h-1}$ & Number of RSS shares of a value held by a party \\[3pt]
				$\Evlset$ & \makecell[l]{Online parties ($P_1, \ldots, P_{t+1}$) that actively carry out the \\computation}\\[3pt]
				$\Hlpset$ & Helper parties ($P_{t+2}, \ldots, P_{n}$)\\[3pt]
				$\va_i$ & $i^{\text{th}}$ element of vector $\vec{a}$\\[3pt]
				$\vec{a} \band \vec{b}$ & dot product of vectors $\vec{a}$ and $\vec{b}$ \\[3pt]
				$\Mat{A} \MatMul \Mat{B}$ & Multiplication of matrices $\Mat{A}$ and $\Mat{B}$ \\[3pt]
				$\arval{\bitb}$ 
				& Arithmetic (Ring) equivalent over $\Z{\ell}$ of bit $\bitb \in \Z{}$\\[3pt]
				$\val[i]$ & $i^{\text{th}}$ bit of $\ell$-bit value $\val \in \Z{\ell}$\\[3pt]
				$\mv{\va}$ = $\va$ + $\lv{\va}$ & Masked value $\mv{\va}$ for $\va \in \Z{\ell}$ with mask $\lv{\va} \in \Z{\ell}$ \\[3pt]
				$\cmv{\va_1 \va_2 \ldots \va_k}$ & $\prod_{i = 1}^{k} \mv{\va_i}$; Product of masked values $\mv{\va_1}, \ldots, \mv{\va_k}$  \\[3pt]
				$\clv{\va_1 \va_2 \ldots \va_k}$ & $\prod_{i = 1}^{k} \lv{\va_i}$; Product of masks $\lv{\va_1}, \ldots, \lv{\va_k}$ \\[3pt]
				\bottomrule
			\end{tabular}
		}
		\vspace{-1mm}
		\caption{\small Notations used in this work\label{tab:notations}}
		\vspace{-8mm}
	\end{table}
\end{footnotesize}

\paragraph{Helper primitives}
We use the primitives described in \tabref{helper} from literature~\cite{BonehBCGI19, BGIN20, BLAZE, CDI05} in our protocols, and their details are deferred to \S\ref{appsec:helper}. The Boolean variants of corresponding primitives are denoted with a superscript {\bf B}.

\section{MPClan Protocol}
\label{sec:ringmpc}
This section details the semi-honest MPC protocol execution performed over the ring $\Z{\ell}$ that comprises three phases-- input sharing, evaluation (linear operations and multiplication), and output reconstruction.

\paragraph{Input sharing and Output Reconstruction}
To enable $P_s \in \Partyset$ to $\shr{\cdot}$-share a value $\val \in \Z{\ell}$, parties first non-interactively sample $\sqr{\cdot}$-shares of $\lv{\val}$, relying on the shared-key setup, such that $P_s$ learns all these shares on clear (via $\PiRandRP$). This enables $P_s$ to compute and send $\mv{\val} = \val + \lv{\val}$ to parties in $\Evlset$, thereby generating $\shr{\val}$.

To reconstruct $\val$ towards all parties given $\shr{\val}$, parties in $\Evlset$ non-interactively generate its additive shares, $\tsgr{\val}$, among themselves (via $\Pictashr$). These parties send their additive shares to $\Pking$, who computes and sends $\val$ to all parties. Reconstruction towards a single party, say $P_s$, proceeds similarly except that the protocol terminates after parties in $\Evlset$ send their additive shares of $\val$ to $\Pking = P_s$, who then computes $\val$.

\paragraph{Evaluation}
Evaluation comprises linear operations of addition and multiplication with public constant, and non-linear operations such as multiplication. Parties can non-interactively compute linear operations owing to the linearity of the $\shr{\cdot}$-sharing. Concretely, given $\shr{\va}, \shr{\vb}$ and public constants $\vc_1, \vc_2$, parties can non-interactively compute $\shr{\vc_1 \va + \vc_2 \vb}$ as $\vc_1 \shr{\va} + \vc_2 \shr{\vb}$. 

To compute $\shr{\cdot}$-shares for non-linear operations such as multiplication, say $\vz = \va \vb$ given $\shr{\va}, \shr{\vb}$, parties proceed as follows. At a high-level, the approach is to enable generation of $\shr{\vz - \vr}$ and $\shr{\vr}$ for a random $\vr \in \Z{\ell}$, which enables parties to non-interactively compute $\shr{\vz} = \shr{\vz - \vr} + \shr{\vr}$. Observe that $\shr{\vr}$ can be generated non-interactively by locally sampling each of its shares. 
To generate $\shr{\vz - \vr}$, we let parties in $\Evlset$ obtain $\vz - \vr$, following which $\shr{\vz - \vr}$ can be generated non-interactively (this is achieved via $\Piptosh$ where all parties set their shares of $\sqr{\lv{\vz - \vr}}$ as 0, and parties in $\Evlset$ set $\mv{\vz - \vr} = \vz - \vr$). Observe that $\vz$ remains private while revealing $\vz - \vr$ to parties in $\Evlset$ since $\vr$ is a random mask not known to adversary. 

To enable parties in $\Evlset$ to obtain $\vz - \vr$, we let $\vz - \vr = \D + \E$, where $\D$ is additively shared among parties in $\Hlpset$ while $\E$ is additively shared among parties in $\Evlset$ ($\D, \E$ are defined in the following paragraphs). Thus, to reconstruct $\vz - \vr$ towards parties in $\Evlset$, parties send their respective additive shares of $\D$ or $\E$ towards $\Pking \in \Partyset$. $\Pking$ reconstructs $\D, \E$, and sends $\vz - \vr = \D + \E$ to parties in $\Evlset$. 
Elaborately, as seen in~\cite{ASTRA, Tetrad}, $\vz - \vr$ can be computed as

\vspace{-6mm}
\begin{align}
	\vz - \vr &=  \va \vb - \vr = \left( \mv{\va} - \lv{\va} \right) \left( \mv{\vb} - \lv{\vb} \right) - \vr \nonumber \\
	&= \cmv{\va \vb} -\mv{\va} \lv{\vb} - \mv{\vb} \lv{\va} + \clv{\va \vb} - \vr \label{maskedVz} \\  
	&= \underbrace{\cmv{\va \vb} -\mv{\va} \lv{\vb} - \mv{\vb} \lv{\va} + (\clv{\va \vb} - \vr)_{\Evlset}}_{\E} + \underbrace{(\clv{\va \vb} -  \vr)_{\Hlpset}}_{\D}  \nonumber
\end{align}
\vspace{-2mm}
where $\clv{\va \vb} - \vr = (\clv{\va \vb} - \vr)_{\Hlpset} + (\clv{\va \vb} - \vr)_{\Evlset}$.

\begin{figure}[htb!]
\centering
\includegraphics[width=\linewidth]{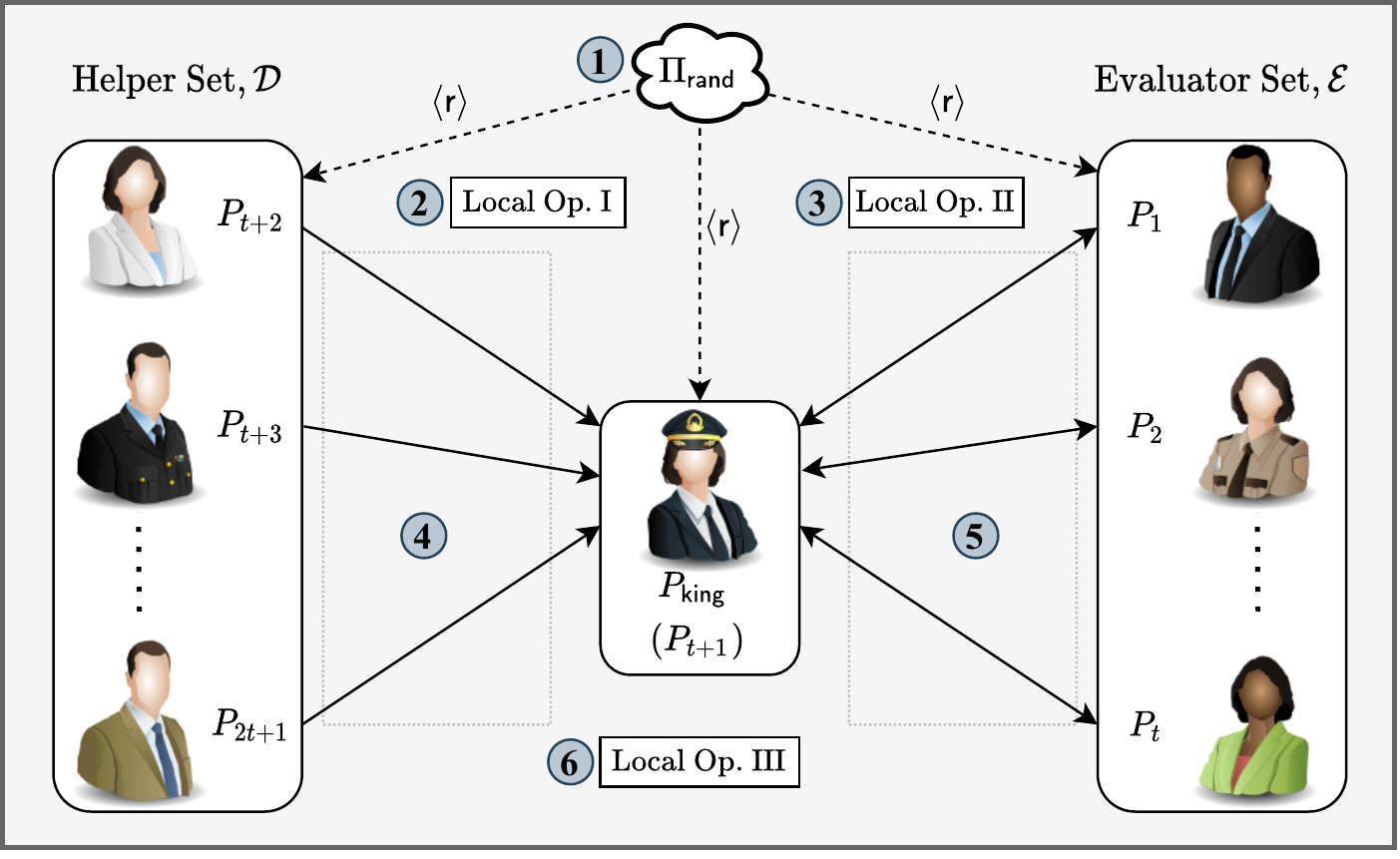}
\negspaceqtr
{\small 
\begin{tabular}{p{0.96\linewidth}}
\rowcolor{LightGray}
\bfcircled{1}~Generation of random $\vr \in \Z{\ell}$
\bfcircled{2}~Computing $\sgr{\vr} \& \shr{\vr}$
\bfcircled{3}~Computing $\tsgr{\lv{\va}}, \tsgr{\lv{\vb}}$
\bfcircled{4}~$\Hlpset$ sending $\{\sgr{\vz - \vr}\}_{\Hlpset}$ to $\Pking$
\bfcircled{5}~$\Evlset$ sending $\{\sgr{\vz - \vr}\}_{\Evlset}$ to $\Pking$ and receiving result from $\Pking$
\bfcircled{6}~Computing $\shr{\vz}$\\
\end{tabular}}
\smallskip
\caption{Steps of multiplication protocol}
\label{figure:multsemi}
\vspace{-3mm}
\end{figure}

We next detail the steps in the multiplication protocol, and its schematic representation is provided in \boxref{figure:multsemi}.

\begin{ccsitem}
    \item[$\blacktriangleright$] {{\em Step}~\bfcircled{1}:}  Parties non-interactively generate $\sqr{\vr}$ by locally sampling each of its shares (via $\PiRandR$). Parties locally compute $\sgr{\vr}$ and $\shr{\vr}$ from $\sqr{\vr}$ using $\Pictaf$ and $\Pisqtosh$, respectively. Looking ahead, $\sgr{\vr}$ aids in generating additive shares of $\D, \E$, while $\shr{\vr}$ aids in computing $\shr{\vz}$ from $\shr{\vz - \vr}$.
    \item[$\blacktriangleright$] {{\em Step}~\bfcircled{2}:} This step involves computing additive shares of $\clv{\va \vb} - \vr$ among all parties. For this, parties non-interactively generate $\sgr{\clv{\va \vb}}$ from $\sqr{\lv{\va}}, \sqr{\lv{\vb}}$ (via $\Picprod$). $P_i \in \Partyset$ sets its additive share of $\clv{\va \vb} - \vr$ as $\sgr{\clv{\va \vb} - \vr}_i = \sgr{\clv{\va \vb}}_i - \sgr{\vr}_i$. Observe that the shares $\sgr{\clv{\va \vb} - \vr}_i$ of $P_i \in \Hlpset$ define the additive shares of $\D = (\clv{\va \vb} - \vr)_{\Hlpset}$ among parties in $\Hlpset$. Similarly, the shares $\sgr{\clv{\va \vb} - \vr}_i$ of $P_i \in \Evlset$ define the additive shares of $(\clv{\va \vb} - \vr)_{\Evlset}$ among parties in $\Evlset$ (i.e. $\tsgr{(\clv{\va \vb} - r)_{\Evlset}}$ ). 
    \item[$\blacktriangleright$] {{\em Step}~\bfcircled{3}:} Parties in $\Evlset$ generate additive shares of $\lv{\va}, \lv{{\vb}}$ among themselves ($\tsgr{\cdot}$-shares, via $\Picta$). Looking ahead, $\tsgr{\lv{\va}}, \tsgr{\lv{\vb}}$ aid in generating additive shares of $\E$ among $\Evlset$.
    \item[$\blacktriangleright$] {{\em Step}~\bfcircled{4}:} Parties in $\Hlpset$ send their additive shares of $\D$ (as defined in step \bfcircled{2}) to $\Pking$, who reconstructs $\D$. 
    \item[$\blacktriangleright$] {{\em Step}~\bfcircled{5}:} $P_i \in \Evlset \setminus \{\Pking\}$ non-interactively generates additive share, $\tsgr{\E}_i$, of $\E$ among parties in $\Evlset$ as $\tsgr{\E}_i = - \mv{\va} \tsgr{\lv{\vb}}_i - \mv{\vb} \tsgr{\lv{\va}}_i + \tsgr{(\clv{\va \vb} - \vr)_{\Evlset}}_i$. Note that it suffices for only one designated party in $\Evlset$ to add $\cmv{\va \vb}$ in its share of $\tsgr{\E}$, and without loss of generality we let this designated party be $\Pking$. For $\Pking = P_{t+1}$ in our case, $\tsgr{\E}_{t+1} = \cmv{\va \vb} - \mv{\va} \tsgr{\lv{\vb}}_{t+1} - \mv{\vb} \tsgr{\lv{\va}}_{t+1} + \tsgr{(\clv{\va \vb} - \vr)_{\Evlset}}_{t+1}$. Parties send their additive shares of $\E$ to $\Pking$, who reconstructs $\E$, and sends $\vz - \vr = \D + \E$ to parties in $\Evlset$.
    \item[$\blacktriangleright$]{{\em Step}~\bfcircled{6}:} Parties non-interactively generate $\shr{\vz - \vr}$ (via $\Piptosh$) as explained earlier. Using $\shr{\vr}$ generated in step \bfcircled{1}, parties compute $\shr{\vz} = \shr{\vz - \vr} + \shr{\vr}$, as required. 
\end{ccsitem}

\begin{protboxgray}{$\PiMult(\Partyset, \shr{\va}, \shr{\vb}, \istr)$}{Semi-honest: Multiplication protocol}{fig:piMultr}
	\justify
	$\istr = 1$ denotes perform truncation, $\istr = 0$ denotes otherwise. 

	\justify 
	\algoHead{Preprocessing:}
	\begin{ccsitem} 
		\item[{\bfcircled{1}}] If $\istr = 0$: invoke $\PiRandR$ to generate $\sqr{\vr}$ where $\vr  \in \Z{\ell}$. Invoke $\Pictaf$ and $\Pisqtosh$ on $\sqr{\vr}$ to generate $\sgr{\vr}$ and $\shr{\vr}$, respectively. 		
		\item Else, invoke $\PiDSBits(\Partyset, 1)$ (\boxref{fig:piDSBits}) to generate $\shr{\vr}, \shr{\trunc{\vr}}$, and $\Pictashrf$ on $\shr{\vr}$ to generate $\sgr{\vr}$ .
		\item[{\bfcircled{2}}] Invoke $\Picprod$ on  $\sqr{\lv{\va}}, \sqr{\lv{\vb}}$ to generate $\sgr{\clv{\va \vb}}$, and compute $\sgr{\clv{\va \vb} - \vr} = \sgr{\clv{\va \vb}} - \sgr{\vr}$.
		\item $P_i \in \Evlset$ sets $\tsgr{(\clv{\va \vb} - r)_{\Evlset}}_i = \sgr{\clv{\va \vb} - r}_i$.
		\item[{\bfcircled{3}}] $P_i \in \Evlset$ invokes $\Picta$ on $\sqr{\lv{\va}}, \sqr{\lv{\vb}}$ to generate $\tsgr{\lv{\va}}_i, \tsgr{\lv{\vb}}_i$, respectively. 
		\item[{\bfcircled{4}}] $P_i \in \Hlpset$ sends $\sgr{\clv{\va \vb} - \vr}_i$ to $\Pking$, who sets $\D = \sum_{i: P_i \in \Hlpset} \sgr{\clv{\va \vb} - \vr}_i$.
	\end{ccsitem} 
	\algoHead{Online:}
	\begin{ccsitem}
		\item[{\bfcircled{5}}] $P_i \in \Evlset$ computes $\tsgr{\zeta}_{i} = -\mv{\va} \tsgr{\lv{\vb}}_{i} - \mv{\vb} \tsgr{\lv{\va}}_{i} + \tsgr{(\clv{\va \vb} - \vr)_{\Evlset}}_i$, and sends $\tsgr{\zeta}_i$ to $\Pking$.
		\item $\Pking$ computes $\E = \cmv{\va \vb} + \sum_{i: P_i \in \Evlset} \tsgr{\zeta}_{i}$ and sends $\vz - \vr = \D + \E$ to all parties in $\Evlset$.
		\item[{\bfcircled{6}}] If $\istr = 0$: invoke $\Piptosh$ on $\vz - \vr$ to generate $\shr{\vz - \vr}$, and compute $\shr{\vz} = \shr{\vz - \vr} + \shr{\vr}$.
		\item Else, invoke $\Piptosh$ on $\trunc{(\vz - \vr)}$ to generate $\shr{\trunc{(\vz - \vr)}}$, and compute $\shr{\trunc{\vz}} = \shr{\trunc{(\vz - \vr)}} + \shr{\trunc{\vr}}$.
	\end{ccsitem}
\end{protboxgray}
\vspace{-2mm}

\begin{lemma}
Protocol $\PiMult$ (\boxref{fig:piMultr}) incurs a communication of $t$ elements in the preprocessing phase and $2t$ elements in $2$ rounds in the online phase for multiplication when $\istr = 0$. 
\end{lemma}

\noindent
{\em Analysis:} Observe that the communication towards $\Pking$ in steps \bfcircled{4} and \bfcircled{5}, can be performed in parallel, resulting in the overall round complexity of the protocol being two. Further, a communication of $t$ elements is required in step \bfcircled{4} and $2t$ elements is required in \bfcircled{5} (since $\Pking \in \Evlset$), thereby having a total communication complexity of $3t$ ring elements. This complexity resembles that of DN07. However, our sharing semantics enables us to push some of the steps mentioned above to a preprocessing phase, resulting in a fast online phase, which is non-trivial to achieve in the case of DN07. Elaborately, observe that since $\vr, \lv{\va}, \lv{\vb}$ are independent of the input (owing to our sharing semantics), computation involving these terms ranging from steps \bfcircled{1} to \bfcircled{4} can thus be moved to a preprocessing phase. This improves the online communication complexity by slashing the inward communication towards $\Pking$ by half. Thus, the online phase requires only $2t$ ring elements of communication while offloading $t$ elements of communication to the preprocessing phase.

Note that a straightforward extension of semi-honest multiplication of \cite{DamgaardN07} to the preprocessing model, which can be derived from \cite{ED20}, does not provide an efficient solution. Although such a protocol has the same online complexity ($2t$ elements) as our online phase, it has the drawback of inflating the overall communication cost by a factor of $1.6\times$ over \cite{DamgaardN07}. Elaborately, the online communication cost of $2t$ elements can be attained by appropriately defining the sharing semantics and using the $\Pking$ approach, similar to our protocol. However, this requires parties to generate the sharing of $\clv{\va \vb} = \lv{\va}\cdot\lv{\vb}$ from the shares of $\lv{\va}$ and $\lv{\vb}$ during the preprocessing phase, and requires a full-fledged multiplication, incurring a cost of $3t$ elements. This yields a protocol with a total cost of $5t$ elements in comparison to the $3t$ cost of the all-online DN07 protocol. Thus, departing from this approach, the novelty of our protocol lies in leveraging the interplay between the sharing semantics and redesigning the communication pattern among the parties to ensure that the total cost of $3t$ does not change.

Furthermore, our protocol design allows parties in $\Hlpset$ to remain shut in the online phase, thereby reducing the system's operational load. This is because parties in $\Hlpset$ only contribute towards the computation of $\D$, which can be completed in the preprocessing phase. However, the preprocessing phase becomes function-dependent due to the linear gates, for which the $\lv{}$ value for the output wires cannot be chosen randomly. Concretely, if $\vc$ is the output of a linear gate, say addition, with inputs $\va, \vb$, then $\lv{\vc}$ cannot be chosen randomly and should be defined as $\lv{\vc} = \lv{\va} + \lv{\vb}$.

Ideal functionality $\Fmpc$ for evaluating function $f$ in the $n$-party setting with semi-honest security appears in \boxref{fig:Fmpc}. 

\smallskip
\begin{systemboxgray}{$\Fmpc$}{Semi-honest: Ideal functionality for function $f$}{fig:Fmpc}
	\justify
	$\Fmpc$ interacts with the parties in $\Partyset$ and the adversary $\simsh$. Let $f$ denote the function to be  computed. Let $\wx_s$ be the input corresponding to the party $P_s$, and $\wy_s$ be the corresponding output, i.e $(\{\wy_s\}_{s=1}^{n}) = f(\{\wx_s\}_{s=1}^{n})$. 
	\begin{myitemize}
		\item[ ] {\bf Step 1:} $\Fmpc$ receives $(\INPUT,\wx_s)$ from $P_s \in \Partyset$, \& computes $(\{\wy_s\}_{s=1}^{n}) = f(\{\wx_s\}_{s=1}^{n})$.
		\item[ ] {\bf Step 2:} $\Fmpc$ sends  $(\OUTPUT, \wy_s)$ to $P_s \in \Partyset$.
	\end{myitemize}
\end{systemboxgray}
\vspace{-2mm}

\paragraph{Incorporating truncation}
To retain FPA semantics, it is required to truncate the result of multiplication, $\vz = \va \vb$, which ends up having $2d$ bits in the fractional part, by $d$ bits, i.e. compute $\trunc{\vz} = \vz / 2^d$. For this, we extend the {\em probabilistic} truncation technique of~\cite{MR18, SWIFT, Tetrad} proposed in the small party domain to the $n$-party setting. Given $(\vr, \trunc{\vr})$-pair, with $\trunc{\vr} = \vr/2^d$, the truncated value of $\vz$ can be obtained as $\trunc{\vz} = \trunc{(\vz - \vr)} + \trunc{\vr}$. Accuracy and correctness of this method follows from~\cite{MR18, MLRG20}. 

\smallskip
\begin{systemboxgray}{$\Ftrgen$}{Ideal functionality $\Ftrgen$}{fig:Ftrgen}
	\medskip
	\justify
	\begin{ccsitemize}
	    \item[--] Samples random $\vr \in \Z{\ell}$, and computes $\trunc{\vr} = \vr/ 2^{d}$.
	    \item[--] Generates $\shr{\cdot}$-shares of $\vr, \trunc{\vr}$ and set output share for $P_s \in \Partyset$ as $\wy_s = \{\shr{\vr}_s, \shr{\trunc{\vr}}_s \}$.
	\end{ccsitemize}
	\begin{description}
		\item {\bf Output: } Send $(\OUTPUT, \wy_s)$ to $P_s \in \Partyset$.
	\end{description}
\end{systemboxgray}
\vspace{-2mm}

Our multiplication protocol can be modified to additionally perform truncation by incorporating the following two changes-- (i) generate $\shr{\trunc{\vr}}$ in step \bfcircled{1}, and (ii) compute $\shr{\trunc{\vz}} = \shr{\trunc{(\vz - \vr)}} + \shr{\trunc{\vr}}$, instead, in step \bfcircled{6}. For (i), we rely on the ideal functionality, $\Ftrgen$~(\boxref{fig:Ftrgen}), for computing $\shr{\vr}, \shr{\trunc{\vr}}$. $\Ftrgen$ can be instantiated using the appropriate MPC protocol which will be used as a black-box in our multiplication. Thus, improvements in the MPC protocol that realizes $\Ftrgen$ can be inherited in our multiplication protocol. In our work, we instantiate $\Ftrgen$ using $\PiDSBits$ (\boxref{fig:piDSBits}), which is a slightly modified version of the doubly-shared random bit generation protocol of~\cite{Damgard0FKSV19}, adapted to our $n$-party setting. Concretely, $\PiDSBits$ generates $\ell$ doubly-shared random bits instead of a single bit, as done in the protocol of~\cite{Damgard0FKSV19}. Here, a doubly-shared random bit is a bit which is arithmetic as well as Boolean shared.  We defer the details of $\PiDSBits$ to \S\ref{app:bbsh} since it follows easily from the protocol of~\cite{Damgard0FKSV19}.
With respect to (ii), observe that it is a local operation, and hence performing truncation does not incur any additional overhead in the online phase.

\paragraph{Dot product}

Given $\shr{\cdot}$-shares of vectors $\vec{x}$ and $\vec{y}$ of size $\nf$, dot product outputs $\shr{\vz}$ where $\vz = \vec{x} \band \vec{y} = \sum_{k = 1}^{\nf} \vx_k \vy_k$ and $\band$ denotes the dot product operation. 
The design of our multiplication protocol enables easy extension to support dot product computation without incurring any overhead. Concretely, similar to multiplication,
\begin{align} 
	\vz - \vr &=  (\vec{x} \band \vec{y}) - \vr  \nonumber\\
	          &= \sum_{k=1}^{\nf}\cmv{\vx_k \vy_k} -\sum_{k=1}^{\nf} \mv{\vx_k} \lv{\vy_k} - \sum_{k=1}^{\nf} \mv{\vy_k} \lv{\vx_k} + \sum_{k=1}^{\nf} \clv{\vx_k \vy_k} - \vr \label{maskeddotp}      
\end{align}
In each of the summands of $\vz - \vr$, each of the $\nf$ product terms can be generated similar to that in the multiplication protocol, which can then be locally summed up before sending it towards $\Pking$. Due to this simple extension, we defer the formal dot product protocol (\boxref{fig:pidotp}) to \S\ref{app:bbsh}.
Looking ahead, for matrix multiplication, each element of the resultant matrix can be computed via a dot product. 

\paragraph{Multi input multiplication}
3-input and 4-input multiplication protocols have showcased their wide applicability in improving the online phase complexity~\cite{Tetrad, aby2, ohata20}. Concretely, computing $\vz = \va \vb \vc$ (3-input) or $\vz = \va \vb \vc \vd$ (4-input) naively requires at least two sequential invocations of 2-input multiplication protocol in the online phase. Instead, 3-input and 4-input multiplication protocol, respectively, enables performing this computation with the same online complexity as that of a {\em single} 2-input multiplication. Thus, we design 3-input and 4-input multiplication protocols by extending the techniques of~\cite{aby2, Tetrad} to the $n$-party setting. 
Designing these protocols require modifications in the preprocessing steps. 
Consider 3-input multiplication where the goal is to generate $\shr{\cdot}$-sharing of $\vz = \va \vb \vc$ given $\shr{\va}, \shr{\vb}, \shr{\vc}$. Note that      
\begin{align*}
	\vz - \vr
	&= \va \vb \vc - \vr = (\mv{\va} - \lv{\va})(\mv{\vb} - \lv{\vb})(\mv{\vc} - \lv{\vc}) - \vr \\
	&= \cmv{\va\vb\vc} - \cmv{\va\vc} \lv{\vb} - \cmv{\vb\vc} \lv{\va} - \cmv{\va\vb} \lv{\vc} \\
	&~~~+ \mv{\va} \clv{\vb \vc} + \mv{\vb} \clv{\va \vc} + \mv{\vc} \clv{\va \vb} - \clv{\va \vb \vc} - \vr
\end{align*}
%
We follow an approach closely related to 2-input multiplication, with the difference being that parties additionally require to generate the additive sharing of $\clv{\vb \vc}, \clv{\va \vc}$ and $\clv{\va \vb \vc}$ during preprocessing. Given these sharings, parties proceed with a similar online phase as in $\piMult$ to compute the 3-input multiplication without inflating the online cost. Similarly, for 4-input multiplication, parties need to generate the additive sharing of $\clv{\va \vd}, \clv{\vb \vd}, \clv{\vc \vd}, \clv{\va \vb \vd}, \clv{\va \vc \vd}, \clv{\vb \vc \vd}, \clv{\va \vb \vc \vd}$ in addition to those required in the case of 3-input multiplication. The generation of these sharings follows a similar approach as the 2-input multiplication, and the details are deferred to \S\ref{app:bbsh}. \tabref{mmultsh} compares the cost of computing $\vz = \va \vb \vc$ via a 2-input multiplication sequentially vs a 3-input multiplication, and computing $\vz = \va \vb \vc \vd$ via a 2-input and 4-input multiplication.

\begin{table}[htb!]
	\centering
	\resizebox{.4\textwidth}{!}{
		\begin{NiceTabular}{r r r r r}
			\toprule
			\Block[c]{2-1}{Multiplication\\type} & \Block[c]{2-1}{Building\\Block}
			& \Block[c]{1-2}{Communication} & & \Block[c]{2-1}{Online\\Rounds}\\
			\cmidrule{3-4}
			& & Prep. & Online & \\
			\midrule
			
			\Block{2-1}{$\vz = \va \vb \vc$} & 2-input mult. & $2t \ell$ & $4t \ell$ & $4$ 
			\\
			 & 3-input mult. & $6t \ell$ & $2t \ell$ &  $2$
			\\
			\midrule 
			
			\Block{2-1}{$\vz = \va \vb \vc \vd$} & 2-input mult. & $3t \ell$ & $6t \ell$ & $4$ 
			\\
			 & 4-input mult. & $15t \ell$  & $2t \ell$ & $2$ 
			\\

			\bottomrule
		\end{NiceTabular}
	}
	\vspace{-1mm}
	\caption{\small 
		Semi-honest: Communication and round complexity for computing multi-input multiplications \label{tab:mmultsh}}
	\vspace{-1mm}
\end{table}

\vspace{-3mm}
The recent work of~\cite{GLOPS21} provides a method to reduce the round complexity of circuit evaluation. They group the (distinct) consecutive layers in the circuit into pairs and perform a parallel evaluation of all gates in the two layers in a group. Consider a multiplication gate with inputs $\vx, \vy$ (obtained as output from a previous layer) and output $\vz$. Their approach considers three cases: (i) if $\vx$ and $\vy$ are not the outputs of a multiplication gate, (ii) exactly one among $\vx,\vy$ is the output of a multiplication gate, and (iii) both $\vx, \vy$ are outputs of a multiplication gate. 
We observe that case (ii) and (iii) in their approach resembles multi-input multiplication, which allows evaluating the second layer of multiplication ($\vz = \vx \cdot \vy$) non-interactively, thereby saving on rounds. 

\vspace{-2mm}
\begin{figure}[htb!]
    \centering	\includegraphics[width=0.85\columnwidth]{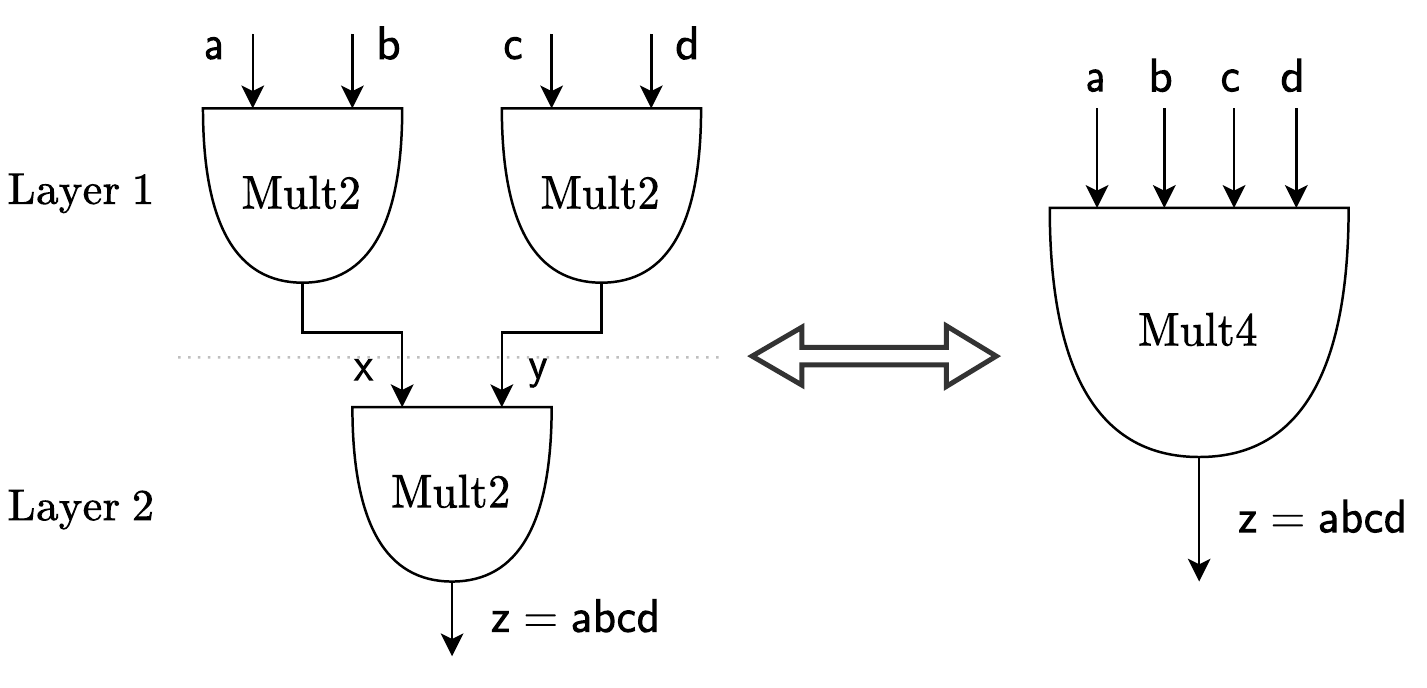}
	\vspace{-2mm}
	\caption{\small 4-input multiplication} \label{fig:multi-input}
	\vspace{-3mm}
\end{figure}


For instance, consider a 2-layer sub-circuit as in \boxref{fig:multi-input} where $\vx = \va \cdot \vb, \vy = \vc \cdot \vd$ are outputs of a multiplication gate which are fed as input to a multiplication gate in the next level. The approach of~\cite{GLOPS21} allows computation of $\vz = (\va \cdot \vb) \cdot (\vc \cdot \vd)$ in a single shot, which is equivalent to computing $\vz$ via a 4-input multiplication in our case. 
Similarly, when only one of the inputs (either $\vx$ or $\vy$) is the output of multiplication, computation of $\vz = \vx \cdot \vy$ resembles a 3-input multiplication. 
Thus, cases (i), (ii), (iii) correspond to 2-input, 3-input, and 4-input multiplication, respectively, in our work and are sufficient to reduce round complexity of any circuit evaluation by half. Hence, we restrict our focus to 3 and 4-input multiplication.

\section{Extending to malicious security}
\label{sec:ringmpcmal}
Using standard approaches~\cite{BLAZE,SWIFT,ED20}, it is straightforward to adapt the semi-honest protocols such as input sharing and output reconstruction to the malicious setting. The details are provided in \S\ref{appsec:malmpc} for completeness. Hence, in this section, we focus on the challenges encountered and their resolutions for obtaining a maliciously secure {\em multiplication} protocol. 

Note that although a maliciously secure multiplication protocol can be achieved by compiling our semi-honest protocol using compiler techniques such as~\cite{AbspoelDEN19,BGIN20}, the resultant protocol has an expensive online phase. For instance, using the compiler of~\cite{AbspoelDEN19} yields a protocol that requires computation over {\em extended} rings and communicating $4t$ extended ring elements in the online phase. This is not favourable compared to working over plain rings, especially in the online phase. Further, compilers such as those in~\cite{BGIN20} require heavy computational machinery like reliance on zero-knowledge proofs in the online phase, which is also not desirable. Thus, to attain a computation and communication efficient online phase, departing from the aforementioned compiler-based approaches, we design a maliciously secure multiplication protocol that requires communicating $3t$ {\em ring} elements in each phase. It is worth noting that we can do this while retaining the benefits of requiring only $t+1$ parties in the online phase (for most of the computation). The remaining $t$ parties are required to come online only for a short one-time verification phase, that is deferred to the end of the computation. Deferring verification may result in a privacy breach~\cite{GoyalLS19}. However, we describe later why the privacy breach does not arise in our protocol. 

To enable generation of $\shr{\vz} = \shr{\va  \vb}$ from $\shr{\va}$ and $\shr{\vb}$, we retain the high-level ideas from the semi-honest protocol. %
Our task reduces to $\mynum{i}$ generating additive shares of $\clv{\va \vb}$ among parties in $\Evlset$ (i.e. $\tsgr{\clv{\va \vb}}$) given $\sqr{\lv{\va}}$ and $\sqr{\lv{\vb}}$, in the preprocessing phase, and $\mynum{ii}$ reconstructing  $\vz - \vr$ in the online phase.
Given $\mynum{i}$, computing $\tsgr{\vz - \vr}$ in the online phase is a local operation. Given  $\mynum{ii}$, parties can invoke $\Piptoshf$ to generate $\shr{\vz - \vr}$, and compute $\shr{\vz} = \shr{\vz - \vr} + \shr{\vr}$, where $\shr{\vr}$ is generated in the preprocessing phase, as discussed in the semi-honest case.

\begin{systemboxgray}{$\FMulPre$}{Ideal functionality $\FMulPre$}{fig:FMulPre}
	\justify
	$\FMulPre$ interacts with the parties in $\Partyset$ and the adversary $\Sim$. Let $\T_{i}$ be the set of the honest parties.
	
	\begin{description}
	    \item {\bf Input:} $\FMulPre$ receives the $\sqr{\cdot}$-shares of $\va, \vb$ from the parties. It also receives $\sqr{\cdot}$-shares of $\vz = \va \vb$ of corrupt parties from $\Sim$. $\Sim$ is also allowed to send a special command, $(\abort, \Ab)$, which indicates that parties in $\Partyset$ with indices in $\Ab$ should $\abort$.
	\end{description}
	
	\noindent $\FMulPre$ proceeds as follows.
	\begin{ccsitemize}
		\item[--] Reconstruct $\va, \vb$ using the shares received from honest parties, and compute $\vz = \va \vb$.
		\item[--] Compute the $\sqr{\cdot}$-share of $\vz$ to be held by the set of honest parties as the difference between $\vz$ and the sum of $\sqr{\cdot}$-shares of $\vz$ received from corrupt parties.
		%
		\item[--] Let $\wy_s$ denote the $\sqr{\cdot}$-shares of $\vz$ 
		for party $P_s \in \Partyset$. If received $(\abort, \Ab)$ from $\Sim$, set  $\wy_s = \abort$ for $P_s$, where $s \in \Ab$.
	\end{ccsitemize}
	\begin{description}
		\item {\bf Output: } Send $(\OUTPUT, \wy_s)$ to every $P_s \in \Partyset$.
	\end{description}
\end{systemboxgray}
\vspace{-2mm}

For task $\mynum{i}$, our idea for the semi-honest case, of making parties in $\Hlpset$ to send their shares to $\Pking$, does not work in the presence of a malicious adversary. To address this, we make black-box use of a maliciously secure multiplication protocol, abstracted as a functionality  $\FMulPre$ in \boxref{fig:FMulPre}, that computes $\sqr{\clv{\va \vb}}$ from $\sqr{\lv{\va}}, \sqr{\lv{\vb}}$. In this work, we instantiate $\FMulPre$ with the state-of-the-art multiplication protocol of~\cite{BGIN20} that provides $\abort$ security. Note that although the protocol of~\cite{BGIN20} relies on zero-knowledge proofs, this computation is carried out in the preprocessing phase of our multiplication protocol. Moreover, since preprocessing is done for many instances in one shot, the zero-knowledge proof can benefit from amortization. 
The parties then invoke $\Picta$ to obtain $\tsgr{\clv{\va \vb}}$ from $\sqr{\clv{\va \vb}}$. Looking ahead, $\sqr{\clv{\va \vb}}$ also aids in performing the online verification check.

For task $\mynum{ii}$,  in the online phase, we retain the idea of parties in $\Evlset$ optimistically reconstructing $\vz - \vr$ from their additive shares ($\tsgr{\cdot}$-share) to ensure that only the parties in $\Evlset$ remain active for most of the computation. Moreover, this optimistic reconstruction requires only $\Order(t)$-element communication rather than the $\Order(t^{2})$ required for reconstruction from $\sqr{\cdot}$-shares (which is what will be used later for performing verification, albeit to perform only one such reconstruction). Thus, similar to the semi-honest protocol, parties in $\Evlset$ optimistically reconstruct $\vz - \vr$ towards $\Pking$, who further sends the reconstructed value to the parties in $\Evlset$. In the malicious setting, this approach requires additional care since a malicious party may send a wrong $\tsgr{\cdot}$-share of $\vz - \vr$ to $\Pking$ or a malicious $\Pking$ may send an incorrectly reconstructed (inconsistent) $\vz - \vr$ to the parties. To account for these behaviours, the protocol is augmented with a short one-off verification phase to verify the consistency and correctness of $\vz - \vr$. This phase is executed in the end of the protocol and requires the presence of {\em all} parties, and hence the possession of $\vz - \vr$ by all. This is in contrast to the semi-honest protocol where $\vz - \vr$ is given to only parties in $\Evlset$. To keep $\Hlpset$ disengaged for most of the online phase, sending $\vz - \vr$ to them is deferred till the end of the protocol. This send is a one-off and can be combined for all multiplication gates. Details of verification protocol  $\Pivrfy$ (\boxref{fig:Pivrfy}) are given next.

Verification comprises two checks-- a {\em consistency} check to first verify that $\Pking$ has indeed sent the same $\vz - \vr$ to all the parties, followed by a {\em correctness} check to verify the correctness of the $\vz - \vr$. For the former, parties perform a hash-based consistency check of $\vz - \vr$,  and abort in case of any inconsistency. If $\vz - \vr$ is consistent, parties verify its correctness.  
The high-level idea for verifying correctness is to {\em robustly} reconstruct $\vz - \vr$, but now from its $\sqr{\cdot}$-shares (can be computed given $\sqr{\lv{\va}}, \sqr{\lv{\va}}, \sqr{\clv{\va \vb}}$ that are generated in the preprocessing phase). Parties can then verify if this reconstructed value equals the value received from $\Pking$. Concretely, this is equivalent to robustly reconstructing $\sqr{\Omega} = \sqr{\vz - \vr - (\cmv{\va \vb} -\mv{\va} \lv{\vb} - \mv{\vb} \lv{\va} + \clv{\va \vb} - \vr)}$, where $\vz - \vr$ is the value received from $\Pking$, and verifying if $\Omega = 0$.
For robust reconstruction of $\sqr{\Omega}$, every party sends its $\sqr{\cdot}$-share to every other party who misses this share, and $\abort$s in case of inconsistencies in the received values. Elaborately, reconstruction of $\Omega$ towards $P_s \in \Partyset$ proceeds as follows. For each missing $\sqr{\cdot}$-share of $\Omega$ at $P_s$, each of the $t+1$ parties holding this share send it to $P_s$. $P_s$ uses this share for reconstruction if all the $t+1$ received values are consistent, else it $\abort$s. Presence of at least one honest party among the $t+1$ guarantees that inconsistency, if any, can be detected. %
Since each share in $\sqr{\Omega}$ is held by $t+1$ parties, comprising at least one honest party, any cheating by up to $t$ corrupt parties is guaranteed to be detected. Note that this reconstruction requires communicating $\Order(t^2)$ ring elements to verify the correct computation of a single multiplication gate, the cost of which can be optimized using standard optimization techniques~\cite{AbspoelDEN19, CGHIKLN18}. Concretely, the correctness of $\vz - \vr$ for several multiplication gates can be verified with a single reconstruction by reconstructing a linear combination of $\Omega$ for several gates and verifying equality with $0$. Thus, only one robust reconstruction from $\sqr{\cdot}$-shares is required for several multiplication gates, whose cost gets amortized due to verification across multiple gates.  

\smallskip
\begin{protboxgray}{$\Pivrfy \left(\Partyset, \{ \shr{\va_i}, \shr{\vb_i}, \vz_i - \vr_i, \sqr{\clv{\va_i \vb_i}}, \sqr{\vr_i} \}_{i=1}^{m} \right)$}{Malicious: Verification protocol for all multiplication gates}{fig:Pivrfy}
	\justify
	
	Let $(\va_1, \vb_1, \vz_1),  \ldots, \allowbreak (\va_m, \vb_m, \vz_m)$ denote the inputs and outputs of the $m$ multiplication gates to be verified. \begin{ccsitemize} 
		\item[--] \textit{Consistency Check.} Invoke $\Picon$ on $\{\vz_1 - \vr_1, \ldots, \vz_m - \vr_m\}$. 
		\item[--] \textit{Correctness Check.} Repeat the following $\kappa$ times.
		\begin{inneritemize}
			\item[-] Generate random $\theta_1, \ldots, \theta_m \in \Z{\ell}$ and compute 
			\begin{scriptsize}
				\begin{align*} 
					\sqr{\Omega} =  \sum_{i=1}^{m} \theta_i \left( \vz_i - \vr_i - \left( \cmv{\va_i \vb_i} -\mv{\va_i} \sqr{\lv{\vb_i}} - \mv{\vb_i} \sqr{\lv{\va_i}} + \sqr{\clv{\va_i \vb_i}} - \sqr{\vr_i} \right) \right)
				\end{align*}
			\end{scriptsize}
			\item[-] For each $\sqr{\cdot}$-share of $\Omega$, the $t+1$ parties possessing this share send it to every party that misses this share. If the recipient party receives inconsistent values for any missing share, it $\abort$s.
			\item[-] Reconstruct $\Omega$ and $\abort$ if $\Omega \neq 0$. 
		\end{inneritemize}
	\end{ccsitemize}
\end{protboxgray}

It is worth noting that this random linear combination technique does not trivially work over rings. This is due to the existence of zero divisors which results in the linear combination being $0$ with a probability $1/2$ (which denotes the cheating probability of the adversary)~\cite{AbspoelDEN19}. Hence, to obtain the desired security, the verification check is repeated $\kappa$ times where $\kappa$ is the security parameter. This bounds the cheating probability of adversary to $1/2^{\kappa}$. Another approach is to perform the verification over extended rings \cite{BonehBCGI19, BGIN19}. Specifically, verification operations are carried out over a ring $\Z{\ell}/f(x)$ which is a ring of all polynomials with coefficients in $\Z{\ell}$ modulo a degree $d$ polynomial $f(x)$ that is irreducible over $\Z{}$. Each element of $\Z{\ell}$ is lifted to a degree $d$ polynomial in $\Z{\ell}[x]/f(x)$, which increases the communication required to perform verification by a factor of $d$.

To summarize, the maliciously secure multiplication protocol (see \boxref{fig:piMultMal}) can be broken down into the following.

\begin{description} 
\item[--] Preprocessing phase which involves generation of $\sqr{\clv{\va \vb}}$ by invoking $\FMulPre$. Malicious behaviour, if any, will be caught by $\FMulPre$. $\sqr{\clv{\va \vb}}$ is non-interactively converted into $\tsgr{\cdot}$-shares of $\lv{\va \vb}$. 
$\tsgr{\lv{\va}}, \tsgr{\lv{\vb}}$ is also generated non-interactively.
\smallskip
\item[--]  Generation of $\tsgr{\cdot}$-shares of $\lv{\va}, \lv{\vb}, \clv{\va \vb}$ during preprocessing enables computation of $\tsgr{\vz - \vr}$ in the online phase, and thereby reconstruction of $\vz - \vr$ via $\Pking$. The crucial point to note here is that this requires the presence of only parties in $\Evlset$ in the online phase. This is followed by non-interactive generation of $\shr{\vz - \vr}$ from which $\shr{\vz}$ is computed as $\shr{\vz} = \shr{\vz - \vr} + \shr{\vr}$, where $\shr{\vr}$ is generated during preprocessing. 
\smallskip
\item[--] Finally, to catch malicious behaviour in the online phase, if any, in the verification phase the correctness of the generated $\shr{\vz}$ is checked simultaneously, for each $\vz$ that is the output of a multiplication gate. This is done by invoking $\Pivrfy$. Note that before this verification begins, $\Pking$ sends $\vz - \vr$ corresponding to all multiplication gates to parties in $\Hlpset$ in a single shot.
\end{description}

As pointed out in~\cite{GoyalLS19}, deferring the correctness check to later may result in a privacy breach when using a sharing scheme that allows for redundancy (such as RSS or Shamir sharing). The details are elaborated in \S\ref{appsec:multattack}. However, the crucial point to note here is that although we rely on a variant of RSS which introduces redundancy, recall that while performing a reconstruction towards $\Pking$, we rely on $\tsgr{\cdot}$-sharing of $\vz - \vr$, which is a $(t+1)$-additive sharing. The use of additive sharing while performing reconstruction towards $\Pking$ eliminates any redundancy in the sharing scheme and thus, helps in overcoming this subtle privacy breach, as also shown in~\cite{GoyalLS19}. This privacy breach persists in~\cite{ED20}, and is discussed in~\S\ref{appsec:multattack}.

\begin{lemma}
Protocol $\PiMultMal$ (\boxref{fig:piMultMal}) incurs a communication of $3t$ elements in the preprocessing phase and $3t$ elements in $2$ rounds in the online phase for multiplication when $\istr = 0$.
\end{lemma}

The ideal functionality $\Fmpc$ for evaluating a function $f$ in the $n$-party setting while providing malicious security (with $\abort$) appears in \boxref{fig:FmpcMal}. 

\smallskip
\begin{systemboxgray}{$\Fmpcmal$}{Malicious: Ideal functionality for evaluating function $f$}{fig:FmpcMal}
	\justify
	$\Fmpc$ interacts with the parties in $\Partyset$ and the adversary $\simmal$. Let $f$ denote the function to be  computed. Let $\wx_s$ be the input of party $P_s$, and $\wy_s$ be the corresponding output, i.e $(\{\wy_s\}_{s=1}^{n}) = f(\{\wx_s\}_{s=1}^{n})$. $\simmal$ is also allowed to send a special command, $(\abort, \Ab)$, which indicates that honest parties in $\Partyset$ with indices in $\Ab$ should $\abort$.
	\begin{myitemize}
		\item[ ] {\bf Step 1:} $\Fmpc$ receives $(\INPUT,\wx_s)$ from $P_s \in \Partyset$. If $(\INPUT, \ast)$ already received from $P_s$, then ignore the current message. Otherwise, record $x_s^{\prime} = x_s$ internally. 
		\item[ ] {\bf Step 2:} Compute $(\{\wy_s\}_{s=1}^{n}) = f(\{\wx_s\}_{s=1}^{n})$ and send the output $\wy_s$ for a corrupt $P_s$ to $\simmal$. 
		\item[ ] {\bf Step 3:} If received $(\SIGNAL, \abort, \Ab)$ from $\simmal$, set $\wy_s = \abort$ for $P_s$, where $s \in \Ab$. Send $(\OUTPUT, \wy_s)$ to honest $P_s \in \Partyset$.
	\end{myitemize}
\end{systemboxgray}

\paragraph{Multiplication with truncation} Similar to the semi-honest protocol, truncation can be incorporated in the malicious multiplication as well without inflating the online communication. For this, we rely on maliciously secure ideal functionality, $\Ftrgenmal$ (\boxref{fig:Ftrgenmal}), to generate the $\shr{\cdot}$-shares of $(\vr, \trunc{\vr})$ and is instantiated using $\PiDSBitsMal$ \cite{Damgard0FKSV19} protocol in our work. On a high level, the semi-honest versions of interactive operations such as multiplication and reconstruction in $\PiDSBits$ are replaced with their maliciously secure counterparts in $\PiDSBitsMal$, and more details are provided in \S\ref{appsec:malmpc}.

\paragraph{Dot product}
Similar to the maliciously secure multiplication protocol that relied on $\FMulPre$ to generate $\sqr{\cdot}$-shares of the multiplicative term, $\clv{\va \vb}$ in the preprocessing phase, the maliciously secure dot product protocol invokes $\FDotPPre$ (\boxref{fig:FDotPPre}) to generate $\sqr{\cdot}$-shares of the multiplicative term, $\sum_{k=1}^{\nf} \clv{\vx_k \vy_k}$, required to compute the dot product as per equation \eqref{maskeddotp}. Given $\sqr{\cdot}$-shares of $\sum_{k=1}^{\nf} \clv{\vx_k \vy_k}$, online phase proceeds similar to that of multiplication.

Observe that a trivial realization of $\FDotPPre$ can be reduced to $\nf$ instances of multiplication. However, we extend the ideas from~\cite{SWIFT} and rely on a distributed zero-knowledge proof~\cite{BGIN20} to eliminate the vector-size dependency in the preprocessing phase. Concretely, we instantiate $\FDotPPre$ using a semi-honest dot product protocol~\cite{GoyalS20} whose cost matches that of semi-honest multiplication~\cite{DamgaardN07} (and thus is independent of the vector-size), followed by a verification phase to verify the correctness of the dot product computation. For the verification, we extend the verification technique for multiplication in~\cite{BGIN20}, to now verify the correctness of dot product, such that the cost due to verification can be amortized away for multiple dot products, thereby resulting in vector-size independent preprocessing. Details of this extension are deferred to~\S\ref{appsec:malmpc}.

\paragraph{Multi input multiplication}
This protocol is similar to its semi-honest counterpart with the difference that the preprocessing phase relies on invoking $\FMulPre$ for generating the required multiplicative terms. The details are deferred to \S\ref{appsec:multiinputmultmal}. \tabref{mmultmal} compares the cost of computing multi-input multiplication via a 2-input multiplication sequentially vs. the multi-input multiplication protocol.

\vspace{-2mm}
\begin{table}[htb!]
	\centering
	\resizebox{.4\textwidth}{!}{
		\begin{NiceTabular}{r r r r r}
			\toprule
			\Block[c]{2-1}{Multiplication\\type} & \Block[c]{2-1}{Building\\Block}
			& \Block[c]{1-2}{Communication} & & \Block[c]{2-1}{Online\\Rounds}\\
			\cmidrule{3-4}
			& & Prep. & Online & \\
			\midrule
			
			\Block{2-1}{$\vz = \va \vb \vc$} & 2-input mult. & $6t \ell$ & $6t \ell$ & $4$ 
			\\
			 & 3-input mult. & $12t \ell$ & $3t \ell$ &  $2$
			\\
			\midrule 
			
			\Block{2-1}{$\vz = \va \vb \vc \vd$} & 2-input mult. & $9t \ell$ & $9t \ell$ & $4$ 
			\\
			 & 4-input mult. & $33t \ell$  & $3t \ell$ & $2$ 
			\\

			\bottomrule
		\end{NiceTabular}
	}
	\caption{\small 
		Malicious: Communication and round complexity for computing multi-input multiplications \label{tab:mmultmal}}
	\vspace{-3mm}
\end{table}

\section{Applications \& benchmarks}
\label{sec:Implementation}
To evaluate the performance of our protocols, we benchmark some of the popular applications such as deep neural networks (NN), graph neural networks (GNN), similar sequence queries (SSQ), and biometric matching where MPC is used to achieve privacy. While these applications have been looked at in the small party setting~\cite{MohasselZ17,SWIFT,SCSDF20,AHLR18,ST19,Falcon,aby2,MR18}, we believe the $n$-party setting is a better fit for reasons described in the introduction. To the best of our knowledge, we are the first to benchmark these in the multiparty honest-majority setting for more than four parties.

\paragraph{Benchmark environment}

\begingroup

The performance of our protocols is analyzed using a prototype implementation building over the ENCRYPTO library~\cite{ENCRYPTO} in C++17.  

We chose $64$ bit ring ($\Z{64}$) for our arithmetic world, and the operations over extended ring were carried out using the NTL library\footnote{\url{https://libntl.org}}. 
Since the correctness and accuracy of the applications considered in the secure computation setting are already established, our benchmark aims to demonstrate our protocols' performance and is not fully functional. Moreover, we believe that incorporating state-of-the-art code optimizations like GPU-assisted computing can enhance the efficiency of our protocols, which is left as future work. Since there is no defined way to capture an adversary's misbehaviour, following standard practice~\cite{MR18, SWIFT, DEK20}, we benchmark honest executions of the protocols, which also include the steps performed for verification in the malicious case.
\setlength{\columnsep}{10pt}%
\setlength{\intextsep}{3pt}%
\begin{wrapfigure}{r}{0.28\textwidth}
	\centering
	\includegraphics[clip, trim=0pt 0pt 0pt 0pt, width=0.24\textwidth]{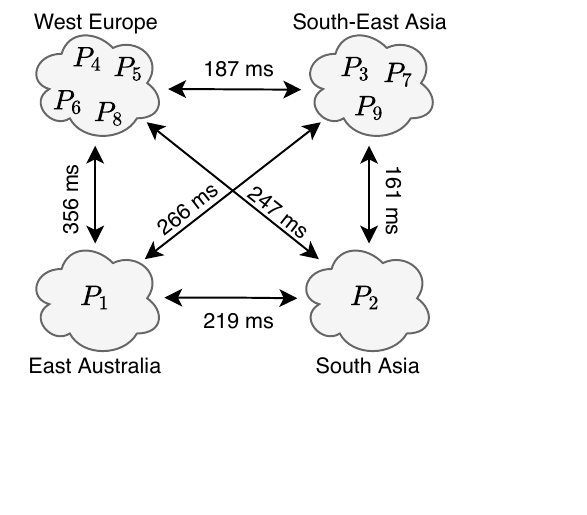}
	\vspace{-1mm}
	\caption{\small Round trip time ($\rtt$)} \label{fig:rtt}
	\vspace{-2mm}
\end{wrapfigure}
We use multi-threading, wherever possible, to facilitate efficient computation and communication among the parties. The parties in the computation are emulated using Google Cloud (n1-standard-64 instances, 2.0 GHz Intel Xeon Skylake, 64 vCPUs, 240 GB RAM) with machines located in East Australia, South Asia, South East Asia, and West Europe. All our experiments are run for 5, 7, and 9 parties, each. 

\endgroup

\paragraph{Benchmark parameters}
We report the run-time and communication of the online phase and total (= preprocessing + online). To capture the effect of online round complexity and communication in one go, we also report the throughput ($\TP$~\cite{AFLNO16,MR18,SWIFT}) of the online phase. $\TP$ denotes the number of operations that can be performed in one minute. Finally, when deployed in the outsourced setting, one pays the price for the communication and up-time of the hired servers. To demonstrate how our protocols fare in this scenario, we additionally report the monetary cost (Cost)~\cite{MP0SY20, Tetrad} for the applications considered. This cost is estimated using Google Cloud Platform~\cite{GCloud} pricing, where $1$ GB and $1$ hour of usage costs USD $0.08$ and USD $3.04$, respectively.

\subsection{Comparison with DN07} \label{bench:dn}
In this section, we benchmark our semi-honest and malicious protocols over synthetic circuits comprising one million multiplications with varying depths of 1, 100, and 1000, and compare against the optimized ring variant of DN07~\cite{BonehBCGI19}. The gates are distributed equally across each level in the circuit. 

\paragraph{Communication}
The communication cost for 1 million multiplications is tabulated in \tabref{cktcomm} for the 5, 7, and 9 party settings. As can be observed, the online phase of our semi-honest protocol enjoys the benefits of pushing $33\%$ communication to a preprocessing phase compared to DN07. The observed values corroborate the claimed improvement in the online complexity of our protocol. Our malicious protocol retains the online communication cost of DN07 while incurring a similar overhead in the preprocessing.

\vspace{-1mm}
\begin{table}[htb!]
	\centering
	\resizebox{.44\textwidth}{!}{
		\begin{NiceTabular}{rrrr}[notes/para] 
			\toprule
			Ref. & $n=5$ & $n=7$ & $n=9$ \\
			\midrule
			DN07 (semi)   & (0, 45.78)    & (0, 68.66)    & (0, 91.55) \\
			$\this$ (semi)            & (15.26, 30.52)    & (22.88, 45.78)    & (30.51, 61.04) \\
			$\this$~~(mal)              & (45.79, 45.78)    & (68.67, 68.67)    & (91.57, 91.57) \\
			\bottomrule
		\end{NiceTabular}
	}
	\vspace{-1mm}
	\caption{\small 
		Communication (Preprocessing, Online) in MB for 1 million multiplications\label{tab:cktcomm}}
	\vspace{-4mm}
\end{table}

%
Note that pushing the communication to the preprocessing phase has several benefits. First, communication with respect to several instances can happen in a single shot and leverage the benefit of serialization. Second, with respect to resource-constrained devices such as mobile phones, the preprocessing communication can occur whenever they have access to a high-bandwidth Wi-Fi network (for instance, when the device is at home overnight). These benefits facilitate a fast online phase, as observed, that may happen over a low-bandwidth network. 

\paragraph{Run-time}
The time taken to evaluate circuits of different depths appears in \tabref{ckttime}. Since the time for the  5, 7, and 9 party settings vary within the range [0, 0.5], we report values only for the 7-party setting in \tabref{ckttime}. With respect to the online run-time, our semi-honest protocol's time is expected to be similar to that of DN07. However, DN07 demonstrates around 1.5$\times$ higher run-time. This difference can be attributed to the asymmetry in the $\rtt$ among parties, which vanished when benchmarked over a symmetric $\rtt$ setting. Compared to the semi-honest protocol, the malicious variant incurs a minimal overhead of less than one second in the online run-time due to the one-time verification phase. However, the overhead is higher for the case of the overall run-time. Concretely, it is around 10 seconds and is due to the distributed zero-knowledge proof computation in the preprocessing phase. Note that this overhead is independent of the circuit depth and gets amortized for deeper circuits as evident from \tabref{ckttime} (depth 1 vs. 1000).

\vspace{-1mm}
\begin{table}[htb!]
	\centering
	\resizebox{.44\textwidth}{!}{
		\begin{NiceTabular}{rrrr}[notes/para] 
			\toprule
			Ref. & $d=1$ & $d=100$ & $d=1000$ \\
			\midrule
			DN07 (semi)   & (0, 0.65)    & (0, 54.97)    & (0, 549.69) \\
			$\this$ (semi)            & (0.47, 0.45)    & (0.47, 30.75)    & (0.47, 307.48) \\
			$\this$~~(mal)              & (10.52, 1.36)    & (10.53, 68.67)    & (10.54, 308.39) \\
			\bottomrule
		\end{NiceTabular}
	}
	\vspace{-2mm}
	\caption{\small 
	 Latency in seconds (Preprocessing, Online) for varying depth ($d$) circuits with 1 million multiplications for $n=7$\label{tab:ckttime}}
	 \vspace{-4mm}
\end{table}

\paragraph{Monetary Cost}
Another key highlight of our protocols is their improved monetary cost, as evident from \boxref{fig:TableIVPlot}. Concretely, for 9 parties (semi-honest), we observe a saving of 17$\%$ over DN07 for a depth-1 circuit, and it increases up to 72$\%$ for circuits with depth 1000. This is primarily due to the reduction in the number of online parties over DN07. 
Comparing our semi-honest and malicious variants, the latter has an overhead of 8$\times$ for depth-1 circuit, and it reduces to 1.14$\times$ for depth-1000 circuit. This is justified because the verification cost is amortized for deeper circuits, as mentioned earlier. Interestingly, our malicious variant outperforms even the semi-honest DN07 upon reaching circuit depths of 100 and above. 
A similar analysis holds in the symmetric $\rtt$ setting as well, where the saving is up to $56\%$ (for $d = 1000$). 
%

\begin{figure}[htb!]
	\centering
		\resizebox{.3\textwidth}{!}{
			\begin{tikzpicture}[
				every axis/.style={ 
					ybar stacked,
					ymin=0,ymax=13,
					xtick={1,2,3}, xticklabels={$d=1$,$d = 100$,$d = 1000$},
					enlarge x limits=0.25,
					cycle list name=exotic, 
					every axis plot/.append style={fill,draw=none,no markers},
					legend style = {anchor = south, legend columns = -1, draw=none, area legend},
					bar width=14pt},]
				
				\begin{axis}[bar shift=-16pt,hide axis, legend style = {at={(0.2, 0.875)}}]
					\addlegendentry{\footnotesize $\this$ (semi)}
					\addplot[fill=darkred] coordinates
					{(1,3.807) (2,6.340) (3,9.430)};
				\end{axis}
				
				\begin{axis}[hide axis, legend style = {at={(0.2, 0.8)}}]
					\addlegendentry{\footnotesize $\this$\hspace*{1.8mm}(mal)}
					\addplot+[fill=UniBlau] coordinates
					{(1,6.807) (2,7.484) (3,9.622)};
				\end{axis}
				
				\begin{axis}[bar shift=16pt, legend style = {at={(0.2, 0.725)}}]
					\addlegendentry{\footnotesize \hspace*{0mm}\cite{DamgaardN07} (semi)}
					\addplot+[] coordinates
					{(1,4.087) (2,8.190) (3,11.274)};
				\end{axis}
			
				\begin{axis}[bar shift=-16pt,hide axis, legend style = {at={(0.2, 0.875)}}]
					\addplot[fill=none,
					postaction={
                        pattern=crosshatch,
                        pattern color=darkred
                    }] coordinates
					{(1,3.907) (2,7.248) (3,10.457)};
				\end{axis}
				
				\begin{axis}[hide axis, legend style = {at={(0.2, 0.8)}}]
					\addplot+[fill=none,
					postaction={
                        pattern=crosshatch,
                        pattern color=UniBlau
                    }] coordinates
					{(1,6.966) (2,8.039) (3,10.566)};
				\end{axis}
				
				\begin{axis}[bar shift=16pt, legend style = {at={(0.2, 0.725)}}]
					\addplot+[fill=none,
					postaction={
                        pattern=crosshatch,
                        pattern color=.
                    }] coordinates
					{(1,4.087) (2,8.401) (3,11.672)};
				\end{axis}

			\end{tikzpicture}

    	}
	\vspace{-2mm}
	\caption{\small Monetary cost (in USD) for evaluating circuits (1000 instances) of various depths ($d$) for $n=9$ parties. The values are reported in $\log_2$ scale. Bars in solid colors denote computation over network given in \boxref{fig:rtt}, while the area represented via crosshatch pattern denotes the additional cost incurred in the symmetric rtt setting (356 ms).}\label{fig:TableIVPlot}
	\vspace{-4mm}
\end{figure}
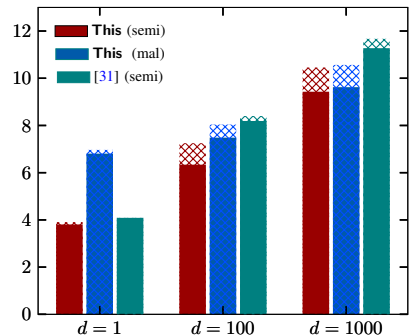

\begin{figure*}[htb!]
	\centering
	\begin{subfigure}[t]{0.32\textwidth}
		\centering
		\resizebox{.85\textwidth}{!}{
			\begin{tikzpicture}
				\begin{axis}[legend style={at={(1,1)}}, anchor=north west, legend pos= north east,
                ylabel={Throughput}, xlabel={ }, 
					ylabel near ticks, 
					xtick={2,3,4,5}, xticklabels={GNN, NN-1, N-2, NN-3}, 
					cycle list name=exotic]
					\addplot plot coordinates { (2,9.05) (3,7.73) (4,6.38) (5,4.65)};
					\addlegendentry{{\footnotesize DN07-5}}
					\addplot plot coordinates { (2,8.99) (3,7.66) (4,6.31) (5,4.58)};
					\addlegendentry{{\footnotesize DN07-7}}
					\addplot plot coordinates { (2,8.99) (3,7.66) (4,6.31) (5,4.58)};
					\addlegendentry{{\footnotesize DN07-9}}
					
					\addplot plot coordinates { (2,10.80) (3,9.70) (4,8.44) (5,6.70)};
					\addlegendentry{{\footnotesize {\bf This} 5, 7, 9}}
					
				\end{axis}
			\end{tikzpicture}
		}
		\vspace{-1mm}
		\caption{\small Online throughput}\label{fig:on_tp}
	\end{subfigure}
	~
	\begin{subfigure}[t]{0.32\textwidth}
		\centering
		\resizebox{.85\textwidth}{!}{
			\begin{tikzpicture}
				\begin{axis}[legend pos= north west, ylabel={End-to-end runtime (seconds)}, xlabel={ }, ylabel near ticks, 
					xtick={2,3,4,5}, xticklabels={GNN, NN-1, N-2, NN-3}, 
					cycle list name=exotic]
					\addplot plot coordinates { (2,4.09) (3,4.42) (4,5.82) (5,7.74)};
					\addlegendentry{{\footnotesize DN07-5}}
					\addplot plot coordinates { (2,4.13) (3,4.48) (4,5.90) (5,7.83)};
					\addlegendentry{{\footnotesize DN07-7}}
					\addplot plot coordinates { (2,4.13) (3,4.48) (4,5.90) (5,7.83)};
					\addlegendentry{{\footnotesize DN07-9}}
					
					\addplot plot coordinates { (2,3.17) (3,3.47) (4,4.66) (5,6.70)};
					\addlegendentry{{\footnotesize {\bf This} 5, 7, 9}}
					
				\end{axis}
			\end{tikzpicture}
		}
		\vspace{-1mm}
		\caption{\small End-to-end runtime}\label{fig:tot_time}
	\end{subfigure}
	~
	\begin{subfigure}[t]{0.32\textwidth}
		\centering
		\resizebox{.85\textwidth}{!}{
			\begin{tikzpicture}
				\begin{axis}[legend pos= north west, ylabel={End-to-end monetary cost (USD)}, xlabel={ }, ylabel near ticks, 
					legend columns=2,
					xtick={2,3,4,5}, xticklabels={GNN, NN-1, N-2, NN-3}, 
					cycle list name=exotic]
					\addplot plot coordinates { (2,7.66) (3,5.98) (4,7.72) (5,10.38)};
					\addlegendentry{{\footnotesize DN07-5}}
					\addplot plot coordinates { (2,7.43) (3,4.64) (4,6.63) (5,9.84)};
					\addlegendentry{{\footnotesize {\bf This} 5}}
					\addplot plot coordinates { (2,8.19) (3,6.58) (4,8.31) (5,10.93)};
					\addlegendentry{{\footnotesize DN07-7}}
					\addplot plot coordinates { (2,8.00) (3,5.11) (4,7.14) (5,10.07)};
					\addlegendentry{{\footnotesize {\bf This} 7}}
					\addplot plot coordinates { (2,8.56) (3,6.99) (4,8.71) (5,11.32)};
					\addlegendentry{{\footnotesize DN07-9}}
					\addplot plot coordinates { (2,8.35) (3,5.46) (4,7.49) (5,10.74)};
					\addlegendentry{{\footnotesize {\bf This} 9}}
					
				\end{axis}
			\end{tikzpicture}
		}
		\vspace{-1mm}
		\caption{\small End-to-end monetary cost (1000 queries)}\label{fig:tot_mon_cost}
	\end{subfigure}
	\vspace{-5mm}
	\caption{\small Comparison for GNN and deep NN between our semi-honest protocol and DN07 (values plotted are logarithmic in base 2) }\label{fig:nngraph}
	\vspace{-5mm}
\end{figure*}
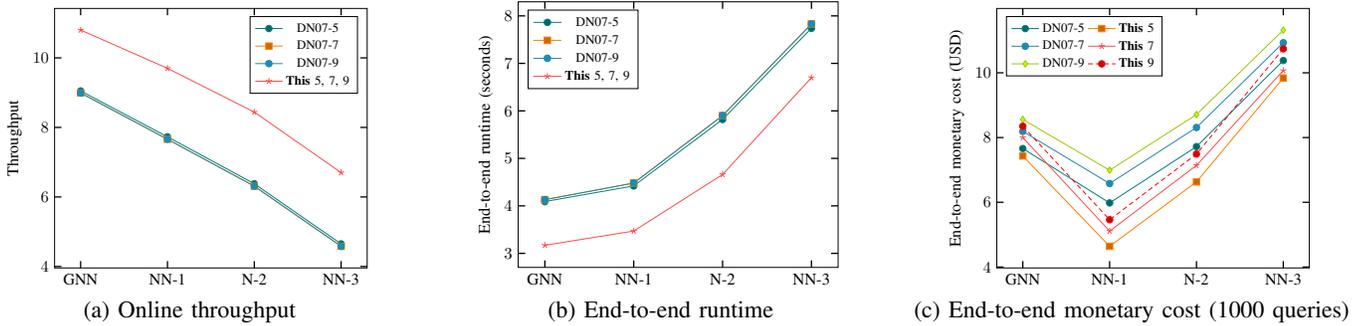

\paragraph{Online Throughput ($\TP$):} 
Owing to the asymmetric $\rtt$ as described earlier, our semi-honest variant witnesses up to $1.78 \times$ improvements in $\TP$ (for a single execution) over DN07, which vanishes in the symmetric $\rtt$ setting. However, recall that our protocol requires only $t+1$ active parties in the online phase, which leaves several channels among the parties underutilized. Hence, we can leverage the load balancing technique where parties' roles are interchanged across various parallel executions. For instance, one approach is to make every party act as $\Pking$, i.e., in 5PC, in one execution, $\Pking = P_1, \Evlset = \{P_1, P_2, P_3\}, \Hlpset = \{P_4, P_5\}$, while in another execution $\Pking = P_2, \Evlset = \{P_2, P_3, P_4\}, \Hlpset = \{P_5, P_1\}$, and so on. To analyse the effect of load balancing, we performed experiments with similar $\rtt$ among the parties and observed a 1.5$\times$ improvement in our semi-honest variant over DN07. This is justified as we communicate over four channels among the parties as opposed to six in DN07.
We note that while enhancing the security from semi-honest to malicious, we observe a significant drop in $\TP$, which is about 3$\times$ for the depth-1 circuit. This is primarily due to increased run time owing to the verification in  online phase for malicious setting. However, this drop tends to zero for deeper circuits (as verification cost gets amortized), making online phase of our maliciously secure protocol on par with semi-honest one.

\subsection{Deep Neural Networks (DNN) and Graph Neural Networks (GNN)}
We benchmark three different neural networks (NN)~\cite{MR18, BLAZE, Falcon} with increasing number of parameters--(i) NN-1: a 3-layer fully connected one from~\cite{MohasselZ17}, (ii) NN-2: the LeNet~\cite{lenet} architecture, and (iii) NN-3: VGG16~\cite{vgg16} architecture (further details are deferred to \S\ref{app:dnn}). We benchmark the inference phase of the above NNs, which comprises computing activation matrices, followed by applying an activation function or pooling operation, depending on the network architecture. NN-1 and NN-2 are benchmarked over MNIST dataset~\cite{MNIST10} while NN-3 is benchmarked using CIFAR-10 dataset~\cite{CIFAR10}. We also benchmark GNN inference, for which we use the simplified architecture of~\cite{defferrard} given in~\cite{SCSDF20}. This architecture~(\S\ref{app:gnn}) is shown to achieve an accuracy of more than $99\%$ on MNIST classification~\cite{SCSDF20}. To analyse the improvement of our protocols, we also benchmark (semi-honest) DN07 for the applications by adapting our building blocks to their setting.

The semi-honest benchmarks for the different NNs and GNN appear in \tabref{comparison-nnsh} (\S\ref{app:dnn}) while the malicious ones appear in \tabref{comparison-nnmal} (\S\ref{app:dnn}).  \boxref{fig:nngraph} gives a pictorial view of the trends observed while comparing the semi-honest variants and are described next. We incur a very minimal overhead in the run-time of our protocols when moving from five to nine parties over all the networks considered. Hence, we use $\pm \delta$ to denote this variation in the table. The trends witnessed in synthetic circuit benchmarks (\S\ref{bench:dn}) carry forward to neural networks as well due to reasons discussed previously. For instance, the improvement in the online run-time for our semi-honest variant is up to 
4.3$\times$ over DN07. The effect of reduced run-time and improved communication results in a significant improvement in online throughput of our protocol over DN07. Concretely, the gain ranges up to 4.3$\times$. 
Further, the improved run-time coupled with the reduced number of online parties for our case brings in a saving of up to 69$\%$ 
in monetary cost for NN-1. However, the improvement drops to 33$\%$ 
for deep network NN-3. The reduction in savings is due to improved run-time getting nullified by increased communication from NN-1 to NN-3, making communication the dominant factor in determining monetary cost. 

Observe that, unlike the case in synthetic circuits (\tabref{cktcomm}), the total communication here is an order of magnitude higher. This is primarily due to the higher communication cost incurred for performing the truncation operation--specifically, generation of the doubly-shared bits ($\PiDSBits$, \boxref{fig:piDSBits}) in the preprocessing phase. It is worth noting that $\PiDSBits$ is used as a black-box, and an improved instantiation for it will lower the communication. Similar trends are observed for GNN as well, where the online run-time of DN07 is up to 3.5$\times$ 
higher than our semi-honest protocol. 
This is reflected in the throughput where we gain up to 3.4$\times$. 
Further, we observe savings of up to 15$\%$ in monetary cost due to the reduced number of active parties and lesser run-time. 

Moving to the malicious setting, we incur an overhead of up to 3$\%$ in online run-time, 6$\%$ in communication, and 13$\%$ in monetary cost over the semi-honest counterpart. Details are deferred to \S\ref{app:dnn}.


\subsection{Genome Sequence Matching}
Given a genome sequence as a query, genome matching aims to identify the most similar sequence from a database of sequences. This task is also known as similar sequence query (SSQ). It requires the computation of Edit Distance (ED), which quantifies how different two sequences are by identifying the minimum number of additions, deletions, and substitutions required to transform one sequence to the other. 
To compute the ED, we extend the (2-party) protocol from \cite{ST19} which builds on top of the approximation from \cite{AHLR18}, to the $n$-party setting. The details of the approximation algorithm for ED computation appear in \S\ref{app:ssq}. The accuracy and correctness of this algorithm follow from~\cite{AHLR18}. Among the two phases of the ED algorithm, where the first phase happens non-interactively, we only focus on the second phase of ED, which requires interaction and benchmark the same. 

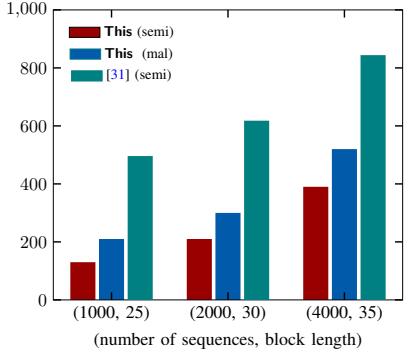
\begin{figure}[htb!]
	\centering
	\resizebox{.3\textwidth}{!}{
		\begin{tikzpicture}[
			every axis/.style={ 
				ybar stacked,
				ymin=0,ymax=1000,
				xtick={1,2,3}, xticklabels={(1000, 25),(2000, 30),(4000, 35)},
				enlarge x limits=0.25,
				xlabel={(number of sequences, block length)}, 
				cycle list name=exotic, 
				every axis plot/.append style={fill,draw=none,no markers},
				legend style = {anchor = south, legend columns = -1, draw=none, area legend},
				bar width=14pt},]
			
			\begin{axis}[bar shift=-16pt,hide axis, legend style = {at={(0.2, 0.875)}}]
				\addlegendentry{\footnotesize $\this$ (semi)}
				\addplot[fill=darkred] coordinates
				{(1,130) (2,210) (3,390)};
			\end{axis}
			
			\begin{axis}[hide axis, legend style = {at={(0.2, 0.8)}}]
				\addlegendentry{\footnotesize $\this$\hspace*{1.8mm}(mal)}
				\addplot+[fill=UniBlau] coordinates
				{(1,210) (2,300) (3,520)};
			\end{axis}
			
			\begin{axis}[bar shift=16pt, legend style = {at={(0.2, 0.725)}}]
				\addlegendentry{\footnotesize \hspace*{0mm}\cite{DamgaardN07} (semi)}
				\addplot+[] coordinates
				{(1,496) (2,618) (3,844)};
			\end{axis}
		\end{tikzpicture}
	}
	\vspace{-2mm}
	\caption{\small Monetary cost for SSQ evaluation for varying number of sequences and block lengths ((1000,25), (2000, 30), (4000,35)) for $n=9$ parties. Costs for 1000 instances are reported in USD.}\label{fig:mon_nine_ssq}
	\vspace{-4mm}
\end{figure}



The benchmarks for genome sequence matching appear in \tabref{comparison-ssqshint}, \tabref{comparison-ssqshext} (\S\ref{app:ssq}). 
Following~\cite{ST19}, we consider three cases with different number of sequences in the database ($\seqn$) and different block lengths ($\seql$). The benchmarks for $\seqn = 2000, \seql = 30$ are reported in \tabref{comparison-ssqshint}, while the ones for $\seqn = 1000, \seql = 25$ and $\seqn = 4000, \seql = 35$ appear in \tabref{comparison-ssqshext}. We witness similar trends here, where our semi-honest protocol has improvements of up to 4$\times$ in both online run-time and throughput over DN07. Our malicious variant incurs a minimal overhead in the range of 5-6$\%$ in online run-time and total communication over the semi-honest counterpart. For the monetary cost (\boxref{fig:mon_nine_ssq}), our semi-honest protocol has up to 66$\%$ 
saving over DN07, and malicious variant has around 42$\%$-54$\%$ overhead over semi-honest counterpart.

\begin{table}[htb!]
	\centering
	\resizebox{.47\textwidth}{!}{
		\begin{NiceTabular}{rr|rrr|rrr}[notes/para] 
			\toprule
			\Block{2-1}{Ref.} & \Block{2-1}{$n$} & \Block[c]{1-3}{Online} && & \Block[c]{1-3}{End-to-end} \\
			\cmidrule{3-8}
			& & Comm\tabularnote{communication in MB} 
			& Time
			& $\TP$\tabularnote{$\TP$ denotes throughput}
			& Comm\tabularnote{communication in GB}  
			& Time\tabularnote{Time in seconds}
			& Cost\tabularnote{monetary cost in USD} 
			\\
			\midrule 

			\Block{3-1}{DN07}     
			& 5 & 25.82 & 66.33 & 57.89 & 0.40 & 82.24 & 0.31 \\
			& 7 & 38.75 & 69.63 & 55.15 & 0.60 & 86.99 & 0.47 \\
			& 9 & 51.67 & 69.66 & 55.09 & 0.80 & 87.39 & 0.62 \\
			\midrule 
			\Block{3-1}{$\this$\\(semi)}     
			& 5 & 15.39  & \Block[c]{3-1}{17.61\\$\pm$.02} & \Block[r]{3-1}{217.93\\$\pm$0.2}  & 0.40 & \Block[c]{3-1}{21.69\\$\pm$.02} & 0.11 \\
			& 7 & 23.08  &  &   & 0.60 &  & 0.16 \\
			& 9 & 30.78  &  &   & 0.80 &  & 0.21 \\
			\midrule 
			\Block{3-1}{$\this$\\(mal)}
			& 5 & 22.79 & \Block[c]{3-1}{18.3\\$\pm$.2} & 209.84 & 0.42 & \Block[c]{3-1}{34.52\\$\pm$.2} & 0.17 \\
			& 7 & 33.88 &  & 209.49 & 0.64 &  & 0.25 \\
			& 9 & 44.06 &  & 207.23 & 0.85 &  & 0.30 \\

			\bottomrule
		\end{NiceTabular}
	}
	\vspace{-2mm}
	\caption{\small 
		Genome sequence matching for $\seqn = 2000, \seql = 30$. \label{tab:comparison-ssqshint}}
		\vspace{-6mm}
\end{table}


\subsection{Biometric Matching}
We extend support for biometric matching, which finds application in many real-world tasks such as face recognition~\cite{ErkinFGKLT09} and fingerprint matching~\cite{Henecka013}. The goal of such computation is to identify a sample from a database of $m$ samples that is ``closest'' to a sample $\vec{u}$ held by a user. We follow the general trend and reduce the biometric matching problem to that of finding the sample from the database which has the least Euclidean Distance ($\ED{}{}$) with the user's sample $\vec{u}$. Details of the protocol are deferred to \S\ref{app:bio}. 

The benchmarks for biometric matching appear in \tabref{comparison-bioshext}, \tabref{comparison-bioshint} (\S\ref{app:bio}). The former table considers the case with 1024 and 65536 sequences in the database, while the latter considers 4096 and 16384 sequences.
As is evident from \tabref{comparison-bioshext}, our semi-honest protocol witnesses a 4.6$\times$ 
improvement over DN07 in both online run-time and throughput. Further, in terms of monetary cost, we observe a saving of around 85$\%$. 
With respect to our maliciously secure protocol, we incur a minimal overhead of around 9.5$\%$ in terms of total communication and around 4$\%$ in online throughput over our semi-honest variant. We note that our malicious variant outperforms semi-honest DN07 in both online run-time and throughput, thereby achieving our goal of a fast online phase. 

\begin{table}[htb!]
	\centering
	\resizebox{.49\textwidth}{!}{
		\begin{NiceTabular}{rrr|rrr|rrr}[notes/para][tabularnote = Communication in MB and time in seconds.] 
			\toprule
			\Block{2-1}{\#seq} & \Block{2-1}{Ref.} & \Block{2-1}{$n$} & \Block[c]{1-3}{Online} & & & \Block[c]{1-3}{End-to-end} \\
			\cmidrule{4-9}
			& & & Comm 
			& Time  
			& $\TP$\tabularnote{$\TP$ denotes throughput} 
			& Comm  
			& Time 
			& Cost\tabularnote{monetary cost in USD} 
			\\
			\midrule 
			

			\Block{9-1}{1024}
			& \Block{3-1}{DN07}    
			& 5 & 0.63 & 55.52 & 69.17 & 6.92  & 66.55 & 0.20 \\
			& & 7 & 0.94 & 58.27 & 65.90 & 10.38 & 69.32 & 0.30 \\
			& & 9 & 1.25 & 58.30 & 65.88 & 13.84 & 69.35 & 0.40 \\
			\cmidrule{2-9}
			& \Block{3-1}{$\this$\\(semi)} 
			& 5 & 0.09  & \Block[c]{3-1}{12.61\\$\pm$.02} & \Block[c]{3-1}{304.62\\$\pm$.03} & 6.93  & \Block[c]{3-1}{14.79\\$\pm$.02} & 0.03 \\
			& & 7 & 0.13  &  &  & 10.40  &  & 0.05 \\
			& & 9 & 0.18  &  &  & 13.86 &  & 0.06 \\
			\cmidrule{2-9}
			& \Block{3-1}{$\this$\\(mal)}
			& 5 & 0.14 & \Block[c]{3-1}{13.43\\$\pm$.02} & \Block[c]{3-1}{285.93\\$\pm$.2} & 7.61  & \Block[c]{3-1}{26.67\\$\pm$.02} & 0.08 \\
			& & 7 	& 0.21 &  &  & 11.42 &  & 0.11 \\
			& & 9 & 0.28 &  &  & 15.22 &  & 0.14 \\
			\midrule 	
			

			\Block{9-1}{65536}
			& \Block{3-1}{DN07}    
			& 5 & 40.14 & 88.64 & 43.32 & 443.34 & 108.53 & 0.40 \\
			& & 7 & 60.23 & 93.04 & 41.27 & 665.00 & 114.45 & 0.59 \\
			& & 9 & 80.31 & 93.10 & 41.16 & 886.67 & 114.55 & 0.79 \\
			\cmidrule{2-9}
			& \Block{3-1}{$\this$\\(semi)} 
			& 5 & 5.62   & \Block[c]{3-1}{19.99\\$\pm$.02}  & \Block[c]{3-1}{192.09\\$\pm$.04}  & 443.99 & \Block[c]{3-1}{24.62\\$\pm$.1}  & 0.13 \\
			& & 7 & 8.44   &   &   & 665.99 &   & 0.18 \\
			& & 9 & 11.25  &   &   & 887.99 &   & 0.23 \\
			\cmidrule{2-9}
			& \Block{3-1}{$\this$\\(mal)}
			& 5 & 8.44  & \Block[c]{3-1}{20.86\\$\pm$.02} & \Block[c]{3-1}{183.88\\$\pm$.07} & 486.85 & \Block[c]{3-1}{37.33\\$\pm$.05} & 0.18 \\
			& & 7 & 12.67 &  &  & 730.28 &  & 0.26 \\
			& & 9 & 16.89 &  &  & 972.72 &  & 0.33 \\ 	
			
			
			\bottomrule
		\end{NiceTabular}
	}
	\vspace{-2mm}
	\caption{\small 
		Benchmarks for biometric matching. \label{tab:comparison-bioshext}}
		\vspace{-4mm}
\end{table}

\section*{Conclusion}
This work improves the practical efficiency of $n$-party honest-majority protocols using {\em function-dependent} preprocessing. While our first construction achieves a fast online phase compared to the semi-honest protocol of DN07, the second enhances security by tolerating malicious adversaries with minimal overhead in the online phase. The active participation of half of the participants in both of our constructions is a major highlight. This reduction in online parties results in monetary benefits in real-world deployments.

\section*{Acknowledgements}
The authors would like to acknowledge support from Centre for Networked Intelligence (a Cisco CSR initiative) at the Indian Institute of Science, Bengaluru, SERB MATRICS (Theoretical Sciences) Grant 2020, Google India AI/ML Research Award 2020, DST National Mission on Interdisciplinary Cyber-Physical Systems (NM-CPS) 2020,   National Security Council, India, and the support from Google Cloud to perform the benchmarking.

This project has received funding from the European Research Council (ERC) under the European Union’s Horizon 2020 research and innovation program (grant agreement No. 850990~(PSOTI). This work was co-funded by the Deutsche Forschungsgemeinschaft~(DFG) -- SFB~1119 CROSSING/236615297.

\clearpage
\bibliographystyle{IEEEtranS} 
\bibliography{main}

\appendices

\section{Preliminaries}
\label{app:prelims}

\paragraph{Shared key setup}
$\FSETUP$~\cite{AFLNO16,MR18,BLAZE} enables establishment of common random keys for a  pseudo-random function (PRF) $F$, among parties. This aids in non-interactively generating correlated randomness. Here $F : \{0, 1\}^{\csec} \times \{0, 1\}^{\csec} \rightarrow X$ is a secure PRF, with co-domain $X$ being $\Z{\ell}$.
The semi-honest functionality, $\FSETUP$ appears in \boxref{fig:FSETUP}. The functionality for the malicious case is similar, except that the adversary now has the capability to $\abort$. 

To sample a random value $\vr \in \Z{\ell}$ among a set of $t+1$ parties $\T = \{P_1, \ldots, P_{t+1}\}$ non-interactively, each $P_i \in \T$ invokes $F_{\Key{\T}}(id_{\T})$ and obtains $\vr$. Here, $id_{\T}$ denotes a counter maintained by the parties in $\T$, and is updated after every PRF invocation. The appropriate keys used to sample is implicit from the context, from the identities of the parties that sample. 

\begin{systemboxgray}{$\FSETUP$}{Ideal functionality for shared-key setup}{fig:FSETUP}
	\justify
	$\FSETUP$ interacts with the parties in $\Partyset$ and the adversary $\Sim$. $\FSETUP$ picks random keys $\Key{\T}, \Key{{\T}^{\prime}}$ for every set $\T, {\T}^{\prime} \subseteq \Partyset$ of $t+1, t+2$ parties, respectively.
	$\FSETUP$ picks random keys $\Key{ij}$ for every pair of parties $P_i,P_j \in \Partyset$ and $i<j$.
	\begin{ccsitemize}
		\item[--] Set $\wx_s = \{\Key{si}, \Key{js}\}_{\forall i: s<i \leq n, \forall j: 1 \leq j<s}$. 
		\item[--] Set $\wy_s = \{\Key{\T}\}_{\forall \T \subset \Partyset: \abs{\T} = t+1}$ when $P_s \in \T$.
		\item[--] Set $\wz_s = \{\Key{{\T}^{\prime}}\}_{\forall {\T}^{\prime} \subseteq \Partyset: \abs{{\T}^{\prime}} = t+2}$ when $P_s \in {\T}^{\prime}$.
	\end{ccsitemize}
	\begin{description}
		\item {\bf Output: } Send $(\OUTPUT, \wx_s, \wy_s, \wz_s)$ to every $P_s \in \Partyset$.
	\end{description}
\end{systemboxgray}

\vspace{-2mm}
\paragraph{Collision-Resistant Hash Function}\label{hashf} 
A family of hash functions~\cite{RogawayS04} $\{\Hash: \mathcal{K} \times \MS \rightarrow \mathcal{Y} \}$ is said to be collision resistant if for all PPT adversaries $\Adv$, given the hash function $\Hash_k$ for $k \in_R \mathcal{K}$, the following holds: $\Prob[(x, x^{\prime}) \leftarrow \Adv(k) : (x \neq x^{\prime}) \wedge \Hash_k(x) = \Hash_k(x^{\prime})] = \negl(\kappa)$, where $x, x^{\prime} \in \{0,1\}^{m}$, $m = \poly(\kappa)$, and $\kappa$ is security parameter.

\paragraph{Commitment Scheme}
Let $\commit(x)$ denote the commitment of a value $x$~\cite{MRZ15}. The commitment scheme $\commit(x)$ possesses two properties; {\em hiding} and {\em binding}. The former ensures privacy of value $x$ given its commitment $\commit(x)$, while the latter prevents a corrupt party from opening the commitment to a different value $x' \ne x$.

\subsection{Helper primitives}
\label{appsec:helper}

\smallskip
\begin{spitemize}
    \item[(1)] {$\pizero \rightarrow \sgr{0}$ (\boxref{fig:pirandrp}):} To generate $\sgr{\cdot}$-shares of $0$, each party non-interactively samples two values, each with one of its neighboring parties. A party's shares of 0 are defined as the difference between these values.
    %
    \begin{protboxgray}{$\pizero$}{Generating $\sgr{\cdot}$-shares of $0$}{fig:pizero}
    	\justify
    	\smallskip
    	\begin{description}
    		\item[1.]  $P_i, P_{i+1}$, for $i \in \{1, \ldots, n-1\}$, sample a random value $\vr_i \in_R \Z{\ell}$, while $P_1, P_n$ sample a random value $\vr_n \in_R \Z{\ell}$, using their respective common PRF keys.
    		\item[2.] $P_i$ for $i \in \{2, \ldots, n\}$ sets  $\sgr{0}_i = \vr_i - \vr_{i-1}$, while $P_1$ sets $\sgr{0}_1 = \vr_{1} - \vr_{n}$.
    	\end{description}
    \end{protboxgray}
    \vspace{-3mm}
    \item[(2)] {$\PiRandR \rightarrow \sqr{\vr}$ (\boxref{fig:pirandr}):}
    To generate $\sqr{\cdot}$-shares of a random $\vr \in \Z{\ell}$, every set of $t+1$ parties non-interactively sample a random value using keys established during the setup phase and define $\vr$ to be the sum of these values. 
    
    \begin{protboxgray}{$\PiRandR$}{Generating $\sqr{\cdot}$-shares of a random value}{fig:pirandr}
    	\justify
    	\smallskip
    	\begin{description}
    		\item[1.]  Every $P_i \in \T_j$ for $j \in \{1, \ldots, \h\}$, samples $\sqr{\vr}_{\T_j} \in_R \Z{\ell}$ using the common PRF key.
    		\item[2.] Define $\vr = \sum_{j = 1}^{\h} \sqr{\vr}_{\T_j}$.
    	\end{description}
    \end{protboxgray}
    \vspace{-3mm}
    \item[(3)] {$\PiRandRP(P_s) \rightarrow \sqr{\vr}$ (\boxref{fig:pirandrp}):}
    This protocol generates $\sqr{\cdot}$-shares of a random value $\vr$ such that $P_s$ learns all the shares. Every set of $t+1$ parties non-interactively samples a random value together with $P_s$, using the keys established (for every set of $t+2$ parties) during the setup phase.  
    
    \begin{protboxgray}{$\PiRandRP(P_s)$}{Generating $\sqr{\cdot}$-shares of a random value along with $P_s$}{fig:pirandrp}
    	\justify
    	\smallskip
    	\begin{description}
    		\item[1.]  Every $P_i \in \T_j$ for $j \in \{1, \ldots, \h\}$, samples $\sqr{\vr}_{\T_j} \in_R \Z{\ell}$, together with $P_s$, using the common PRF key.
    		\item[2.] Define $\vr = \sum_{j = 1}^{\h} \sqr{\vr}_{\T_j}$.
    	\end{description}
    \end{protboxgray}
    
    \vspace{-2mm}
    \item[(4)] {$\Piptosh(\va) \rightarrow \shr{\va}$:}
    \label{subsubsec:shrlocal}
    This protocol generates $\shr{\va}$ when $\va \in \Z{\ell}$ is held by at least $t+1$ parties, say parties in $\Evlset$. For this, $P_i \in \Evlset$ sets $\mv{\va} = \va$ and $\sqr{\cdot}$-shares of $\lv{\va}$ as $0$. To generate $\shr{\va}$ in the malicious case where all parties hold $\va$, we let parties set $\mv{\va} = \va$ and shares of $\lv{\va}$ as $0$.
    \smallskip
    \item[(5)] {$\Pictaa(\sqr{\va}) \rightarrow \tsgra{\va}{\T}$ (\boxref{fig:pirss2tadd}):}
    This protocol enables parties in $\T = \{E_1, E_2, \ldots, E_{t+1} \}$ to generate $\tsgra{\va}{\T}$ from $\sgr{\va}$. To generate $\tsgra{\va}{\T}_i$, the idea is to sum up the shares in $\sqr{\va}_{\T_1}, \ldots, \sqr{\va}_{\T_{\h}}$, while ensuring that every share is accounted for and no share is incorporated more than once. Concretely, for share $\sqr{\va}_{\T_j}$ held by parties in $\T_j$ for $j \in \{1, \ldots, \h \} $, $E_i \in \T_j$ incorporates $\sqr{\va}_{\T_j}$ in its share of $\tsgr{\va}_i$ if $E_i$ has the least index in $\T_j$. 
    \begin{protboxgray}{$\Pictaa(\sqr{\va})$}{Conversion from $\sqr{\cdot}$-share to $\tsgra{\cdot}{\T}$-share}{fig:pirss2tadd}
    	\justify
    	\smallskip
    	\begin{description}
    		\item[1.]  Let $\T = \{E_1, \ldots, E_{t+1}\}$. 
    		\item[2.] $E_i \in \T$ computes $\tsgr{\va}_i = \sum_{j = 1}^{\h} \sqr{\va}_{\T_j} \cdot \e_{j}^i$, where
    		$\e_{j}^i = 1$ if $E_i$ has the least index in $\T_j$, and $0$, otherwise.
    	\end{description} 
    \end{protboxgray}
    \vspace{-2mm}
    \item[(6)] {$\Pictaf(\sqr{\va}) \rightarrow \sgr{\va}$:}
    $\sqr{\cdot}$-share can be converted to $\sgr{\cdot}$-share following similar procedure as $\Pictaa$, and is denoted as $\Pictaf(\sqr{\va})$. We omit the details due to similarity. 
    \smallskip
    \item[(7)] {$\Pictashra(\shr{\va}) \rightarrow \tsgra{\va}{\T}$:}
    \label{subsubsec:shr2tsgr}
    Parties in $\T$ invoke $\Pictaa$ on $-\lv{\va}$ to generate $\tsgra{-\lv{\va}}{\T}$, followed by a designated $P_i \in \T$ that holds $\mv{\va}$ setting $\tsgra{\va}{\T}_i = \mv{\va} + \tsgr{-\lv{\va}}_i$. 
    \smallskip
    \item[(8)] {$\Pictashrf(\shr{\va}) \rightarrow \sgr{\va}$:}
    \label{subsubsec:shr2sgr}
    $\sgr{\va}$ can be generated from $\shr{\va}$ similar to $\Pictashra$, and is denoted as $\Pictashrf(\shr{\va})$.
    \smallskip
    \item[(9)] {$\Pisqtosh(\sqr{\va}) \rightarrow \shr{\va}$:}
    \label{subsubsec:sqr2shr}
    To convert $\sqr{\va}$, to $\shr{\va}$, set $\mv{\va} = 0$ and set $\sqr{\lv{\va}} = - \sqr{\va}$.
    \smallskip
    \item[(10)] {$\Pishtosq(\shr{\va}) \rightarrow \sqr{\va}$:}
    \label{subsubsec:shr2sqr}
    To convert $\shr{\va}$ to $\sqr{\va}$, set $\sqr{\va}_{\T_j} = - \sqr{\lv{\va}}_{\T_j}$ for $j \in \{1, \ldots, \h-1\}$ and $\sqr{\va}_{\T_{\h}} = \mv{\va} - \sqr{\lv{\va}}_{\T_{\h}}$, where $\T_{\h} = \Evlset$.
    \smallskip
    \item[(11)] {$\Picprod(\sqr{\va}, \sqr{\vb}) \rightarrow \sgr{\va \vb}$ (\boxref{fig:pirss2prod}):}
    Given $\sqr{\va}, \sqr{\vb}$, parties non-interactively compute $\sgr{\va \vb}$ as follows. Observe that $\sgr{\va \vb} = \sum_{j=1}^{\h} \sgr{\sqr{\va}_{\T_j} \vb}$. To generate $\sgr{\sqr{\va}_{\T_j} \vb}$, the idea is to generate $\tsgra{\sqr{\va}_{\T_j} \vb}{\T_j}$ and perform a conversion. %
    Parties in $\T_j$ generate $\tsgra{\sqr{\va}_{\T_j} \vb}{\T_j}$ as $\tsgra{\sqr{\va}_{\T_j} \vb}{\T_j} = \left( \sqr{\va}_{\T_j} \right) \cdot \left( \tsgra{\vb}{\T_j} \right)$. To obtain $\sgr{\sqr{\va}_{\T_j} \vb}$ from $\tsgra{\sqr{\va}_{\T_j} \vb}{\T_j}$, $P_i \in \Partyset$ sets $\sgr{\sqr{\va}_{\T_j} \vb}_i = \tsgra{\sqr{\va}_{\T_j} \vb}{\T_j}_i$ if $P_i \in \T_j$ and $\sgr{\sqr{\va}_{\T_j} \vb}_i =0$, otherwise.
    \begin{protboxgray}{$\Picprod(\Partyset, \sqr{\va}, \sqr{\vb})$}{$\sqr{\va}, \sqr{\vb}$ to $\sgr{\va \vb}$}{fig:pirss2prod}
    	\justify
    	\begin{description} 
    		\item[1.] For $j \in \{1, \ldots, \h\}$:
    		\begin{ccsitem} 
    			\item $P_i \in \T_j$ invokes $\Pictaa$ on $\sqr{\vb}$ to generate $\tsgra{\vb}{\T_j}_i$. 
    			\item Set $\sgr{\sqr{\va}_{\T_j} \vb}_i = \left( \sqr{\va}_{\T_j} \right) \cdot \left( \tsgra{\vb}{\T_j}_i \right)$  if $P_i \in \T_j$, and $\sgr{\sqr{\va}_{\T_j} \vb}_i = 0$, otherwise.
    		\end{ccsitem} 
    		\item[2.] $P_i \in \Partyset$ computes $\sgr{\va \vb}_i = \sum_{j=1}^{\h} \sgr{\sqr{\va}_{\T_j} \vb}_i$.
    	\end{description} 
    \end{protboxgray}
    \vspace{-3mm}
    \item[(12)] {$\Picon(\Partyset, \{\vec{\val_1}, \ldots, \vec{\val_n}\}) \rightarrow \continue/ \abort$:}\label{subsubsec:consistency}
    Allows parties to check if they hold the same set of values $\vec{\val} = (\val_1, \ldots, \val_m)$, where parties $\continue$ if the values are same, and $\abort$ otherwise. We denote the version of $\vec{\val}$ held by $P_i \in \Partyset$ as $\vec{\val_i}$. To check for consistency of $\vec{\val}$, parties compute hash, $\Hash = \Hash(\val_1 \concat \ldots \concat \val_m)$, of the concatenation of all values $\val_1, \ldots, \val_m$, and exchange $\Hash$ among themselves. If any party receives inconsistent hashes, it $\abort$s; else it $\continue$s. 
    \smallskip
    \item[(13)] {$\PiRSh(P_s, \va)  \rightarrow \sqr{\va}$: }
    To enable $P_s$ to generate $\sqr{\va}$, parties generate $\sqr{\va}_{\T_j}$ for $j \in \{1, \ldots, \h-1 \}$ using $\PiRandRP$, with $P_s$ learning $\sqr{\va}_{\T_j}$ (i.e., $\sqr{\va}_{\T_j}$ are sampled using common key amongst $t+2$ parties).  
    $P_s$ sets $\sqr{\va}_{\T_
    {\h}} = \va - \sum_{j=1}^{\h-1} \sqr{\va}_{\T_j}$ and sends $\sqr{\va}_{\T_{\h}}$ to parties in $\T_{\h}$. 
    For malicious case, this is followed by invoking $\Picon(\Partyset, \{\sqr{\va}_{\T_{\h}}\})$ to check consistency of value sent by $P_s$.
\end{spitemize}

\section{MPClan protocols}
\label{app:mpclan}

\subsection{Semi-honest protocols}
\label{app:bbsh}

\paragraph{Input sharing}
The protocol for input sharing appears in \boxref{fig:piInr}.

\begin{protboxgray}{$\PiSh(P_s,\va)$}{Semi-honest: Input sharing protocol}{fig:piInr}
	\justify
	\algoHead{Preprocessing:} Invoke $\PiRandRP(P_s)$ to generate $\sqr{\lv{\va}}$, with $P_s$ learning $\lv{\va}$ where $\lv{\va} \in \Z{\ell}$.
	\vspace{-2mm}
	\justify
	\algoHead{Online:} $P_s$ computes and sends $\mv{\va} = \va + \lv{\va}$ to all $P_i \in \Evlset$.
\end{protboxgray}
 
\vspace{-3mm}
\paragraph{Truncation - Instantiating $\Ftrgen$}
We rely on a modified version of the doubly shared random bit (a bit that is arithmetic as well as Boolean shared) generation protocol of~\cite{Damgard0FKSV19}, extended to our $n$-party setting, to generate $\shr{\vr}, \shr{\trunc{\vr}}$ as required to perform truncation. Here, $\trunc{\vr}$ represents the truncated (by $d$ bits) version of $\vr \in \Z{\ell}$. The resulting protocol is referred to as $\PiDSBits$ (\boxref{fig:piDSBits}). 

\begin{protboxgray}{$\PiDSBits(\Partyset, \istr)$}{Semi-honest: Doubly shared bits}{fig:piDSBits}
	\justify
	If $\istr = 1$, set $\kbits = \ell$ else set $\kbits = 1$. For $i \in \{0, \ldots, \kbits-1\}$:
	
	\begin{description}
		\item[1.] Invoke $\PiRandR$ to generate $\sqre{\vu_i}{\ell+2}$ for $\vu_i \in \Z{\ell+2}$, and $\pizero$ to generate $\sgre{0}{\ell+2}$.
		\item[2.] Compute $\sqre{\va_i}{\ell+2} = 2\sqre{\vu_i}{\ell+2} + 1$.
		\item[3.] Invoke $\Picprod$ on $\sqre{\va_i}{\ell+2}$ to generate $\sgre{\ve_i}{\ell+2}$ where $\ve_i = \va_i^2$.
		\item[4.] Send $\sgre{\ve_i}{\ell+2} + \sgre{0}{\ell+2}$ to $\Pking$, who reconstructs $\ve_i + 0 = \ve_i$ and sends to all. 
		\item[5.] Let $\vc_i$ be the smallest root of $\ve_i$ modulo $2^{\ell+2}$, and $\vc_i^{-1}$ its inverse. Compute $\sqre{\vd_i}{\ell+2} =\vc_i^{-1} \sqre{\va_i}{\ell+2} + 1$. 
		\item[6.] $P_j$ sets $\sqre{\bitb_i}{\ell+2}_j = \sqre{\vd_i}{\ell+2}_j/2$, and  $\sqr{\arval{\bitb_i}}_j, \sqrB{\bitb_i}_j$ as the least significant $\ell$ bits and the least significant bit of $\sqre{\bitb_i}{\ell+2}_j$, respectively. 
		\item[7.] Invoke $\Pisqtosh$ on $\sqr{\arval{\bitb_i}}, \sqrB{\bitb_i}$ to generate $\shr{\arval{\bitb_i}}, \shrB{\bitb_i}$.
	\end{description}
	If $\istr = 1$, set:
	\begin{footnotesize} 
		\begin{align*}
			(\shr{\vr}, \shr{\trunc{\vr}}) = \left( \sum_{i=0}^{\kbits-1} 2^i \shr{\arval{\bitb_i}}, \sum_{i=d}^{\kbits-1} 2^{i-d} \shr{\arval{\bitb_i}} \right)  
		\end{align*}
	\end{footnotesize}
	
\end{protboxgray}

\vspace{-3mm}
At a high-level, generation of doubly shared bits relies on the property that every non-zero quadratic residue has exactly one root when working over fields. The work of~\cite{Damgard0FKSV19}, operating over rings, shows that something similar holds over rings as well. Concretely, according to lemma 4.1 of~\cite{Damgard0FKSV19}: {\em if $\va$ is such that $\va^{2} \equiv_{\ell} 1$, then $\va$ is congruent mod $2^{\ell}$ to either $1, -1, -1+2^{\ell-1}, 1+2^{\ell-1}$
}. Thus, the doubly shared bit generation protocol of~\cite{Damgard0FKSV19} proceeds as follows. Generate $\va^{2}$ for $\va \in \Z{\ell+2}$ such that $\va^{2} \equiv_{\ell+2} 1$, and compute its smallest root $\vc$ mod $2^{\ell+2}$. Compute $(\vc^{-1} \va)$, and by lemma 4.1 of~\cite{Damgard0FKSV19} it follows that $\vc^{-1} \va \in \{\pm 1, \pm 1 + 2^{\ell+1}\}$. That is, $(\vc^{-1} \va)$ is congruent to $\pm 1$ modulo $2^{\ell + 1}$. Thus, $\vd = \vc^{-1} \va + 1$ is congruent to $0$ or $2$ modulo $2^{\ell +1}$ with equal probability. Hence, setting $\bitb = \vd /2$ outputs bit $\bitb = 0$ or bit $\bitb = 1$ with equal probability. 
Observe that the computation has to be performed over $\Z{\ell+2}$. Hence, in the protocol description, we use $\ell+2$ in the superscript to distinguish shares of $\vx$ over $\Z{\ell+2}$ from its shares over $\Z{\ell}$.

The main change in $\PiDSBits$ from that of the protocol in~\cite{Damgard0FKSV19} is that to generate $\shr{\vr}, \shr{\trunc{\vr}}$ $\PiDSBits$ generates $\ell$ random doubly shared bits $\bitb_0, \ldots, \bitb_{\ell-1} \in \Z{}$ instead of a single one, and composes these $\ell$ bits to generate $\vr$, and composes the higher $\ell - d$ bits to generate $\trunc{\vr}$, as follows.
\begin{align}
	\left( \shr{\vr} , \shr{\trunc{\vr}} \right) = \left( \sum_{i=0}^{\ell-1} 2^i \shr{\arval{\bitb_i}}, \sum_{i=d}^{\ell-1} 2^{i-d} \shr{\arval{\bitb_i}} \right) 
\end{align}
Looking ahead, $\PiDSBits$ can also be used only to generate a single doubly shared random bit, which finds use in other building blocks such as bit to arithmetic conversion and arithmetic to Boolean conversion. Thus, to distinguish the case when $(\shr{\vr} , \shr{\trunc{\vr}})$ has to be generated versus when only a single doubly shared bit is to be generated, $\PiDSBits$ takes a bit $\istr$ as input and gives as output a doubly shared bit $\shr{\arval{\bitb}}, \shrB{\bitb}$ if $\istr = 0$, and $(\shr{\vr}, \shr{\trunc{\vr}})$ otherwise.
The protocol appears in \boxref{fig:piDSBits}.

A final thing to note is that the computation in $\PiDSBits$ proceeds over secret-shared data. Thus, to generate shares of the doubly shared bit $\bitb$, one should be able to divide each share of $\vd$ by $2$, which necessitates $\vd$ and its shares to be even. This holds true since $\sqre{\vd}{\ell+2} = \vc^{-1} \sqre{\va}{\ell+2} + 1 = \vc^{-1} \left( 2\sqre{\vu}{\ell+2} + 1 \right) + 1 = 2 \vc^{-1} \sqre{\vu}{\ell+2} + \vc^{-1} + 1$. Here, $2 \vc^{-1} \sqre{\vu}{\ell+2}$ is even due to multiplication by $2$, while  $\vc^{-1} + 1$ is even since $\vc^{-1}$ is odd by definition.

\paragraph{Dot product}
As described before, a dot product can be viewed as $\nf$ instances of multiplication such that the communication for all the instances is aggregated and performed in a single shot to eliminate the vector-size dependency. Consequently, the dot product protocol follows along the lines of the multiplication, and the formal details appear in \boxref{fig:pidotp}.

\begin{protboxgray}{$\PiDotP(\Partyset, \shr{\vec{x}}, \shr{\vec{y}})$}{Semi-honest: Dot product protocol}{fig:pidotp}
	\justify
	\algoHead{Preprocessing:} \vspace{-1mm}
	\begin{description}
		\item[1.] Invoke $\PiRandR$ to generate $\sqr{\vr}$ where $\vr  \in \Z{\ell}$, followed by $\Pictaf$ to generate $\sgr{\vr}$. 		
		\item[2.] Invoke $\Picprod$ on  $\sqr{\lv{\vx_k}}, \sqr{\lv{\vy_k}}$ to generate $\sgr{\clv{\vx_k \vy_k}}$ for $k \in \{1, \ldots, \allowbreak \nf\}$, and compute $\sgr{\sum_{k=1}^{\nf}\clv{\vx_k \vy_k} - \vr} = \sum_{k=1}^{\nf}\sgr{\clv{\vx_k \vy_k}} - \sgr{\vr}$.
		\item[3.] $P_i \in \Evlset$ invokes $\Picta$ on $\sqr{\lv{\vx_k}}, \sqr{\lv{\vy_k}}$ to generate $\tsgr{\lv{\vx_k}}_i, \allowbreak \tsgr{\lv{\vy_k}}_i$, respectively, for $k \in \{1, \ldots, \nf\}$. 
		\item[4.] $P_i \in \Hlpset$ sends $\sgr{\sum_{k=1}^{\nf}\clv{\vx_k \vy_k} - \vr}_i$ to $\Pking$, who sets $\D = \sum_{i: P_i \in \Hlpset} \allowbreak \sgr{\sum_{k=1}^{\nf}\clv{\vx_k \vy_k} - \vr}_i$.
	\end{description} 
	\algoHead{Online:} \vspace{-1mm}
	\begin{description}
		\item[1.] $P_i \in \Evlset$ computes and sends $\tsgr{\zeta}_{i} = \sum_{k=1}^{\nf}\left(-\mv{\vx_k} \tsgr{\lv{\vy_k}}_{i} - \mv{\vy_k} \tsgr{\lv{\vx_k}}_{i} \right) + \sgr{\sum_{k=1}^{\nf}\clv{\vx_k \vy_k} - \vr}_i $ to $\Pking$.
		\item[2.] $\Pking$ computes $\E = \sum_{k=1}^{\nf} \cmv{\vx_k \vy_k} + \sum_{i: P_i \in \Evlset} \tsgr{\zeta}_{i}$ and sends $\vz - \vr = \D + \E$ to all parties in $\Evlset$.
		\item[3.] Invoke $\Piptosh$ on $\vz - \vr$ to generate $\shr{\vz - \vr}$, and compute $\shr{\vz} = \shr{\vz - \vr} + \shr{\vr}$.
	\end{description}
\end{protboxgray}

\paragraph{Multi-input multiplication} 
\label{subsec:multiinputmultshap}     
The goal of 3-input multiplication (\boxref{fig:pitMult}) is to generate $\shr{\cdot}$-sharing of $\vz = \va \vb \vc$ given $\shr{\va}, \shr{\vb}, \shr{\vc}$, in a single shot. Observe that   

\vspace{-2mm}
\begin{align*}
	\vz - \vr
	&= \va \vb \vc - \vr = (\mv{\va} - \lv{\va})(\mv{\vb} - \lv{\vb})(\mv{\vc} - \lv{\vc}) - \vr \\
	&= \cmv{\va\vb\vc} - \cmv{\va\vc} \lv{\vb} - \cmv{\vb\vc} \lv{\va} - \cmv{\va\vb} \lv{\vc} \\
	&~~~+ \mv{\va} \clv{\vb \vc} + \mv{\vb} \clv{\va \vc} + \mv{\vc} \clv{\va \vb} - \clv{\va \vb \vc} - \vr
\end{align*}
%

Given that $\tsgr{\clv{\va \vb}}, \tsgr{\clv{\va \vc}}, \tsgr{\clv{\vb \vc}}, \tsgr{\clv{\va \vb \vc} + \vr}$ can be generated in the preprocessing among the parties in $\Evlset$, parties proceed with a similar online phase as in $\piMult$ to compute the 3-input multiplication without inflating the online cost. With respect to the preprocessing phase, 

-- For generating $\tsgr{\clv{\va \vc}}, \tsgr{\clv{\vb \vc}}$ parties first compute the respective additive sharings ($\sgr{\cdot}$) using $\sqr{\lv{\va}}, \sqr{\lv{\vb}}$ and $\sqr{\lv{\vc}}$ (via two invocations of $\Picprod$, \boxref{fig:pirss2prod}). Following this parties in $\Hlpset$ communicate their share of $\sgr{\clv{\va \vc}}$ and $\sgr{\clv{\vb \vc}}$ to $\Pking$, each masked with a random $\sgr{\cdot}$-sharing of $0$ (generated using $\pizero$, \boxref{fig:pizero}). This establishes $\tsgr{\clv{\va \vc}}, \tsgr{\clv{\vb \vc}}$ among parties in $\Evlset$.

-- For generating $\tsgr{\clv{\va \vb}}$, a slightly different approach is taken where parties first generate $\sqr{\clv{\va \vb}}$ using $\sqr{\lv{\va}}, \sqr{\lv{\vb}}$ (as explained later), followed by non-interactively generating $\tsgr{\clv{\va \vb}}$ (via $\Pictaa$, \boxref{fig:pirss2tadd}). The reason for generating $\sqr{\clv{\va \vb}}$ (instead of directly generating $\tsgr{\clv{\va \vb}}$) is to facilitate generation of $\tsgr{\clv{\va \vb \vc} - \vr}$ from $\sqr{\clv{\va \vb}}$,  $\sqr{\lv{\vc}}$ and $\sgr{\vr}$, which closely follows the preprocessing phase of the 2-input multiplication. Specifically, parties can generate $\sgr{\clv{\va \vb \vc}}$ using $\Picprod$ (\boxref{fig:pirss2prod}) on $\sqr{\clv{\va \vb}}$,  $\sqr{\lv{\vc}}$, followed by parties in $\Hlpset$ communicating their $\sgr{\clv{\va \vb \vc}}$ shares masked with $\sgr{\cdot}$-sharing of a random $\vr$ to $\Pking$. This generates $\tsgr{\clv{\va \vb \vc} + \vr}$-sharing required during online phase.

-- Regarding generation of $\sqr{\clv{\va \vb}}$, all parties generate $\sqr{\cdot}$-sharing of a random $\gamma \in \Z{\ell}$ non-interactively and convert it to $\sgr{\gamma}$. Parties then compute $\sgr{\clv{\va \vb} + \gamma}$ by computing $\sgr{\clv{\va \vb}}$ from $ \sqr{\lv{\va}}, \sqr{\lv{\vb}}$ followed by summing it up with $\sgr{\gamma}$. Parties reconstruct this value towards $\Pking$, who then generates $\sqr{\clv{\va \vb} + \gamma}$, from which parties compute $\sqr{\clv{\va \vb}} = \sqr{\clv{\va \vb} + \gamma} - \sqr{\gamma}$.
On obtaining $\sqr{\clv{\va \vb}}$, parties generate $\tsgr{\clv{\va \vb}}$ by invoking $\Picta$.

Similarly, for the 4-input multiplication, to obtain $\shr{\cdot}$-sharing of $\vz = \va \vb \vc \vd$ given the $\shr{\cdot}$-sharing of $\va, \vb, \vc, \vd$, we can write $\vz + \vr$ as 
\begin{align}\label{eqmultinputmult}
	\vz - \vr &=(\mv{\va} - \lv{\va})(\mv{\vb} - \lv{\vb})(\mv{\vc} - \lv{\vc})(\mv{\vd} - \lv{\vd}) - \vr \\ \nonumber 
	&= \cmv{\va\vb\vc\vd} - \cmv{\vb\vc\vd} \lv{\va} - \cmv{\va\vc\vd}  \lv{\vb} - \cmv{\va\vb\vd} \lv{\vc} - \cmv{\va\vb\vc} \lv{\vd} \\  \nonumber 
	&+ \cmv{\va\vb} \clv{\vc \vd} + \cmv{\va\vc} \clv{\vb \vd} + \cmv{\va\vd} \clv{\vb \vc} + \cmv{\vb\vc} \clv{\va \vd} + \cmv{\vb\vd} \clv{\va \vc} \\ \nonumber
	&+ \cmv{\vc\vd} \clv{\va \vb} - \mv{\va}\clv{\vb\vc\vd} - \mv{\vb}\clv{\va\vc\vd} - \mv{\vc}\clv{\va\vb\vd} - \mv{\vd}\clv{\va\vb\vc}  \\ \nonumber
	&+ \clv{\va \vb \vc \vd} - \vr \nonumber
\end{align}
Here, parties need to generate the $\tsgr{\cdot}$-sharing of $\clv{\va \vb}, \clv{\va \vc}, \clv{\va \vd}, \allowbreak \clv{\vb \vc}, \allowbreak \clv{\vb \vd}, \clv{\vc \vd}, \clv{\va \vb \vc}, \clv{\va \vb \vd}, \clv{\va \vc \vd}, \clv{\vb \vc \vd}, \clv{\va \vb \vc \vd}$. Generation of $\tsgr{\cdot}$-sharing of $\clv{\va \vc}, \clv{\va \vd}, \clv{\vb \vc}, \clv{\vb \vd}$ can proceed similar to generation of $\tsgr{\clv{\va \vc}}$ in $\PitMult$. Generation of $\tsgr{\cdot}$-sharing of $\clv{\va \vb}, \clv{\vc \vd}$ is carried out by first generating its $\sqr{\cdot}$-sharing. This enables generation of $\tsgr{\cdot}$-sharing of  $\clv{\va \vb \vc}, \clv{\va \vb \vd}, \clv{\va \vc \vd}, \clv{\vb \vc \vd}$ following steps similar to generation of $\tsgr{\clv{\va \vc}}$ in $\PitMult$. Finally, $\tsgr{\clv{\va \vb \vc \vd} - \vr}$ is generated similar to generating $\tsgr{\clv{\va \vb \vc} + \vr}$ in $\PitMult$. We omit formal details of 4-input multiplication protocol, $\PifMult$, as it is very close to $\PitMult$.

\begin{protboxgray}{$\PitMult(\Partyset, \shr{\va}, \shr{\vb}, \shr{\vc})$}{Semi-honest: 3-input multiplication protocol}{fig:pitMult}
	\justify
	\algoHead{Preprocessing:}
	\begin{description}
		\item[1.] Invoke $\PiRandR$ to generate $\sqr{\vr}$ and $\sqr{\gamma}$ where $\vr, \gamma  \in \Z{\ell}$. Invoke $\Pictaf$ to generate $\sgr{\vr}, \sgr{\gamma}$.
		\item[2.] Invoke $\pizero$ to generate two different $\sgr{\cdot}$-shares of $0$: $\sgr{0_1}$, $\sgr{0_2}$.
		\item[3.] Generation of $\tsgr{\clv{\va \vc}}, \tsgr{\clv{\vb \vc}}$.  
		\begin{ccsitem}
			\item Invoke $\Picprod$ on $\sqr{\lv{\va}}, \sqr{\lv{\vc}}$ to generate $\sgr{\clv{\va \vc}}_i$, and compute $\sgr{\clv{\va \vc} + 0_1}_i = \sgr{\clv{\va \vc}}_i + \sgr{0_1}_i$.
			\item $P_i \in \Hlpset$ sends $\sgr{\clv{\va \vc} + 0_1}_i$ to $\Pking (= P_{t+1})$.
			\item Analogous steps are carried out to generate $\sgr{\clv{\vb \vc} + 0_2}$.
			\item $P_i \in \Evlset \setminus P_{t+1}$ sets $\tsgr{\clv{\vb \vc}}_i = \sgr{\clv{\vb \vc} + 0_2}_i$ and $\tsgr{\clv{\va \vc}}_i = \sgr{\clv{\va \vc} + 0_1}_i$.
			\item $P_{t+1}$ sets $\tsgr{\clv{\vb \vc}}_{t+1} = \sgr{\clv{\vb \vc} + 0_2}_{t+1} + \sum_{i: P_i \in \Hlpset} \sgr{\clv{\vb \vc} + 0_2}_{i}$ and $\tsgr{\clv{\va \vc}}_{t+1} = \sgr{\clv{\va \vc} + 0_1}_{t+1} + \sum_{i: P_i \in \Hlpset} \sgr{\clv{\va \vc} + 0_1}_{i}$.
		\end{ccsitem}		
		\item[4.] Generation of $\tsgr{\clv{\va \vb}}$.
		\begin{ccsitem}
			\item Invoke $\Picprod$ on $\sqr{\lv{\va}}, \sqr{\lv{\vb}}$ to generate $\sgr{\clv{\va \vb}}_i$, set $\sgr{\clv{\va \vb} +  \gamma}_i = \sgr{\clv{\va \vb}}_i + \sgr{\gamma}_i$, and send $\sgr{\clv{\va \vb} + \gamma}_i$ to $\Pking$.
			\item $\Pking$ reconstructs $\clv{\va \vb} + \gamma$, and sends $\clv{\va \vb} + \gamma$ to $P_i \in \Evlset$. Parties non-interactively generate $\sqr{\clv{\va \vb} + \gamma}$ via $\Piptosh$ and $\Pishtosq$. 
			\item Compute $\sqr{\clv{\va \vb}} = \sqr{\clv{\va \vb} + \gamma} - \sqr{\gamma}$ and invoke $\Picta$ on $\sqr{\clv{\va \vb}}$ to generate $\tsgr{\clv{\va \vb}}$.
		\end{ccsitem} 
		\item[5.] Generation of $\tsgr{\clv{\va \vb \vc} + \vr}$.
		\begin{ccsitem}
			\item Invoke $\Picprod$ on $\sqr{\clv{\va \vb}}, \sqr{\lv{\vc}}$ to generate $\sgr{\clv{\va \vb \vc}}_i$,  and compute $\sgr{\clv{\va \vb \vc} + \vr}_i = \sgr{\clv{\va \vb \vc}}_i + \sgr{\vr}_i$.
			\item $P_i \in \Hlpset$ sends $\sgr{\clv{\va \vb \vc} + \vr}_i$ to $\Pking$.
			\item $P_i \in \Evlset \setminus P_{t+1}$ sets $\tsgr{\clv{\va \vb \vc} + \vr}_i = \sgr{\clv{\va \vb \vc} + \vr}_i$.
			\item $P_{t+1}$ sets $\tsgr{\clv{\va \vb \vc} + \vr}_{t+1} = \sgr{\clv{\va \vb \vc} + \vr}_{t+1} + \sum_{i: P_i \in \Hlpset} \sgr{\clv{\va \vb \vc} + \vr}_{i}$.
		\end{ccsitem}
		\item[6.] $P_i \in \Evlset$ invoke $\Picta$ on $\sqr{\lv{\va}}, \sqr{\lv{\vb}}$ and $\sqr{\lv{\vc}}$ to generate $\tsgr{\lv{\va}}_i, \allowbreak \tsgr{\lv{\vb}}_i, \tsgr{\lv{\vc}}_i$, respectively.
	\end{description} 
	\algoHead{Online:}
	\begin{description}
		\item[1.] $P_i \in \Evlset$ computes and sends $\tsgr{\zeta}_i = - \cmv{\va\vc} \tsgr{\lv{\vb}}_i - \cmv{\vb\vc} \tsgr{\lv{\va}}_i - \cmv{\va\vb} \tsgr{\lv{\vc}}_i + \mv{\va} \tsgr{\clv{\vb \vc}}_i + \mv{\vb} \tsgr{\clv{\va \vc}}_i + \mv{\vc} \tsgr{\clv{\va \vb}}_i - \tsgr{\clv{\va \vb \vc} + \vr}_i$ to $\Pking$.
		\item[2.]  $\Pking$ computes and sends $\vz - \vr = \cmv{\va \vb \vc} + \sum_{i: P_i \in \Evlset} \tsgr{\zeta}_{i}$ to $P_i \in \Evlset$.
		\item[3.] Invoke $\Piptosh$ on $\vz - \vr$ to generate $\shr{\vz - \vr}$, and compute $\shr{\vz} = \shr{(\vz - \vr)} + \shr{\vr}$.
	\end{description}
\end{protboxgray}

\vspace{-4mm}
 
\subsection{Malicious protocols}
\label{appsec:malmpc}

\paragraph{Input sharing}
This protocol ($\PiShMal(P_s,\va)$) is similar to the semi-honest one, where to enable $P_s$ to generate $\shr{\va}$, parties generate $\sqr{\lv{\va}}$ such that $P_s$ learns $\lv{\va}$, followed by $P_s$ sending the masked value $\mv{\va} = \va + \lv{\va}$ to all. 
However, note that a corrupt $P_s$ can cause inconsistency among the honest parties by sending different masked values. To ensure the same value is received by all, parties perform a hash-based consistency check, denoted by $\Picon$ (\S\ref{sec:prelim}), where each party sends a hash of the received masked value(s) to every other party and $\abort$s if it receives inconsistent hashes. Note that this check for all the inputs can be combined, thereby amortizing the cost.

\begin{protboxgray}{$\PiShMal(P_s,\va)$}{Malicious: Input sharing protocol}{fig:piInMal}
	\justify
	\algoHead{Preprocessing:} Invoke $\PiRandRP(P_s)$ to generate $\sqr{\lv{\va}}$, with $P_s$ learning $\lv{\va}$ where $\lv{\va} \in \Z{\ell}$.
	\justify
	\vspace{-2mm}
	\algoHead{Online:} $P_s$ computes and sends $\mv{\va} = \va + \lv{\va}$ to all $P_i \in \Partyset$.
	\justify 
	\vspace{-2mm}
	\algoHead{Verification:}  Invoke $\Picon$ on $\{\mv{\va}\}$. 
\end{protboxgray}

\paragraph{Reconstruction}
To reconstruct $\shr{\cdot}$-shared value $\va$ towards $P_s \in \Partyset$, observe that each share that $P_s$ misses is held by $t+1$ other parties. Each of these parties sends the missing share to $P_s$. If the received values for a share are consistent, $P_s$ uses this value to perform reconstruction, and $\abort$s otherwise. As an optimization, one party can send the missing share while reconstructing several values, and $t$ others can send its hash.

\medskip
\begin{protboxgray}{$\PiRecfair(\shr{\vz})$}{Fair: Reconstruction protocol}{fig:pirecfair}
	\justify
	\algoHead{Preprocessing:} 
	\begin{description}
	    \item[1.] Invoke $\PiRandR$ to generate $\sqr{\lv{\vz}}$ where $\lv{\vz} \in_R \Z{\ell}$. 
	    \item[2.] For $j \in \{1, \ldots, \h\}$:
	    \begin{ccsitem}
	        \item Each $P_i \in \T_j$ generates commitments on $\sqr{\lv{\vz}}_{\T_j}$ using the common randomness, and sends to all other parties. 
	        \item $P_i \notin \T_j$ $\abort$s if commitments for $\sqr{\lv{\vz}}_{\T_j}$ are inconsistent.
	    \end{ccsitem}
	\end{description}
	%
	\justify
	\vspace{-2mm}
	\algoHead{Online:} 
	\begin{description}
	    \item[1.] Parties broadcast an $\alive$ bit, indicating that they did not $\abort$. 
	    \item[2.] If all parties are alive, $P_i \in \Partyset$ sends the decommitment to the shares in $\sqr{\lv{\vz}}_i$ to the respective parties. 
	    \item[3.] Parties use the valid decommitment to obtain the missing share of $\lv{\vz}$, reconstruct $\lv{\vz}$, and compute $\vz = \mv{\vz} - \lv{\vz}$.
	\end{description}
\end{protboxgray}

\vspace{-2mm}
Fairness is a stronger security notion than security with abort, where, during reconstruction, either all parties learn the output or none do. For fair reconstruction, we extend the techniques in~\cite{BLAZE} to the $n$-party setting, where commitments are generated on each share of the mask (required to reconstruct $\vz$) by $t+1$ parties in the preprocessing phase. During the online phase, these are decommitted towards the respective parties if all parties are $\alive$ (did not $\abort$). Since there is at least one honest party among every set of $t+1$ parties, if all honest parties are alive, then parties are guaranteed to obtain the correct decommitment of the missing share from the honest party, and all honest parties can reconstruct the output. Else, none of the parties will obtain the output. 

\paragraph{Multiplication}
\label{appsec:multattack}
The maliciously secure multiplication protocol ($\PiMultMal$) appears in \boxref{fig:piMultMal}.

\noindent {\em Overcoming the privacy breach described in~\cite{GoyalLS19}}
We elaborate on the privacy breach that arises due to deferring the correctness check and how it is overcome in our case. 
We first explain the attack that a malicious adversary can launch if reconstruction towards $\Pking$ is performed by relying on RSS (or Shamir sharing) naively and further justify why it gets bypassed in our protocol.
Consider a circuit with two sequential multiplication gates with the output of the first gate, say $\va$, going as input to the second gate. Let $\vb$ denote the other input to the second multiplication gate, and $\vz$ denote its output. In a $\Pking$ based approach for multiplication, $t$ parties send their respective (RSS/Shamir) share of a masked value to $\Pking$. In particular, for the first multiplication gate in the circuit mentioned above, $t$ parties send their corresponding share of $\va - \vr_{\va}$ to $\Pking$, who reconstructs it and sends it back to all. Delaying the verification allows a malicious $\Pking$ to send an inconsistent value of $\va - \vr_{\va}$ to the parties, using which it can learn the private input $\vb$, as follows. Suppose $\Pking$ sends the correct $\va - \vr_{\va}$ to all but one out of the remaining $t$ online parties, to which it sends $\va - \vr_{\va} + \delta$. Owing to this, for the next multiplication gate $\Pking$ receives the shares of $\vz - \vr_{\vz}$ from the former $t-1$ parties and a share of $(\va +  \delta) \vb - \vr_{\vz} = \vz + \delta \vb - \vr_{\vz}$ from the latter party. Having obtained these and additionally using the shares of $\vz - \vr_{\vz}$ and $\vz + \delta \vb - \vr_{\vz}$ corresponding to the $t$ corrupt parties including itself, a malicious $\Pking$ can reconstruct $\vz - \vr_{\vz}$ as well as $\vz + \delta \vb - \vr_{\vz}$, thus learning $\vb$ in clear. The crux of this attack lies in the fact that a malicious adversary corrupting $t$ parties including $\Pking$ already possesses $t$ shares each of $\vz - \vr_{\vz}$ and $\vz + \delta \vb - \vr_{\vz} $. Thus, an additional share of these obtained from the online parties allows it to carry out the attack successfully. However, the same does not hold for the case of additive ($\tsgr{\cdot}$) sharing.

\begin{protboxgray}{$\PiMultMal(\Partyset, \shr{\va}, \shr{\vb}, \istr)$}{Malicious: Multiplication protocol}{fig:piMultMal}
	\justify
	$\istr = 1$ denotes that truncation is required and $\istr = 0$ denotes otherwise. 
	
	\justify 
	\algoHead{Preprocessing:}
	\begin{description}
		\item[1.] If $\istr = 0$: invoke $\PiRandR$ to generate $\sqr{\vr}$ where $\vr  \in \Z{\ell}$. Invoke $\Pictaf$ and $\Pisqtosh$ on $\sqr{\vr}$ to generate $\sgr{\vr}$ and $\shr{\vr}$, respectively. 
		\item[2.] Else, invoke $\PiDSBitsMal(\Partyset, 1)$ (\boxref{fig:piDSBits}) to generate $\shr{\vr}, \shr{\trunc{\vr}}$, and $\Pictashrf$ on $\shr{\vr}$ to generate $\sgr{\vr}$ .
		\item[3.] Invoke $\PiMultPre$ on $\sqr{\lv{\va}}, \sqr{\lv{\vb}}$ to generate $\sqr{\clv{\va\vb}}$.
		\item[4.] $P_i \in \Evlset$ invokes $\Picta$ on $\sqr{\clv{\va\vb}}$, $\sqr{\lv{\va}}$, $\sqr{\lv{\vb}}$ and $\sqr{\vr}$ to generate $\tsgr{\clv{\va \vb}}$, $\tsgr{\lv{\va}}$, $\tsgr{\lv{\vb}}$ and $\tsgr{\vr}$, respectively.
	\end{description} 
	\algoHead{Online:}
	\begin{description}
		\item[1.] $P_i \in \Evlset$ computes $\tsgr{\zeta}_{i} = -\mv{\va} \tsgr{\lv{\vb}}_{i} - \mv{\vb} \tsgr{\lv{\va}}_{i} + \tsgr{\clv{\va \vb} - \vr}_i$, and sends $\tsgr{\zeta}_i$ to $\Pking$.
		\item[2.] $\Pking$ reconstructs $\zeta$, computes and sends $\vz - \vr = \zeta + \cmv{\va \vb}$ to all parties\footnote{$\vz - \vr$ is sent to parties in $\Evlset$ during the online phase computation whereas it is sent to parties in $\Hlpset$ in a single shot before verification begins.}.
		\item[3.] If $\istr = 0$: invoke $\Piptoshf$ on $\vz - \vr$ to generate $\shr{\vz - \vr}$, and compute $\shr{\vz} = \shr{\vz - \vr} + \shr{\vr}$. 
		\item[4.] Else, invoke $\Piptoshf$ on $\trunc{(\vz - \vr)}$ to generate $\shr{\trunc{(\vz - \vr)}}$, and compute $\shr{\trunc{\vz}} = \shr{\trunc{(\vz - \vr)}} + \shr{\trunc{\vr}}$.
	\end{description}
	\algoHead{Verification for all multiplication gates:} Invoke $\Pivrfy$ on $\shr{\cdot}$-shares of $(\va_1, \vb_1, \vz_1),  \ldots, \allowbreak (\va_m, \vb_m, \vz_m)$ which denote the inputs and outputs of the $m$ multiplication gates whose correctness is to be verified. 
\end{protboxgray}

Notice that in our protocol, during reconstruction towards $\Pking$, any redundancy due to $\shrd$-sharing is eliminated with parties switching to $\tsgr{\cdot}$-sharing (additive sharing among parties in $\Evlset$). Due to this, even if $\Pking$ sends inconsistent values to the parties, the $\tsgr{\cdot}$-share of $\vz - \vr_{\vz}$ or $\vz + \delta \vb - \vr_{\vz}$ that it receives, corresponds to an additive share defined with respect to parties in $\Evlset$. 
Hence, this additionally received additive share cannot be combined with the shares held by the $t$ corrupt parties to perform the reconstruction. Thus, the earlier strategy of $\Pking$ of using these additional shares in conjunction with the $t$ corrupt shares to reconstruct $\vz - \vr_{\vz}$ and $\vz + \delta \vb - \vr_{\vz}$ does not hold. The primary reason which prevents the attack is the elimination of redundancy in the sharing scheme by switching to $(t+1)$-out-of-$(t+1)$ additive sharing ($\tsgr{\cdot}$-sharing) for the set of parties in $\Evlset$, which is known to withstand this attack~\cite{GoyalLS19}.

\noindent {\em Discussion about \cite{ED20}}
The above attack can be circumvented by making $\Pking$ broadcast the reconstructed value to all the parties, as discussed in~\cite{ED20}. To further optimize the protocol by requiring only $t+1$ parties to be active in the online phase, they rely on broadcast with abort, which comprises two phases--(i) \textit{send}:  where $\Pking$ sends the value to the recipients, and (ii) \textit{verification}: where the recipients exchange hash of the received value among themselves, and abort in case of inconsistency. However, for amortization, they defer the verification (even with respect to broadcast) towards the end of the protocol, thus making their protocol susceptible to the aforementioned attack. We observe that one fix is to perform the verification with respect to broadcast after each level in the circuit. This, however, requires all the parties to be online. An optimization to let only the $t+1$ parties in the online phase to perform this verification after each level, thereby allowing the remaining $t$ parties to be shut off. Specifically, this involves performing {\em verification} where the online parties exchange the hash of the received value and abort in case of inconsistency. When the remainder $t$ (offline) parties come online towards the end of the protocol for verifying the correctness of the multiplication gates, this verification should be preceded by first verifying the consistency of the values broadcast by $\Pking$ to the offline parties (and involves participation of all $n$ parties). 
Since the online phase involves broadcasting the reconstructed value to $t$ other online parties, this amounts to an exchange of $\Order(t^2)$ hashes after each level, thereby incurring a circuit depth-dependent overhead in the communication cost as well as the rounds. In order for the communication cost to get amortized, it is required that the circuit has $\Order(t^2)$ gates at each level. However, the overhead in terms of number of rounds persists.

\noindent {\em Multiplication with truncation -- Instantiating $\Ftrgenmal$ with maliciously secure doubly shared bits generation protocol:} 
As mentioned earlier, $\Ftrgenmal$ (\boxref{fig:Ftrgenmal}) can be realized using the maliciously secure variant of $\PiDSBits$, denoted as $\PiDSBitsMal$. This protocol is similar to the semi-honest protocol except with the following differences to account for malicious behaviour. The $\sqr{\cdot}$-shares of $\ve_i = \va^2$ are generated by invoking $\PiMultPre$ instead of relying on $\Picprod$. This ensures generation of correct $\sqr{\cdot}$-shares of $\ve_i$, and malicious behaviour, if any, will lead to an $\abort$. Following this, $\ve_i$ is either correctly reconstructed towards all or parties $\abort$. This ensures that an adversary cannot lead to reconstruction of an incorrect $\ve_i$. Concretely, for reconstruction, similar to multiplication, every party sends its $\sqr{\cdot}$-share to every other party, and $\abort$s in case of inconsistencies in the received values\footnote{This can be optimized similar to the online phase of multiplication, where the value is first reconstructed towards $\Pking$ who sends the reconstructed value to all, followed by verifying its correctness via the verification check.}. 
The rest of the protocol steps (which are non-interactive) remain unchanged, and hence a formal protocol is omitted. 

\begin{systemboxgray}{$\Ftrgenmal$}{Ideal functionality $\Ftrgenmal$}{fig:Ftrgenmal}
	\justify
	
	$\Ftrgenmal$ interacts with the parties in $\Partyset$ and the adversary $\Sim$. 
	
	\begin{description}
		\item {\bf Input:} $\Ftrgenmal$ optionally receives a special command, $(\abort, \Ab)$, from $\Sim$ indicating that honest parties in $\Partyset$ with indices in $\Ab$ should $\abort$.
	\end{description}
	
	\noindent $\Ftrgenmal$ proceeds as follows.
	\begin{ccsitemize}
		\item[--] Samples random $\vr \in \Z{\ell}$, and computes $\trunc{\vr} = \vr/ 2^{d}$.
		\item[--] Generates $\shr{\cdot}$-shares of $\vr, \trunc{\vr}$. 
		\item[--] Let $\wy_s$ denote the $\shr{\cdot}$-shares of $\vr, \trunc{\vr}$ for party $P_s \in \Partyset$. If received $(\abort, \Ab)$ from $\Sim$, set  $\wy_s = \abort$ for $P_s$, where $s \in \Ab$.
	\end{ccsitemize}
	\begin{description}
		\item {\bf Output: } Send $(\OUTPUT, \wy_s)$ to every $P_s \in \Partyset$.
	\end{description}
\end{systemboxgray}

\vspace{-3mm}

\paragraph{Multi-input multiplication}
\label{appsec:multiinputmultmal}
The malicious variant of multi-input multiplication protocol, at a high level, can be viewed as an amalgamation of the semi-honest multi-input multiplication and the malicious multiplication protocol. For the case of 3-input multiplication, recall that the semi-honest protocol to compute $\shr{\vz}$ given $\shr{\va}, \shr{\vb}$ and $\shr{\vc}$ where $\vz = \va \vb \vc$ requires parties to obtain $\tsgr{\clv{\va \vb}}, \tsgr{\clv{\va \vc}}, \tsgr{\clv{\vb \vc}}$ and $\tsgr{\clv{\va \vb \vc}}$ in the preprocessing phase, which is then used to reconstruct $\mv{\vz}$ in the online phase.

Since parties in $\Evlset$ are required to hold the correct $\tsgr{\cdot}$-sharings before the online phase begins, as in the case of multiplication, the techniques from semi-honest protocol fail in this setting. Hence, our protocol uses $4$ instances of a maliciously secure multiplication protocol in the preprocessing phase, one each to compute $\sqr{\clv{\va \vb}}, \sqr{\clv{\va \vc}}, \sqr{\clv{\vb \vc}}$ and $\sqr{\clv{\va \vb \vc}}$. Each of the $\sqr{\cdot}$-sharing is further converted to $\tsgr{\cdot}$-sharing using $\Picta$ to ensure active participation of only $t+1$ parties in the online phase for reconstruction of $\vz - \vr$ . Further, to detect malicious behaviour during reconstruction of $\vz - \vr$, a verification check similar to the multiplication protocol is performed such that parties $\abort$ if the check fails.

For $4$-input multiplication, parties obtain $\shr{\cdot}$-sharing of $\vz = \va \vb \vc \vd$ using $\vz - \vr = (\mv{\va} - \lv{\va})(\mv{\vb} - \lv{\vb})(\mv{\vc} - \lv{\vc})(\mv{\vd} - \lv{\vd}) - \vr$. The protocol proceeds in a similar manner as the $3$-input case by delegating the computation of product terms to the preprocessing phase.

\paragraph{Dot product}
To generate $\shr{\vz}$ for $\vz = \vec{\vx} \band \vec{\vy}$ where $\vec{\vx}$ and $\vec{\vy}$ are vectors of size $\nf$ and are $\shrd$-shared, the protocol proceeds similar to the semi-honest variant. That is, in the preprocessing phase parties in $\Evlset$ obtain $\tsgr{\cdot}$-shares of $\sigma = \sum_{k=1}^{\nf} \lv{\vx_k} \lv{\vy_k}$ and $\lv{\vx_k}, \lv{\vy_k}$ for $k \in \{1, \ldots, \nf \}$. 
Although the latter two can be computed by parties locally with an invocation of $\Picta$ (\boxref{fig:pirss2tadd}), computation of the former differs significantly from the semi-honest protocol. For this, we extend the ideas from SWIFT~\cite{SWIFT} and generate 
$\sigma$, by executing a maliciously secure dot product protocol $\PiDotPre$ (abstracted as a functionality $\FDotPPre$ in \boxref{fig:FDotPPre}). 
Specifically, parties invoke $\PiDotPre$ on $\sqr{\cdot}$-shares of $\vec{\lv{\vx}} = (\lv{\vx_1}, \ldots, \lv{\vx_{\nf}})$ and $\vec{\lv{\vy}} = (\lv{\vy_1}, \ldots, \lv{\vy_{\nf}})$ to compute 
$\sqr{\sigma}$, followed by an invocation of $\Picta$ to obtain 
$\tsgr{\sigma}$.
Having computed the necessary preprocessing data, the online phase proceeds similarly to the semi-honest protocol where parties reconstruct $\vz - \vr$ via $\Pking$. To account for misbehaviour, the protocol is augmented with a verification phase similar to that in malicious multiplication. 

Observe that a trivial realization of $\FDotPPre$ can be reduced to $\nf$ instances of multiplication. However, we extend the ideas from \cite{BGIN19, BGIN20, SWIFT} to eliminate the vector-size dependency in the preprocessing phase. For this, we instantiate $\PiDotPre$ using a semi-honest dot product protocol~\cite{GoyalS20} whose cost matches that of semi-honest multiplication~\cite{DamgaardN07}, followed by a verification phase where the cost of verification can be amortized away for multiple dot products, thereby resulting in vector-size independent preprocessing. 

Elaborately, the semi-honest dot product~\cite{GoyalS20} protocol takes as input $\sqr{\vec{x}}, \sqr{\vec{y}}$ where $\vec{x}, \vec{y}$ are vectors of size $\nf$, and outputs $\sqr{\vz} = \sqr{\vec{x} \band \vec{y}}$. For this, parties invoke $\Picprod$ on each element in $\vec{x}, \vec{y}$ and sum these up to generate $\sgr{\rho} = \sgr{\vec{x} \band \vec{y}}$. These shares are randomized by summing with $\sgr{\vr}$ (converted from $\sqr{\vr}$) for a random $\vr$, and the sum $\vz + \vr = (\vec{x} \band \vec{y} ) + \vr$ is reconstructed towards $\Pking$, who sends the reconstructed $\vz + \vr$ to parties in $\Evlset$. All parties then non-interactively generate $\sqr{\vz + \vr}$ by setting one of its share as $\vz + \vr$ and the others as 0. Given $\sqr{\vz + \vr}, \sqr{\vr}$, parties can compute $\sqr{\vz} = \sqr{\vz + \vr} - \sqr{\vr}$. Observe that communication of $\sgr{\vz + \vr}$ to $\Pking$ requires $2t$ elements, while communicating $\vz + \vr$ to parties in $\Evlset$ requires $t$ elements, resulting in a matching cost of $3t$ elements as that required for semi-honest multiplication~\cite{DamgaardN07}.

\begin{systemboxgray}{$\FDotPPre$}{Ideal functionality for $\PiDotPre$}{fig:FDotPPre}
	\justify
	$\FDotPPre$ interacts with the parties in $\Partyset$ and the adversary $\Sim$. Let $\T_{i}$ be the set of the honest parties.
	
	\begin{description}
	    \item {\bf Input:} $\FDotPPre$ receives the $\sqr{\cdot}$-shares of the vectors $\vec{\va} = (\va_1, \ldots, \va_{\nf})$ and  $\vec{\vb} = (\vb_1, \ldots, \vb_{\nf})$ from the parties. 
	    $\FDotPPre$ also receives $\sqr{\cdot}$-shares of $\vz = \vec{\va} \band \vec{\vb}$ of corrupt parties from $\Sim$. $\Sim$ is also allowed to send a special command, $(\abort, \Ab)$, which indicates that parties in $\Partyset$ with indices in $\Ab$ should $\abort$.
	\end{description}
	
	\noindent $\FDotPPre$ proceeds as follows.
	\begin{ccsitemize}
		\item[--] Reconstruct $\va_k, \vb_k$ for $k \in \{1, \ldots, \nf \}$ using the shares received from honest parties and compute $\vz = \sum_{k=1}^{\nf} \va_k \cdot \vb_k$.
		\item[--] Compute the $\sqr{\cdot}$-share of $\vz$ to be held by the set of honest parties as the difference between $\vz$ and the sum of $\sqr{\cdot}$-shares of $\vz$ received from corrupt parties. 
		\item[--] Let $\wy_s$ denote the $\sqr{\cdot}$-shares of $\vz$ 
		for party $P_s \in \Partyset$. If received $(\abort, \Ab)$ from $\Sim$, set  $\wy_s = \abort$ for $P_s$, where $s \in \Ab$.
	\end{ccsitemize}
	\begin{description}
		\item {\bf Output: } Send $(\OUTPUT, \wy_s)$ to every $P_s \in \Partyset$.
	\end{description}
\end{systemboxgray}

To verify the correctness of this dot product computation, we extend the verification technique for multiplication in~\cite{BGIN20}, to verify the correctness of dot product. 
We give a high level idea of how the verification of $m$ dot product triples $(\vec{x_1}, \vec{y_1}, \vz_1), \ldots, (\vec{x_m}, \vec{y_m}, \vz_m)$, can be performed. For this, correctness of the dot product triples can be verified by taking a random linear combination,

\begin{align*}
	\beta = \sum_{k =1}^{m} \theta_k \cdot \left(\vz_k - \sum_{j=1}^{\nf} \vx_{kj} \cdot \vy_{kj} \right)
\end{align*}
where $\{\theta_k\}_{k=1}^{m}$ is randomly chosen by all the parties and checking if $\beta = 0$. Given $\sqr{\cdot}$-shares of $\vec{x}_{k}, \vec{y}_{k}, \vz_k$ for $k \in \{1, \ldots, m\}$, 
parties can compute an additive share ($\sgr{\cdot}$-share) of $\beta$ by invoking $\Picprod$. However, since $\sgr{\cdot}$-sharing does not allow for robust reconstruction, the approach is to generate $\sqr{\beta}$ and then robustly reconstruct it and check equality with 0. To generate $\sqr{\beta}$, parties first $\sqr{\cdot}$-share (via $\PiRSh$, \S\ref{appsec:helper}) their $\sgr{\cdot}$-share of
\begin{align*} 
\psi = \sum_{k=1}^{m} \theta_k \cdot \sum_{j=1}^{\nf} \vx_{kj} \cdot \vy_{kj}. 
\end{align*}
Let $\psi^i$ denote the $\sgr{\cdot}$-share of $\psi$ held by $P_i$. %
Given $\sqr{\psi^i}$ for $i \in \{1, \ldots, n\}$, parties can compute
\begin{align*}
	\sqr{\beta} = \sum_{k=1}^{m} \theta_k \cdot \sqr{\vz_k} - \sum_{i=1}^{n} \sqr{\psi^i}
\end{align*}
and reconstruct $\beta$. It is, however, required to ensure that every party $P_i$ $\sqr{\cdot}$-shares the correct $\psi^i$. 
To check the correctness of $\psi^{i}$, parties need to verify if
\begin{align} \label{eq:psizk}
    \psi^i - \sum_{k=1}^{m} \theta_k \left( \sum_{j=1}^{\nf} \vx^{i}_{kj} \cdot \vy^{i}_{kj} \right) = 0
\end{align}
where $\vx^{i}_{kj}, \vy^{i}_{kj}$ denote the $\sqr{\cdot}$-share of $\vx_{kj}, \vy_{kj}$ held by $P_i$. 
Note that following along the lines of $\Piptosh$, parties can generate these $\sqr{\cdot}$-share of $\vx^{i}_{kj}, \vy^{i}_{kj}$ 
from $\sqr{\cdot}$-shares of $\vx_{kj}, \vy_{kj}$, non-interactively. Now, setting $\va_{kj} = \theta_k \vx^{i}_{kj}, \vb_{kj} = \vy^{i}_{kj}, \vc = \psi^i$, for $k \in \{1, \ldots, m \}$, \refeqn{eq:psizk}, can re-written as 
\begin{align} \label{eq:deg2eq}
    &\vc - \sum_{k=1}^{m} \sum_{j=1}^{\nf} \va_{kj} \vb_{kj} = 0 \nonumber \\ 
    \implies &\vc - \sum_{l=1}^{m\nf}  \Tilde{\va}_{l} \Tilde{\vb}_{l} = 0
\end{align}
The correctness of \refeqn{eq:deg2eq} can be verified by invoking $\Fzk$ (see section 3 of~\cite{BGIN20} for the definition and its instantiation), which takes as input $\sqr{\cdot}$-shares of 
$\Tilde{\va}_{l}, \Tilde{\vb}_{l}, \vc$ for $l \in \{1, \ldots, m\nf \}$,
which are known in clear to party $P_i$, and verifies if \refeqn{eq:deg2eq} holds. The protocol realizing $\Fzk$ for all $n$ parties requires communicating $\Order(n \log(m \nf) + n)$ extended ring elements per party. Further, since steps other than $\Fzk$ require sharing and reconstructing one element, it adds a small constant cost, resulting in the communication cost for verifying $m$ dot products for vector size $\nf$ being $\Order(n \log(m \nf) + n)$ extended ring elements per party.

\section{Building blocks}
\label{app:bb}
For completeness, we discuss the building blocks used in our framework. These blocks are known from the literature~\cite{Tetrad, aby2} and we show how these can be extended to $n$-party setting. 


\subsection{Semi-honest building blocks} \label{BB_semi}

\paragraph{Bit to arithmetic }
Given Boolean shares $\shrB{\bitb}$ of bit $\bitb$, protocol $\PiBitA$ generates its arithmetic shares, $\shr{\arval{\bitb}}$ over $\Z{\ell}$ (\boxref{fig:piBit2A}). Here, $\arval{\bitb}$ denotes the arithmetic values of $\bitb$ over the ring $\Z{\ell}$. The idea is to generate a randomized version, $\zeta = \bitb \xor \vr$ of $\bitb$, and then recover arithmetic shares of $\bitb$ as $\bitb = \zeta \xor \vr$ by performing arithmetic equivalent of XOR  
($\vx \xor \vy = \arval{\vx} + \arval{\vy} - 2 \arval{\vx} \arval{\vy}$).

\paragraph{Bit injection}
$\PiBitInj$ facilitates generation of $\shr{\arval{\bitb} \cdot \val}$ given $\shrB{\bitb}, \shr{\val}$ for $\bitb \in \Z{}$ and $\val \in \Z{\ell}$. 
As seen in~\cite{Tetrad}, 
	\begin{align*}
		\arval{\bitb} \val &= \arval{(\mv{\bitb} \xor \lv{\bitb})} (\mv{\val} - \lv{{\val}}) \\
		&= \arval{\mv{\bitb}} \mv{\val} - \arval{\mv{\bitb}} \lv{\val} + (2 \arval{\mv{\bitb}} - 1)(\arval{\lv{\bitb}} \lv{\val} - \mv{\val} \arval{\lv{\bitb}})
	\end{align*}
Given $\tsgr{\cdot}$-shares of $\lv{\val}, \arval{\lv{\bitb}}, \arval{\lv{\bitb}} \lv{\val}, \vr$ and $\shr{\vr}$ for $\vr \in \Z{\ell}$, and the knowledge that $\mv{\val}, \arval{\mv{\bitb}}$ is held by all parties in $\Evlset$, parties can compute $\tsgr{\arval{\bitb} \val + \vr}$, reconstruct it via $\Pking$ and generate $\shr{\arval{\bitb} \val + \vr}$. $\shr{\arval{\bitb} \val}$ can then be computed as $\shr{\arval{\bitb} \val} = \shr{\arval{\bitb} \val + \vr} - \shr{\vr}$. 
To facilitate this, in the preprocessing phase parties generate $\tsgr{\cdot}$-shares of $\vr, \lv{\val}, \arval{\lv{\bitb}}, \arval{\lv{\bitb}} \lv{\val}$, and $\shr{\vr}$.
Here, $\tsgr{\vr}, \tsgr{\lv{\val}}$ and $\shr{\vr}$ are generated as in the preprocessing of multiplication, $\tsgr{\arval{\lv{\bitb}}}$ is generated via $\PiBitA$ followed by $\Pictashr$ (\S\ref{sec:prelim}), and $\tsgr{\arval{\lv{\bitb}} \lv{\val}}$  is generated as in the preprocessing of multiplication. 
%

\begin{protboxgray}{$\PiBitA(\Partyset,\shrB{\vb})$}{Semi-honest: Bit to arithmetic}{fig:piBit2A}
	\justify
	\algoHead{Preprocessing:} \vspace{-2mm}
	\begin{description}
		\item[1.] Invoke $\PiDSBits$ to generate $\shr{\arval{\vr}}, \shrB{\vr}$ for $\vr \in \Z{}$.
		\item[2.] Invoke preprocessing phase of $\PiMult$.
	\end{description}
	\algoHead{Online:} \vspace{-2mm}
	\begin{description}
		\item[1.] Compute $\shrB{\zeta} = \shrB{\bitb} \xor \shrB{\vr}$. 
		\item[2.] $P_i \in \Evlset$ invokes $\PictashrB$ to generate $\tsgrB{\zeta}_i$ and sends $\tsgrB{\zeta}_i$ to $\Pking$, who reconstructs $\zeta$ and generates $\shr{\arval{\zeta}}$.
		\item[3.] Invoke online phase of $\PiMult$ to generate $\shr{\arval{\zeta}\arval{\vr}}$, and compute $\shr{\arval{\bitb}} = \shr{\arval{\zeta}} + \shr{\arval{\vr}} - 2 \shr{\arval{\zeta} \arval{\vr}}$.
	\end{description}
\end{protboxgray}

\vspace{-2mm}
\paragraph{Arithmetic to Boolean sharing}
\label{sec:atob}
Extending the techniques from~\cite{Tetrad}, protocol $\piab$ generates $\shrB{\vx}$ from $\shr{\vx}$ for $\vx \in \Z{\ell}$. 
For this, given arithmetic and Boolean shares of $\vr \in \Z{\ell}$, Boolean shares of $\vx$ are computed as $(\vx+\vr) - \vr$ by evaluating a parallel prefix adder (PPA) circuit~\cite{aby2, MR18}. The PPA circuit inputs two Boolean values ($\vx+\vr$, $-\vr$ in this case) and outputs their sum. 
The protocol appears in \boxref{fig:piab}. Looking ahead, $\piab$ is used in the preprocessing phase in the applications considered. Hence, we rely on PPA circuit from~\cite{MR18} as it provides a good trade-off between rounds and communication as opposed to the circuit from~\cite{aby2} which is optimized to provide a fast online phase at the expense of a higher preprocessing cost (yielding higher total cost than~\cite{MR18}). 
%

\paragraph{Boolean to arithmetic sharing}
This protocol generates $\shr{\vx}$ from $\shrB{\vx}$ where $\vx \in \Z{\ell}$. Inspired from~\cite{Tetrad,SWIFT}, observe that $\vx = \sum_{i=0}^{\ell -1} 2^i \arval{(\vx[i])}$. Thus, we invoke $\PiBitA$ on $\vx[i]$ for $i \in \{0, \ldots, \ell-1 \}$ to generate $\shr{\arval{\vx[i]}}$ followed by locally combining it as per the above equation to generate $\shr{\vx}$. Optimizations in~\cite{Tetrad} carry forward to our setting as well.

\begin{protboxgray}{$\piab(\Partyset,\shr{\vx})$}{Semi-honest: Arithmetic to Boolean}{fig:piab}
	\justify
	\algoHead{Preprocessing:} \vspace{-2mm}
	\begin{description}
		\item[1.]  Invoke $\PiDSBits(\Partyset, 0)$ to generate $\shr{\arval{(\vr[i])}}$ and $\shrB{\vr[i]}$ where $\vr[i] \in \Z{}$ for $i \in \{0, \ldots, \ell-1\}$, and set $\shr{\vr} = \sum_{i=0}^{\ell-1} 2^i \shr{\arval{(\vr[i])}}$.
		\item[2.] Execute the preprocessing phase for the PPA circuit which computes $\shrB{\vx} = \shrB{\vx + \vr} - \shrB{\vr}$.
	\end{description}
	\algoHead{Online:} \vspace{-2mm}
	\begin{description}
		\item[1.] Compute $\shr{\vx + \vr} = \shr{\vx} + \shr{\vr}$
		\item[2.] Parties in $\Evlset$ invoke $\Picta$ on $\shr{\vx + \vr}$ to generate $\tsgr{\vx + \vr}$ and send their share to $\Pking$. 
		\item[3.] $\Pking$ reconstructs and sends $\vx + \vr$ to all parties in $\Evlset$. 
		\item[4.] Invoke $\PiptoshB$ to generate $\shrB{\vx + \vr}$, and execute the online phase of PPA circuit to compute $\shrB{\vx} = \shrB{\vx + \vr} - \shrB{\vr}$.
	\end{description}
\end{protboxgray}

\vspace{-3mm}
\paragraph{Comparison}
To compare $\vx, \vy \in \Z{\ell}$ in FPA, we extend the technique of~\cite{MR18, BLAZE, SWIFT, Trident,Tetrad, aby2}, where checking $\vx < \vy$ is equivalent to checking if the most significant bit ($\MSB$) of $\val = \vx - \vy$ is $1$.
To extract the $\MSB$ from $\shr{\val}$, we rely on $\piBitExt$ which takes as input $\shr{\val}$ and outputs the $\shrB{\cdot}$-share of the $\MSB$ of $\val$, denoted as $\shrB{\MSB(\val)}$. The optimized bit extraction circuit from ~\cite{aby2} is used for computing the $\MSB$ whose inputs are two $\shrB{\cdot}$-shared values and output is the $\shrB{\cdot}$-shared $\MSB$ of the sum of these two inputs. 
Observe that, given $\shr{\val}$, $\val$ can be written as $\val = \mv{\val} - \lv{\val}$, and hence $\shrB{\cdot}$-shares of $\mv{\val}$ and $\lv{\val}$ constitute the two inputs to the circuit. While $\shrB{\mv{\val}}$ can be generated non-interactively by invoking $\PiptoshB$ in the online phase, $\shrB{\lv{\val}}$ is generated by performing an arithmetic to boolean conversion in the preprocessing phase.  Evaluation of bit extraction circuit then gives $\shrB{\MSB(\val)}$.

\begin{table*}[htb!]
	\centering
	\resizebox{.95\textwidth}{!}{
		\begin{NiceTabular}{r| r r| r| r r| r}[notes/para][ tabularnote = $\ell$ - size of ring in bits.] 
			\toprule
			\Block{3-1}{Building\\Block} & \Block[c]{1-3}{Semi-honest} & & &
			\Block[c]{1-3}{Malicious} & &\\
			\cmidrule{2-7}
			& \Block[c]{1-2}{Communication} & & \Block[c]{2-1}{Rounds\\Online} & \Block[c]{1-2}{Communication} & & \Block[c]{2-1}{Rounds\\Online}\\ 
			\cmidrule{2-3}
			\cmidrule{5-6}
			& Preprocessing & Online & & Preprocessing & Online & \\
			\midrule
			
			Sharing & -- & $(t+1) \ell$ & $1$ & -- & $2t \ell$ & $1$  
			\\
			
			Reconstruction\tabularnote{Accounts for reconstruction towards all; $\h = \binom{n}{h}$, $\g = \binom{n-1}{h-1}$.} & -- & $3t \ell$ &  $2$ & -- & $n(\h - \g) \ell$ & $1$ 
			\\
			
			Multiplication & $t \ell$ & $2t \ell$ &$2$ & $3 t \ell$ & $3 t \ell$ & $2$ 
			\\
			
			3-input multiplication & $6t \ell$ & $2t \ell$ &$2$ & $12 t \ell$ & $3 t \ell$ & $2$ 
			\\
			
			4-input multiplication & $15t \ell$ & $2t \ell$ &$2$ & $33 t \ell$ & $3 t \ell$ & $2$ 
			\\
			
			Doubly shared bits & $4t(\ell+2)$ & -- & -- & $6t(\ell+2)$ & -- & -- 
			\\
			
			\Block{}{Multiplication \\with truncation} & $4t(\ell+2)\ell + t \ell$ & $2t \ell$ &$2$ & $3 t \ell + 6t(\ell+2)\ell$ & $3 t \ell$ & $2$ 
			\\
			
			Dot product &  $t \ell$ & $2t \ell$ &$2$ & $3 t \ell$ & $3 t \ell$ & $2$ 
			\\
			
			Bit to arithmetic & $4t(\ell+2) + t\ell$ & $4t \ell$ & $4$ & $6t(\ell+2) + 3t\ell$ & $6 t \ell$ & $4$ 
			\\
			
			Bit injection & $4t(\ell+2) + 6t\ell$ & $2t \ell$ & $2$ & $6t(\ell+2) + 12t\ell$ & $3 t \ell$ & $2$ 
			\\
			
			Arithmetic to Boolean & \makecell[r]{$4t(\ell+2)\ell$\\ $+ t\ell \log_2 \ell$} & $2t \ell (1 + \log_2 \ell)$ & $2 + 2 \log_2 \ell$ & \makecell[r]{$6t(\ell+2)\ell$\\ $+ 3t \ell \log_2 \ell$} & $3t \ell (1 + \log_2 \ell)$ & $2 + 2 \log_2 \ell$ 
			\\
			
			Boolean to arithmetic & $4t(\ell+2)\ell$ & $2t \ell$ & $2$ & $6t(\ell+2)\ell$ & $3t \ell$ & $2$ 
			\\
			
			Comparison\tabularnote{$\sfu_1 = 3t\nf_2+12t\nf_3+33t\nf_4$, $\sfu_2 = \nf_2 + \nf_3 + \nf_4$, $\nf_2=41, \nf_3=27, \nf_4=47$ denote the number of AND gates in the bit extraction circuit of ABY2~\cite{aby2} with $2, 3, 4$ inputs, respectively.} & \makecell[r]{$\sfu_1 + 4t(\ell+2)\ell +$\\$3t \ell \log_2 \ell + 2 t\ell $} & $2t \sfu_2$ & $2 \log_4 \ell $ & \makecell[r]{ $6t(\ell+2)\ell + 6t \ell \log_2 \ell$\\ $ + 3t \ell +  \sfu_1$} & $3t \sfu_2$ & $2 \log_4 \ell$ 
			\\

			\bottomrule
		\end{NiceTabular}
	}
	\caption{\small 
		Communication and round complexity of protocols: semi-honest and malicious \label{tab:costshmal}}
		\vspace{-3mm}
\end{table*}

\begin{protboxgray}{$\PiEq(\Partyset, \shr{\vx}, \shr{\vy})$}{Semi-honest: Equality check protocol}{fig:pieq}
	\justify
	\algoHead{Preprocessing:} \vspace{-2mm}
	\begin{description}
		\item[1.] Perform preprocessing phase of $\piab$ and the preprocessing of $4$-input multiplications.
	\end{description} 
	\algoHead{Online:} \vspace{-2mm}
	\begin{description}
		\item[1.] Compute $\shr{\sfv} = \shr{\vx} - \shr{\vy}$ and invoke $\piab$ to generate $\shrB{\val}$.
		\item[2.] Generate $\shrB{\bar{\sfv}}$ by setting $\mv{\bar{\sfv}} = 1 \xor \mv{\sfv}$ and $\lv{\bar{\sfv}} = \lv{\sfv}$.
		\item[3.] Perform AND of all the bits in $\bar{\sfv}$ following the tree based approach by invoking the online phase of $4$-input multiplication to generate $\shrB{\bitb}$.
	\end{description}
\end{protboxgray}

\vspace{-3mm}
\paragraph{Equality Check}
Given $\shr{\cdot}$-shared $\vx, \vy \in \Z{\ell}$, this protocol outputs a $\shrB{\cdot}$-shared bit, which is set to $1$ if $\vx = \vy$, and $0$ otherwise. 
The approach is to obtain the bit decomposition of $\sfv = \vx -\vy$ by performing $\piab$, and check if all bits of $\sfv$ are $0$. 
For this, parties non-interactively obtain $1$'s complement of the bits of $\sfv$, denoted as $\bar{\sfv}$, by setting the corresponding $\mv{\bar{\sfv}} = 1 \xor \mv{\sfv}$ and $\lv{\bar{\sfv}} = \lv{\sfv}$. Parties proceed to compute an AND of all the bits in $\bar{\sfv}$ following the standard-tree based approach where we use the 4-input multiplication to save on rounds and communication. If $\sfv = 0$, then the AND outputs $1$ else it outputs a $0$. The protocol appears in \boxref{fig:pieq}.

\paragraph{Maxpool / Minpool}
Maxpool allows parties to compute $\shr{\cdot}$-share of the maximum value $\vx_{\max}$ among a vector of values $\vec{x} = (\vx_1, \ldots, \vx_{\nf})$. For this, we proceed along the lines of~\cite{Tetrad}. Observe that the maximum among two values $\vx_i, \vx_j$ can be computed by first using the secure comparison protocol to obtain $\shrB{\vb}$ such that $\vb = 0$ if $\vx_i \geq \vx_j$ and $1$ otherwise. Following this, parties can compute $\vb (\vx_j - \vx_i) +\vx_i$ using the bit injection protocol, to obtain the maximum value as the output. To compute the maximum among a vector of values, parties follow the standard binary tree-based approach where consecutive pairs of values are compared in a level-by-level manner. 
We refer to the resulting protocol as $\Pi_{\max}$. A protocol $\Pi_{\min}$ for minpool can be worked out similarly.

\paragraph{ReLU}
The ReLU function, $\ReLU(\val) = \maxv(0, \val)$, can be written as $\ReLU(\val) = \overline{\bitb} \cdot \val$, where bit $\bitb = 1$ if $\val < 0$ and $0$ otherwise. Here $\overline{\bitb}$ denotes the complement of $\bitb$. 
Given $\shr{\val}$, parties invoke $\piBitExt$ on $\shr{\val}$ to obtain $\shrB{\bitb}$. $\shrB{\cdot}$-sharing of $\overline{\bitb}$ is then computed, non-interactively, by setting $\mv{\overline{\bitb}} = 1 \xor \mv{\bitb}$. Given $\shrB{\overline{\bitb}}$ and $\shr{\val}$, $\ReLU$ is computed using $\PiBitInj$.



\subsection{Malicious blocks} \label{BB_mal}
Note that the malicious variants for the building blocks such as bit to arithmetic, Boolean to arithmetic, and arithmetic to Boolean conversion, bit extraction, secure comparison, secure equality check, ReLU, maxpool, and convolutions, follow along similar lines to that of the semi-honest protocols with the difference that the underlying protocols used are replaced with their maliciously secure variants. Moreover, for steps that involve opening values via $\Pking$, the reconstructed values are sent to all and are accompanied by a verification check similar to the one in the multiplication protocol.


\subsection{Communication cost}
\tabref{costshmal} summarises communication cost and online round complexity of semi-honest and maliciously secure protocols.

\section{Additional Benchmarks}
\label{app:bench}

\subsection{Deep NN and GNN}

\paragraph{NN architecture}\label{app:dnn}

Among NNs, the first, NN-1, is a 3-layered fully connected network with ReLU activation after each layer, as considered in~\cite{MR18, BLAZE, SWIFT}. The second, NN-2, is LeNet~\cite{lenet} architecture, which contains two convolutional layers and two fully connected layers with ReLU activation after each layer. Additionally, for convolutional layers, this is followed by maxpool operation. Finally, NN-3 is VGG16~\cite{vgg16} architecture that comprises 16 layers in total, which includes fully connected, convolutional, ReLU activation, and maxpool layers. Last 2 NNs were considered in~\cite{Falcon}.

\paragraph{GNN architecture}\label{app:gnn}
The goal of spectral-based GNNs~\cite{defferrard, KW17} is to learn a function of signals $\vec{\vx_1}, \ldots, \vec{\vx_m}$ each of length $n$, on a graph $\graph = (\vertices, \edges, \Mat{M})$, where $\vertices$ is the set of $n$ vertices of the graph, $\edges$ is the set of edges and $\Mat{M}$ is the the graph description in terms of an $n \times n$ adjacency matrix. The $j$\textsuperscript{th} component of every signal $\vec{\vx_i}$ corresponds to $j$\textsuperscript{th} node of the graph. Training data is used to compute graph description $\Mat{M}$ , which is common for all signals considered. 

The approximation of graph filters using a truncated expansion in terms of Chebyshev polynomials was put forth in \cite{defferrard}. 
Chebyshev polynomials are recursively defined as follows:
$$
T_k(x) = 
\begin{cases}
	1 & \text{if } k = 0\\
	x & \text{if } k = 1\\
	2x T_{k-1}(x) - T_{k-2}(x) & \text{otherwise}
\end{cases}
$$
and the inference phase for a $n \times c$ signal matrix $\Mat{X}$ with $f$ feature maps, where $c$ represents the dimension of feature vector for each node, with a $K$-localized filter matrix $\Theta_k$  can be performed as $\Mat{Y} = \sum_{k=0}^{K-1} T_k(\tilde{\Mat{L}}) \Mat{X} \Theta_k$. Here, $\tilde{\Mat{L}} = \frac{2}{\lambda_{max}} \cdot \Mat{L} - \Mat{I} \cdot \lambda_{max}$, and $\lambda_{max}$ is the largest eigenvalue of the normalized graph Laplacian $\Mat{L}$, $\Mat{Y}$ is an $n \times f$ dimensional matrix and the trainable parameter for the $k$\textsuperscript{th} layer $\Theta_k$ is of dimension $c \times f$. 

We use the simplified architecture of~\cite{defferrard} given in~\cite{SCSDF20}. The GNN architecture in the latter uses one graph convolution layer without pooling operation instead of the original model with two graph convolution layers, each of which is followed by a pooling operation. Further, $K$ is set to $5$ instead of $25$. This architecture is shown to achieve an accuracy of more than $99\%$ on MNIST classification in~\cite{SCSDF20}. 
%
%
\vspace{-2mm}
\begin{itemize}
	\item[--] Graph convolution layer: 
	\begin{itemize}
		\item[-] \textit{Input:} $T_k(\tilde{\Mat{L}})$ with dimensions $784 \times 784$, $\Theta_k$ with dimensions $1 \times 32$, for $k \in \{0, \ldots, K-1\}$, and $28 \times 28$ image transformed into a vector $\vec{{\vx}}$ of dimension $784$.
		\item[-] \textit{Output:} $\sum_{k=0}^{K-1} T_k(\tilde{\Mat{L}}) \vec{\vx} \Theta_k$ with dimensions $784 \times 32$.
	\end{itemize}
	\item[--] ReLU activation: Calculates the ReLU for each input.
	\item[--] Fully connected layer (FC): with 10 nodes. 
\end{itemize}

Benchmarks for the semi-honest and maliciously secure protocol appear in \tabref{comparison-nnsh}, \tabref{comparison-nnmal}.
%
\begin{table}[htb!]
	\centering
	\resizebox{.46\textwidth}{!}{
		\begin{NiceTabular}{rrr|rrr|rrr}[notes/para][tabularnote = Communication in MB and time in seconds.] 
			\toprule
			\Block{2-1}{NN\\Type} & \Block{2-1}{Ref.} & \Block{2-1}{$n$} & \Block[c]{1-3}{Online} & & & \Block[c]{1-3}{End-to-end} \\
			\cmidrule{4-9}
			& & & Comm
			& Time
			& $\TP$\tabularnote{$\TP$ denotes throughput} 
			& Comm  
			& Time 
			& Cost\tabularnote{monetary cost in USD} 
			\\
			\midrule 
			
			
			\Block{6-1}{NN-1}
			& \Block{3-1}{DN07 }    
			& 5 & 0.16 & \Block[c]{3-1}{18.55\\$\pm$.4} & 211.69 & 3.41 & 21.46 & 0.06 \\
			& & 7 & 0.24 &  & 202.48 & 5.11 & 22.29 & 0.10 \\ 
			& & 9 & 0.33 &  & 202.49 & 6.81 & 22.31 & 0.13 \\
			\cmidrule{2-9}
			& \Block{3-1}{$\this$} 
			& 5 & 0.02 & \Block[c]{3-1}{4.61\\$\pm$.02}  & \Block[c]{3-1}{832.61\\$\pm$.04} & 3.41 & \Block[c]{3-1}{11.09\\$\pm$.02} & 0.02 \\
			& & 7 & 0.03 &  &  & 5.11 &  & 0.03 \\
			& & 9 & 0.05 &  &  & 6.81 &  & 0.04 \\
			\midrule 	
			

			\Block{6-1}{NN-2}
			& \Block{3-1}{DN07}    
			& 5 & 15.58 & 46.20 & 83.12 & 269.23 & 56.44 & 0.21 \\
			& & 7 & 23.39 & 48.39 & 79.35 & 403.85 & 59.60 & 0.32 \\
			& & 9 & 31.18 & 48.40 & 79.35 & 538.47 & 59.61 & 0.42 \\
			\cmidrule{2-9}
			& \Block{3-1}{$\this$} 
			& 5 & 1.92   & \Block[c]{3-1}{11.08\\$\pm$.02} & \Block[c]{3-1}{346.60\\$\pm$.3} & 269.50 & \Block[c]{3-1}{25.34\\$\pm$.03} & 0.10 \\
			& & 7 & 2.88   &  &  & 404.25 &  & 0.14 \\
			& & 9 & 3.84   &  &  & 539.00 &  & 0.18 \\			
			\midrule 				
			
			
			\Block{6-1}{NN-3}
			& \Block{3-1}{DN07}    
			& 5 & 228.07 & 152.95 & 25.11 & 4288.26 & 213.77 & 1.34 \\
			& & 7 & 342.24 & 160.10 & 23.99 & 6432.39 & 227.28 & 1.96 \\
			& & 9 & 456.33 & 160.14 & 23.99 & 8576.52 & 227.33 & 2.56 \\
			\cmidrule{2-9}
			& \Block{3-1}{$\this$} 
			& 5 & 29.70   & \Block[c]{3-1}{36.92\\$\pm$.02}  & \Block[c]{3-1}{104.01\\$\pm$.04} & 4292.06 & \Block[c]{3-1}{104.09\\$\pm$.03} & 0.91 \\
			& & 7 & 44.55   &   &  & 6438.09 &  & 1.08 \\
			& & 9 & 59.40   &   &  & 8584.12 &  & 1.71 \\		
			\midrule 				
			

			\Block{6-1}{GNN} 
			& \Block[c]{3-1}{DN07} 
			& 5 & 20.14 & 7.26 & 528.66 & 956.21  & 17.00 & 0.20 \\
			& & 7 & 30.22 & 7.54 & 509.38 & 1434.31 & 17.54 & 0.29 \\
			& & 9 & 40.29 & 7.56 & 509.38 & 1912.41 & 17.57 & 0.38 \\
			\cmidrule{2-9}
			& \Block[c]{3-1}{$\this$} 
			& 5 & 5.34    & \Block[c]{3-1}{2.16\\$\pm$.02}  & \Block[c]{3-1}{1777.78\\$\pm$.06} & 956.46  & \Block[c]{3-1}{8.97\\$\pm$.02} & 0.17 \\
			& & 7 & 8.00    &   &  & 1434.69 &  & 0.26 \\
			& & 9 & 10.67   &   &  & 1912.92 &  & 0.33 \\
			
			\bottomrule
		\end{NiceTabular}
	}
	\caption{\small 
		Semi-honest: Benchmarks for deep NN and GNN. \label{tab:comparison-nnsh}}
	\vspace{-6mm}
\end{table}

Compared to our semi-honest variant for evaluating NNs, the malicious variant incurs a 2$\times$ higher online communication cost for NN-1 and NN-2. However, this difference closes in with deeper NNs, with the communication being 1.5$\times$ for NN-3. The drop in the difference can be attributed to the one-time cost of verification required in the malicious variant, which gets amortized over deeper circuits. Due to the same reason, in comparison to the semi-honest case, the malicious variant has an overhead of around $1$ second in the online run-time, which in turn reflects in the reduced throughput. Similar to the semi-honest evaluation of NNs, the overall communication is an order of magnitude higher than the online communication due to the cost incurred for truncation during preprocessing. Also, analogous to the trend observed for synthetic circuits, the overhead in overall run-time is approximately $11$ seconds owing to the distributed zero-knowledge proof verification required in the preprocessing phase. For GNN, the trend follows closely to that of NN-3,where malicious variant incurs 1.5$\times$ higher communication than its semi-honest counterpart. 
%

\begin{table}[htb!]
	\centering
	\resizebox{.48\textwidth}{!}{
		\begin{NiceTabular}{rr|rrr|rrr}[notes/para][ tabularnote = Communication in MB and time in seconds.] 
			\toprule
			\Block{2-1}{NN\\Type} & \Block{2-1}{$n$} & \Block[c]{1-3}{Online} && & \Block[c]{1-3}{End-to-end} \\
			\cmidrule{3-8}
			& & Comm
			& Time
			& $\TP$\tabularnote{$\TP$ denotes throughput} 
			& Comm
			& Time
			& Cost\tabularnote{monetary cost in USD} 
			\\
			\midrule 
			
			
			\Block{3-1}{NN-1}    
			& 5 & 0.04  & \Block[c]{3-1}{5.44\\$\pm$.02}  & \Block[c]{3-1}{706.40\\$\pm$.04} & 3.59    & \Block[c]{3-1}{22.96\\$\pm$.02}  & 0.07 \\
			& 7 & 0.06  &   &  & 5.39    &   & 0.10 \\
			& 9 & 0.08  &   &  & 7.20    &   & 0.11 \\
			\midrule 
			
			
			\Block{3-1}{NN-2}     
			& 5 & 2.88  & \Block[c]{3-1}{11.93\\$\pm$.03} & \Block[c]{3-1}{322.63\\$\pm$.2} & 286.18  & \Block[c]{3-1}{37.71\\$\pm$.04}  & 0.15 \\
			& 7 & 4.32  &  &  & 429.28  &   & 0.22 \\
			& 9 & 5.77  &  &  & 571.98  &   & 0.27 \\
			\midrule 
			
			
			\Block{3-1}{NN-3}     
			& 5 & 44.56 & \Block[c]{3-1}{37.91\\$\pm$.02} & \Block[c]{3-1}{101.27\\$\pm$.04} & 4535.95 & 124.54 & 1.04 \\
			& 7 & 66.84 &  &  & 6804.06 & 126.69 & 1.53 \\
			& 9 & 89.12 &  &  & 9066.43 & 129.42 & 1.94 \\
			\midrule

			
			\Block{3-1}{GNN}     
			& 5 & 8.01  & \Block[c]{3-1}{3.02\\$\pm$.02} & 1275.75 & 977.65  & \Block[c]{3-1}{22.39\\$\pm$.03} & 0.23 \\
			& 7 & 12.01 &  & 1267.35 & 1466.49 &  & 0.34 \\
			& 9 & 16.02 &  & 1267.32 & 1954.95 &  & 0.42 \\
			
			\bottomrule
		\end{NiceTabular}
	}
	\vspace{-2mm}
	\caption{\small 
		Malicious: Benchmarks for deep NN and GNN. \label{tab:comparison-nnmal}}
	\vspace{-6mm}
\end{table}

\subsection{Biometric Matching}\label{app:bio}
Given a database of $m$ biometric samples $(\vec{s}_1, \ldots, \vec{s}_m)$ each of size $\nf$, and a user holding its sample $\vec{u}$, the goal of biometric matching is to identify the sample from the database that is ``closest'' to $\vec{u}$. The notion of ``closeness'' can be formalized by various distance metrics, of which Euclidean Distance ($\ED{}{}$) is the most widely used. 
Following the general trend, we reduce our biometric matching problem to that of finding the sample from the database 
which has the least $\ED{}{}$ with the user's sample $\vec{u}$.
We follow \cite{MohasselZ17, aby2} where $\ED{}{}$ between two vectors $\vec{x}, \vec{y}$ of length $\nf$ is given as 
\begin{align} \label{EucDist}
	\ED{\vec{x}}{\vec{y}} = \sum_{i=1}^{i=\nf}(\vx_i-\vy_i)^2 = \vec{z} \band \vec{z}
\end{align}
where $\vec{z} = ((\vx_1-\vy_1),\ldots,(\vx_\nf-\vy_\nf))$. 

\begin{table}[htb!]
	\centering
	\resizebox{.48\textwidth}{!}{
		\begin{NiceTabular}{rrr|rrr|rrr}[notes/para][tabularnote = Communication in MB and time in seconds.] 
			\toprule
			\Block{2-1}{\#seq} & \Block{2-1}{Ref.} & \Block{2-1}{$n$} & \Block[c]{1-3}{Online} & & & \Block[c]{1-3}{End-to-end} \\
			\cmidrule{4-9}
			& & & Comm 
			& Time  
			& $\TP$\tabularnote{$\TP$ denotes throughput} 
			& Comm  
			& Time 
			& Cost\tabularnote{monetary cost in USD} 
			\\
			\midrule 
			

			\Block{9-1}{4096}
			& \Block{3-1}{DN07}    
			& 5 & 2.51 & 66.51 & 57.73 & 27.70 & 79.85 & 0.24 \\
			& & 7 & 3.76 & 69.81 & 55.00 & 41.55 & 83.24 & 0.36 \\
			& & 9 & 5.02 & 69.87 & 54.97 & 55.40 & 83.30 & 0.48 \\
			\cmidrule{2-9}
			& \Block{3-1}{$\this$ (semi)} 
			& 5 & 0.35  & \Block[c]{3-1}{15.07\\$\pm$.02} & \Block[c]{3-1}{254.86\\$\pm$.03} & 27.74 & \Block[c]{3-1}{17.35\\$\pm$.03} & 0.04 \\
			& & 7 & 0.53  &  &  & 41.61 &  & 0.06 \\
			& & 9 & 0.70  &  &  & 55.49 &  & 0.08 \\
			\cmidrule{2-9}
			& \Block{3-1}{$\this$ (mal)}	
			& 5 & 0.53 & \Block[c]{3-1}{15.89\\$\pm$.02} & \Block[c]{3-1}{241.66\\$\pm$.1} & 30.43 & \Block[c]{3-1}{29.27\\$\pm$.02} & 0.09 \\
			& & 7 & 0.80 &  &  & 45.65 &  & 0.13 \\
			& & 9 & 1.07 &  &  & 60.81 &  & 0.16 \\
			\midrule 		
			

			\Block{9-1}{16384}
			& \Block{3-1}{DN07}    
			& 5 & 10.03 & 77.51 & 49.54 & 110.83 & 93.47 & 0.29 \\
			& & 7 & 15.06 & 81.35 & 47.20 & 166.24 & 97.70 & 0.45 \\
			& & 9 & 20.07 & 81.36 & 47.14 & 221.66 & 97.71 & 0.59 \\
			\cmidrule{2-9}
			& \Block{3-1}{$\this$ (semi)} 
			& 5 & 1.41   & \Block[c]{3-1}{17.53\\$\pm$.02}  & \Block[r]{3-1}{219.16\\$\pm$.04} & 110.99  & \Block[c]{3-1}{20.24\\$\pm$.05}  & 0.06 \\
			& & 7 & 2.11   &   &  & 166.49 &   & 0.09 \\
			& & 9 & 2.81   &   &  & 221.99 &   & 0.11 \\
			\cmidrule{2-9}
			& \Block{3-1}{$\this$ (mal)}	
			& 5 & 2.11  & \Block[c]{3-1}{18.35\\$\pm$.02} & \Block[c]{3-1}{209.24\\$\pm$.06} & 121.71 & \Block[c]{3-1}{32.30\\$\pm$.03} & 0.11 \\
			& & 7 & 3.17  &  &  & 182.58 &  & 0.16 \\
			& & 9 & 4.23  &  &  & 243.19 & & 0.20 \\
			
			
			\bottomrule
		\end{NiceTabular}
	}
	\caption{\small 
		Benchmarks for biometric matching for varying number of sequences in the database. \label{tab:comparison-bioshint}}
\end{table}

To achieve this goal of performing biometric matching securely, each $\vec{s}_i$, for all $i \in \{1, \ldots, m\}$ in the database is $\shrd$-shared among the $n$ parties participating in the computation. Specifically, each component $\vec{s}_{i_j}$, for all $j \in \{1, \ldots, \nf\}$ is $\shrd$-shared among all the parties. Similarly, the user also $\shrd$-shares its sample $\vec{u}$. The parties compute a $\shrd$-shared distance vector $\DV$ of size $m$, where the $i$\textsuperscript{th} component corresponds to the $\ED{}{}$ between $\vec{u}$ and $\vec{s}_i$. For this, each party locally obtains $\shr{\vec{z}_i} = \shr{\vec{s}_i} - \shr{\vec{u}}$ and computes $\shr{\DV_i}$ according to Eq.~\ref{EucDist} using the dot product operation. The final step is then to identify the minimum of these $m$ components of $\DV$, which can be performed using the protocol $\Pi_{\min}$ for minpool operation. \tabref{comparison-bioshint} tabulates the benchmarks when the database has 4096 and 16384 samples.

The trend observed for 4096 and 16384 samples adheres to that observed from \tabref{comparison-bioshext} for the case of 1024 and 65536 samples. Specifically, these settings also enjoy a 4.6$\times$ 
improvement in online run-time and throughput and around 83$\%$ saving in the monetary cost compared to DN07. Moreover, similar to the prior case, the malicious variant incurs a minimal overhead of 4$\%$ in the online throughput and 9.5$\%$ in the total communication compared to our semi-honest setting.


\vspace{-3mm}
\subsection{Genome Sequence Matching}\label{app:ssq}
Given a genome sequence as a query, genome matching aims to identify the most similar sequence from a database of sequences. This task is also commonly referred to as Similar Sequence Query (SSQ) identification and has implications in the advancing field of medical science. 
An SSQ algorithm on two sequences $s$ and $q$, requires the computation of Edit Distance (ED), which quantifies how different two sequences are by identifying the minimum number of additions, deletions, and substitutions needed to transform one sequence to the other. 
To compute the ED, we extend the (2-party) protocol from \cite{ST19} which builds on top of the approximation from \cite{AHLR18}, to the $n$-party setting. We describe high-level idea of the approximation algorithm for ED computation for a query sequence $\query$ against a database of sequences $\{\seq_1, \ldots, \seq_\seqn\}$.

\begin{protboxgray}{$\PiED(\Partyset,\shr{\lut_\seq},\shr{\query})$}{Edit distance between query $\query$ and sequence $\seq$ with respect to a database of $\seqn$ sequences and $\seql$ blocks}{fig:piED}
	\justify
	\smallskip 
	\begin{description}
		\item[1.] For $i = 1$ to $\seql$
		\begin{ccsitem}
			\item For $j = 1$ to $\seqn$
			\begin{ccsitem}
				\item[-] Invoke $\PiEq$ on $\shr{\lut_\seq[i][j]}$ and $\shr{\query[i]}$ to generate $\shrB{\vb_j}$.
				\item[-] Invoke $\PiBitA$ on $\shrB{\vb_j}$ and generate $\shr{\vb_j}$.
			\end{ccsitem}
			\item Let $\vec{\vb} = \{\vb_1, \ldots, \vb_\seqn\}$. Compute $\shr{\dist_i} = \PiDotP(\shr{\vec{\vb}}, \shr{\lut_\seq[i][\cdot]} )$. 
		\end{ccsitem}
		\item[2.] Compute $\shr{\dist} = \sum_{i=1}^{\seql} \shr{\dist_i}$.
	\end{description}
	
\end{protboxgray}

The ED approximation algorithm has a non-interactive phase, during which the database owner with the sequences $\seq_1, \ldots, \seq_\seqn$, generates a Look-Up-Table ($\lut$) for each sequence. These $\lut$s are then secret-shared among all the parties. To generate the $\lut$, the sequences in the database are aligned with respect to a common reference genome sequence (using the Wagner-Fischer algorithm \cite{WF74}), and divided into blocks of a fixed, predetermined size. Based on the most frequently occurring block sequences in the database, an $\lut$ is constructed consisting of these block values and their distance from each other. Specifically, for a database of $\seqn$ sequences $\{\seq_1, \ldots, \seq_\seqn\}$, each of length $\seql$ blocks, an $\lut_i$ is constructed for each $\seq_i$. Each $\lut$ has $\seqn$ columns, one corresponding to each $\seq_i$ in the database, and $\seql$ rows, one corresponding to each block of a sequence, where $\lut_\seq[i][j]$ corresponds to the ED between block $i$ of the sequence $\seq$ and $\seq_j$ . This completes the non-interactive phase of the ED approximation algorithm. 

Given the $\lut$s, when a new query $\query$ has to be processed, its ED must be computed from every sequence $\seq$ in the database. For this, similar to the non-interactive phase, the query is first aligned with the reference sequence and broken down into blocks of the same fixed size. Then, the $i$\textsuperscript{th} block from the query is matched with the $i$\textsuperscript{th} block of each sequence in the $\lut$ for a sequence $\seq$. If the block values match, then the precomputed distance is taken as the output for that block; otherwise, the output is taken to be $0$. Finally, the resultant sum of distances for all the blocks is taken to be the approximated ED between $\query$ and the sequence $\seq$. Computing the ED to all such sequences $\seq$ in the database then allows the identification of the most similar sequence for the query using the minpool operation. Algorithms for ED computation between two sequences, and SSQ appear in \boxref{fig:piED}, \boxref{fig:piSSQ}, respectively, where accuracy and correctness follow from~\cite{AHLR18}. 

\begin{protboxgray}{$\PiSSQ(\Partyset,\{\shr{\lut_\seq}\}_{s=1}^{\seqn},\shr{\query})$}{Similar sequence queries}{fig:piSSQ}
	\justify
	\smallskip 
	\begin{description}
	    \item[1.] For $\seq= 1$ to $\seqn$
	    \begin{ccsitem}
		    \item Invoke $\PiED$ on $\shr{\lut_\seq}$ and $\shr{\query}$ to generate $\shr{\dist_\seq}$.
	    \end{ccsitem}
	    \item[2.] Invoke $\Pi_{\min}$ on $\shr{\dist_1}, \ldots, \shr{\dist_\seqn}$ to generate $\shr{\seq_{\min}} \in \{\dist_1, \ldots, \dist_\seqn\}$.
	\end{description}
\end{protboxgray}

Since the generation of $\lut$s happens non-interactively, we only focus on the computation of ED with respect to the new query $\query$, which requires interaction, and benchmark the same.
\tabref{comparison-ssqshext} provides the benchmarks when the database consists of $\seqn = 1000, 4000$ for block length $\seql = 25, 35$ respectively. 
As expected, the observations tabulated for the varying sequence lengths follows closely to the ones for the case of $\seqn = 2000$ and $\seql = 30$ given in \tabref{comparison-ssqshint}.

\begin{table}[htb!]
	\centering
	\resizebox{.48\textwidth}{!}{
		\begin{NiceTabular}{rrr|rrr|rrr}[notes/para][tabularnote = Time in seconds.] 
			\toprule
			\Block{2-1}{$\seqn, \seql$} & \Block{2-1}{Ref.} & \Block{2-1}{$n$} & \Block[c]{1-3}{Online} && & \Block[c]{1-3}{End-to-end} \\
			\cmidrule{4-9}
			& & & Comm\tabularnote{communication in MB} 
			& Time
			& $\TP$\tabularnote{$\TP$ denotes throughput}
			& Comm\tabularnote{communication in GB}  
			& Time
			& Cost\tabularnote{monetary cost in USD} 
			\\
			\midrule 
			
			\Block{9-1}{$\seqn = 1000$ \\ $\seql = 25$}
			& \Block{3-1}{DN07}     
			& 5 & 10.85 & 60.58 & 63.39 & 0.17 & 74.13 & 0.25 \\
			& & 7 & 16.28 & 63.60 & 60.38 & 0.25 & 77.76 & 0.37 \\
			& & 9 & 21.71 & 63.62 & 60.37 & 0.33 & 77.79 & 0.50 \\
			\cmidrule{2-9}
			&\Block{3-1}{$\this$ (semi)}     
			& 5 & 6.42   & \Block[c]{3-1}{16.12\\$\pm$.01} & \Block[r]{3-1}{236.21\\$\pm$.15} & 0.17 & \Block[c]{3-1}{19.08\\$\pm$.02} & 0.07 \\
			& & 7 & 9.63   &  &  & 0.25 &  & 0.10 \\
			& & 9 & 12.84  &  &  & 0.33 &  & 0.13 \\
			\cmidrule{2-9}
			&\Block{3-1}{$\this$ (mal)}     
			& 5 & 9.51  & \Block[c]{3-1}{16.8\\$\pm$.1} & 228.71 & 0.18 & \Block[c]{3-1}{31.21\\$\pm$.08} & 0.12 \\
			& & 7 & 14.14 &  & 228.44 & 0.27 &  & 0.16 \\
			& & 9 & 18.40 &  & 226.82 & 0.36 &  & 0.21 \\
			\midrule 
			
			
			\Block{9-1}{$\seqn = 4000$ \\ $\seql = 35$}
			& \Block{3-1}{DN07}     
			& 5 & 59.87  & 72.08 & 53.27 & 0.92 & 92.04 & 0.43 \\
			& & 7 & 89.86  & 75.65 & 50.76 & 1.39 & 98.90 & 0.64 \\
			& & 9 & 119.81 & 75.67 & 50.72 & 1.85 & 98.93 & 0.84 \\
			\cmidrule{2-9}
			&\Block{3-1}{$\this$ (semi)}     
			& 5 & 35.87   & \Block[c]{3-1}{19.34\\$\pm$.02}  & \Block[r]{3-1}{198.55\\$\pm$.35} & 0.92 & \Block[c]{3-1}{25.97\\$\pm$.03}  & 0.21 \\
			& & 7 & 53.80   &   &   & 1.39 &   & 0.31 \\
			& & 9 & 71.74   &   &   & 1.85 &   & 0.39 \\
			\cmidrule{2-9}
			&\Block{3-1}{$\this$ (mal)}     	
			& 5 & 53.11  & \Block[c]{3-1}{20.11\\$\pm$.06} & \Block[c]{3-1}{190.95\\$\pm$.4} & 0.99 & \Block[c]{3-1}{41.83\\$\pm$.06} & 0.29 \\
			& & 7 & 78.96  &  &  & 1.48 &  & 0.42 \\
			& & 9 & 102.68 &  &  & 1.97 &  & 0.52 \\
			
			\bottomrule
		\end{NiceTabular}
	}
	\vspace{-2mm}
	\caption{\small 
		Benchmarks for genome sequence matching for varying number of sequences ($\seqn$) and block length ($\seql$). \label{tab:comparison-ssqshext}}
		\vspace{-4mm}
\end{table}


\section{Security Proofs}
\label{app:SecurityMPC}
Security proofs are given in the real-world/ideal-world simulation-based paradigm~\cite{Lindell17}. 
Let $\Advsh, \Advmal$ denote the real-world semi-honest, malicious adversary, respectively, corrupting at most $t$ parties in $\Partyset$, denoted by $\Cor$. Let $\simsh, \simmal$ denote the corresponding ideal world semi-honest, malicious adversary, respectively. Security proofs are given in the $\FSETUP, \Ftrgen$-hybrid (and $\Ftrgenmal, \FMulPre, \allowbreak \FDotPPre$-hybrid for malicious setting) model. For modularity, we provide simulation steps for each protocol separately. 

The following is the strategy for simulating the computation of function $f$ (represented by a circuit $\ckt$). The simulator $\simsh$ knows the input and output of the adversary $\Advsh$, and sets the inputs of the honest parties to be $0$. $\simsh$ emulates $\FSETUP$ and gives the respective keys to the $\Advsh$. Knowing all the inputs and randomness, $\simsh$ can compute all the intermediate values for each building block in the clear. Thus, $\simsh$ proceeds to simulate each building block in topological order using the aforementioned values (input and output of $\Advsh$, randomness and intermediate values). 
We provide the simulation steps for each of the sub-protocols separately for modularity. When carried out in the respective order, these steps result in the simulation steps for the entire computation. To distinguish the simulators for various protocols, we use the corresponding protocol name as the subscript of $\simsh$. 

\paragraph{Sharing and Reconstruction}
Simulation for input sharing (\boxref{fig:piInr}) and reconstruction appears in \boxref{fig:ShRSim}, \boxref{fig:RecRSim}, respectively.

\begin{simulatorboxgray}{$\simsh_{\sf Sh}$}{Semi-honest: Simulation for $\PiSh(P_s,\va)$}{fig:ShRSim}
	\justify
	\algoHead{Preprocessing:} 
	\begin{ccsitem}
		\item[--] Emulate $\FSETUP$ and give the respective shared keys to $\Advsh$.
		\item[--] Samples shares of $\lv{\va}$ commonly held with $\Advsh$ using the respective PRF keys while other values are sampled randomly.
	\end{ccsitem}
	\algoHead{Online:} 
	\begin{ccsitem} 
		\item[--] If $P_s \in \Cor$, receive $\mv{\va}$ from $\Advsh$ on behalf of honest parties in $\Evlset$. Else, set $\va = 0$, $\mv{\va} = \lv{\va}$ and sends $\mv{\va}$ to $\Advsh$ on behalf of $P_s$ if there exists a corrupt party in $\Evlset$.
	\end{ccsitem}
\end{simulatorboxgray}

\begin{simulatorboxgray}{$\simsh_{\sf Rec}$}{Semi-honest: Simulation for reconstruction}{fig:RecRSim}
	\justify
	\medskip 
	\begin{ccsitem}
	    \item[--] If $\Pking \in \Cor$, use the output $\va$, and $\mv{\va}$ and $\tsgr{\lv{\va}}_j$ held by corrupt $P_j \in \Cor \cap \Evlset$ to compute the shares $\tsgr{\lv{\va}}_i$ of each honest $P_i \in \Evlset$ such that $\mv{\va} - \va = \sum_{P_i \in \Evlset \backslash \Cor} \tsgr{\lv{\va}}_i + \sum_{P_j \in \Cor \cap \Evlset} \tsgr{\lv{\va}}_j$. Send the shares of the honest parties in $\Evlset$ to $\Advsh$.
		\item[--] If $\Pking$ is honest, send output $\va$ to $\Advsh$ on behalf of $\Pking$. 
	\end{ccsitem}
\end{simulatorboxgray}

\paragraph{Multiplication}
Simulation steps for multiplication (\boxref{fig:piMultr}) are provided in \boxref{fig:MultRSim0}.  
Observe that the adversary's view in the simulation is indistinguishable from its view in the real world since it only receives random value in each step of the protocol. 

\begin{simulatorboxgray}{$\simsh_{\sf mult}$}{Semi-honest: Simulation for $\PiMult$}{fig:MultRSim0}
	\justify
	\algoHead{Preprocessing:}
	\begin{ccsitem}
		\item[--] If $\istr = 0$: Sample $\sqr{\cdot}$-shares of $\vr$ commonly held with $\Advsh$ using the respective shared keys while other values are sampled randomly.
		\item[--] Else if $\istr = 1$: Emulate $\Ftrgen$ to generate $\shr{\vr}, \shr{\trunc{\vr}}$.
		\item[--] On behalf of every honest $P_i \in \Hlpset$, send a random value for $\sgr{\clv{\va \vb} - \vr}_i$ to $\Advsh$ if $\Pking \in \Cor$. 
	\end{ccsitem}
	\algoHead{Online:} 
	\begin{ccsitem} 
		\item[--] If $\Pking \in \Cor$, send random value for $\tsgr{\zeta}_i$ to $\Advsh$ on behalf of the honest $P_i \in \Evlset$. 
		\item[--] If $\Pking \notin \Cor$, send a random $\vz - \vr$ to $\Advsh$, if there exists a corrupt party in $\Evlset$.
	\end{ccsitem}
\end{simulatorboxgray}

\vspace{-3mm}
\paragraph{Other building blocks}
Simulation steps for the remaining building blocks can be obtained analogously by simulating the steps for the respective underlying protocols in their order of invocations.

\vspace{-2mm}
\subsection{Malicious security}

The following is the strategy for simulating the computation of function $f$ (represented by a circuit $\ckt$). The simulator emulates $\FSETUP$ and gives the respective keys to the malicious adversary, $\Advmal$. This is followed by the input sharing phase in which $\simmal$ extracts the input of $\Advmal$, using the known keys, and sets the inputs of the honest parties to be $0$. Knowing all the inputs, $\simmal$ can compute all the intermediate values for each building block in the clear. Further, $\simmal$ invokes $\Fmpcmal$ and obtains the function output on clear. $\simmal$ proceeds to simulate each building block in topological order using the aforementioned values (inputs of $\Advmal$, intermediate values, and function output). 
As before, we provide the simulation steps for each of the sub-protocols separately for modularity. When carried out in the respective order, these steps result in the simulation steps for the entire computation. To distinguish the simulators for various protocols, the corresponding protocol name appears as the subscript of $\simmal$.

\paragraph{Sharing}
Simulation for sharing appears in \boxref{fig:ShRSimMal}.

\begin{simulatorboxgray}{$\simmal_{\sf Sh}$}{Malicious: Simulation for $\PiShMal(P_s,\va)$}{fig:ShRSimMal}
	\justify
	\algoHead{Preprocessing:}
	\begin{ccsitem}
		\item[--] Emulate $\FSETUP$ and give the respective shared keys to $\Advmal$.
		\item[--] Samples shares of $\lv{\va}$ commonly held with $\Advmal$ using the respective PRF keys while other values are sampled randomly.
	\end{ccsitem}
	\algoHead{Online:} 
	\begin{ccsitem} 
		\item[--] For $P_s \in \Cor$, receive $\mv{\va}$ from $\Advmal$ on behalf of honest parties in $\Evlset$, and obtain $\va = \mv{\va} - \lv{\va}$ (since $\simmal$ knows all the PRF keys, it knows $\lv{\va}$). Invoke $\Fmpcmal$ with $(\INPUT, \va)$ on behalf of $\Advmal$. 
		\item[--] On behalf of the honest parties, set its input $\va = 0$, $\mv{\va} = \lv{\va}$ and send $\mv{\va}$ to $\Advmal$ if there exists a corrupt party in $\Evlset$.
	\end{ccsitem}
	\algoHead{Verification:} Send $\Hash(\mv{\va})$ to $\Advmal$ on behalf of the honest parties. 
	If inconsistent $\mv{\va}$s were received with respect to a corrupt party, invoke $\Fmpcmal$ with $(\SIGNAL, \abort)$. 
\end{simulatorboxgray}

\paragraph{Reconstruction}
Simulation for reconstruction (with $\abort$) appears in \boxref{fig:RecRSimMal}.

\begin{simulatorboxgray}{$\simmal_{\sf Rec}$}{Malicious: Simulation for reconstruction}{fig:RecRSimMal}
	\justify
	\smallskip
	\begin{ccsitem}
		\item[--] Use output $\va$ obtained from $\Fmpcmal$, $\mv{\va}$ and $\sqr{\lv{\va}}_j$ held by corrupt $P_j \in \Cor$ to compute the shares $\sqr{\lv{\va}}_i$ of each honest $P_i \in \Evlset$ such that $\mv{\va} - \va = \lv{\va}$. Send the shares of the honest parties in $\Evlset$ to $\Advmal$, and receive shares from $\Advmal$ on behalf of honest parties.
		\item[--] If any honest party $P_i$ is unable to reconstruct the output, add $P_i$ to set $P$. Send $(\SIGNAL, \abort, P)$ to $\Fmpcmal$. 
	\end{ccsitem}
\end{simulatorboxgray}

\paragraph{Multiplication}
Simulation steps for multiplication (\boxref{fig:piMultMal}) are provided in \boxref{fig:MultRSimMal0}.

\begin{simulatorboxgray}{$\simmal_{\sf mult}$}{Malicious: Simulation for $\PiMultMal$}{fig:MultRSimMal0}
	\justify
	\algoHead{Preprocessing:}
	\begin{ccsitem}
		\item[--] If $\istr = 0$: Sample $\sqr{\cdot}$-shares of $\vr$ commonly held with $\Advmal$ using the respective shared keys while other values are sampled randomly.
		\item[--] Else if $\istr = 1$: Emulate $\Ftrgenmal$ to generate $\shr{\vr}, \shr{\trunc{\vr}}$. 
		\item[--] Emulate $\FMulPre$ to generate $\sqr{\cdot}$-shares of $\clv{\va 
		\vb}$.
	\end{ccsitem}
	\algoHead{Online:} 
	\begin{ccsitem} 
		\item[--] If $\Pking \in \Cor$, send random value for $\tsgr{\zeta}_i$ to $\Advmal$ on behalf of the honest $P_i \in \Evlset$. 
		\item[--] If $\Pking \notin \Cor$, send a random $\vz - \vr$ to $\Advmal$.
	\end{ccsitem}
	\algoHead{Verification:} 
	\begin{ccsitem}
	    \item[--] Send $\Hash(\vz_1 - \vr_1 || \ldots || \vz_m - \vr_m)$ with respect to $m$ multiplications, to $\Advmal$ on behalf of the honest parties. If the hash values received from $\Advmal$ are inconsistent, invoke $\Fmpcmal$ with $(\SIGNAL, \abort)$.  
	    \item[--] If $\Advmal$ has sent incorrect $\vz - \vr$ for any multiplication ($\simmal$ can detect this since it knows all inputs and randomness that should be used by $\Advmal$), generate random shares for $\Omega$ and simulate reconstruction steps of $\simmal_{\sf Rec}$. Invoke $\Fmpcmal$ with $(\SIGNAL, \abort)$.
	    \item[--] Else, if $\Advmal$ has behaved honestly throughout, simulate reconstruction of $\Omega = 0$ using steps from $\simmal_{\sf Rec}$. Invoke $\Fmpcmal$ with $(\SIGNAL, \abort)$. 
	\end{ccsitem}
\end{simulatorboxgray}
\vspace{-2mm}
Observe that since $\Advmal$ sees random shares in both the real-world protocol and in the simulation, indistinguishability of the simulation follows.

\paragraph{Other building blocks}

Simulations for the remaining building blocks can be obtained analogously and using the steps for the underlying protocols.

\end{document}